\documentclass[11pt]{article}
\usepackage{amsfonts, amsbsy, amssymb, amsmath, graphicx, float, subfigure, natbib, color}

\usepackage{geometry}                % See geometry.pdf to learn the layout options. There are lots.
\usepackage{graphicx}
\usepackage{amssymb}
\usepackage{epstopdf}
\usepackage{ifpdf}
\usepackage{url}

\begin{document}

\bibliographystyle{natbib}

\title{Lagrangian Descriptors: A Method for Revealing Phase Space Structures of
General Time Dependent Dynamical Systems}
\author{Ana M. Mancho$^1$, Stephen Wiggins$^2$, Jezabel Curbelo$^{1,3}$, Carolina Mendoza$^{1,4}$\\
$^1$Instituto de Ciencias Matem\'aticas, CSIC-UAM-UC3M-UCM, \\ C/ Nicol\'as Cabrera 15, Campus Cantoblanco UAM, 28049 Madrid, Spain\\
$^2$School of Mathematics, University of Bristol, \\Bristol BS8 1TW, United Kingdom\\
$^3$Departamento de Matem\'aticas, Facultad de Ciencias, \\Universidad Aut\'onoma de Madrid, 28049 Madrid, Spain\\
$^4$ETSI Navales, U. Polit\'ecnica de Madrid, \\
Av. Arco de la Victoria 4, 28040 Madrid, Spain}
\maketitle
\begin{abstract}
In this paper we develop new techniques for revealing geometrical structures in phase space that are valid  for aperiodically time dependent dynamical systems, which we refer to as {\em Lagrangian descriptors}. These quantities are based on
the integration, for a finite time, along trajectories of  an intrinsic bounded,  positive geometrical and/or physical property of the trajectory itself.
We discuss a general methodology for constructing Lagrangian descriptors,  and we discuss  a ``heuristic argument'' that explains why this method is successful for revealing geometrical structures in the phase space of a dynamical system. We support this argument by explicit calculations on a benchmark problem having a hyperbolic fixed point with stable and unstable manifolds that are known analytically.  Several other benchmark examples are considered that allow us the assess the performance of Lagrangian descriptors in revealing invariant tori and regions of shear.
Throughout the paper ``side-by-side'' comparisons of the performance of Lagrangian descriptors with both finite time Lyapunov exponents (FTLEs) and finite time averages of certain components of the vector field  (`time averages'') are carried out and discussed. In all cases Lagrangian descriptors are shown to be both more accurate and computationally efficient than these methods. We also perform computations for an explicitly  three dimensional, aperiodically time-dependent vector field  and an aperiodically time dependent vector field defined as a data set. Comparisons with FTLEs and time averages for these examples are also carried out, with similar conclusions as for the benchmark examples.
\end{abstract}

{\bf Keywords:} Invariant manifolds, nonautonomous systems, aperiodically time-dependent vector fields, transport barriers.

\section{Introduction}

Since the insights of \cite{poincare} geometrical  structures in phase space  have played a central role in characterizing  the global behaviours of dynamical systems. This point of view gives insight into the evolution of qualitatively distinct  classes of trajectories without having to explicitly integrate and compare the behaviour of trajectories. This point of view has also been shown to provide much insight into fluid transport and mixing after the recognition that the equations for incompressible  fluid particle motion (in the absence of molecular diffusion) are formally equivalent to  Hamilton's equations where, in the fluid mechanics context, the stream function plays the  role of the Hamiltonian function and the  physical space  in which the fluid moves plays the role of the phase space of the corresponding Hamilton's equations \cite{aref1}. This framework for studying fluid transport and mixing is often referred to as the `dynamical systems approach to Lagrangian transport'' since the focus is on understanding the ``organising structures'' in phase space for fluid particle trajectories. References  describing this approach for general  fluid mechanics are \cite{ottbook, wo}, and references specific to geophysical fluid dynamics are \cite{jw2, warfm, physrep, samwig}.

In this paper we further develop a technique  for revealing phase space structure that was proposed and used in \cite{prl, nlpg2, jfm, amism11}.
This method is based on the computation of arc length of particle trajectories and it is extended by considering  the integration of a bounded, positive quantity that is an intrinsic geometrical and/or physical property of the dynamical system along trajectories of the dynamical system, for a finite time. Techniques based on the integration of properties along trajectories have been used in the past. \cite{rypina} have proposed methods based on the idea of {\em complexity of isolated trajectories}. Recently   in the computer graphics community,  \cite{acar12} have also used the time-normalized arc length   of trajectories for human action recognition. Their work is inspired on the {\em line integral convolution}  discussed by \cite{shi08, cabral93} that advect conserved  physical fields along trajectories. The advection of 
conserved physical fields  has been also used in the atmospheric science community since the early 90s.   \cite{oneill94, sutton94} proposed 
 the {\em reverse domain filling}  method  for visualizing transport structures in geophysical flows. Recently in the context of realistic oceanic flows the works by \cite{prants11a, prants11b} have highlighted Lagrangian features  by computing the time of exit of particles from a given region or the number of changes of the sign of zonal and meridional velocities.  \cite{huhn12} and \cite{prants13a} have addressed similar  purposes  by measuring the absolute displacements of particles and \cite{prants13b} by counting the number of cyclonic and anticyclonic rotations of particles. 

The techniques explained in the current article  have been shown to be effective for revealing phase space structures  for aperiodically time dependent flows. In Section \ref{secLd} we describe the general methodology for constructing Lagrangian descriptors, and we emphasise the role of the norm in quantifying the integral of the chosen positive quantity along trajectories. Section \ref{sec:hyp} describes the performance of Lagrangian descriptors in hyperbolic regions, and Section \ref{sec:linsad} discusses the benchmark problem of a linear, autonomous vector field in the plane having a hyperbolic saddle point at the origin. This is an excellent example,  not only for discussing the issues associate with the performance of Lagrangian descriptors, but also for comparing their performance with finite time Lyapunov exponents (FTLEs) and time averages of particular components  of the vector field along trajectories since the stable and unstable manifolds of the hyperbolic point, as well as all trajectories, are known exactly. In this example we are able to illustrate the effect of the integration time and make precise the idea that ``singular contours of the Lagrangian descriptors correspond to invariant manifolds'', both numerically, and analytically in Section \ref{sec:asymp}.  This example also illustrates why the $L^\gamma$ norms, $\gamma >1$ are not successful in revealing ''singular contours'' in the same way as the $L^\gamma$ norms, $\gamma \le 1$.  In Section \ref{sec:ftle_ed} we show that for this example FTLEs completely fail to reveal any phase space. Since the system is linear  almost all FTLEs, {\em for any time}, are equal for every trajectory. A similar failure for finite time averages of a component of the vector field along trajectories is reported in Section \ref{sec:ergod_decomp}.
In Section \ref{sec:ell} we consider another benchmark example --a linear vector field having an elliptic equilibrium point at the origin.  As in the example of the linear saddle point, since this system is linear almost all FTLEs are zero, for any time. Hence the FTLE field fails to reveal any phase space structure. The Lagrangian descriptors, on the other hand, reveal the expected phase space structure, consistent with the trajectories. Since this example is linear elliptic point, effects of shear are not considered. In Section \ref{sec:twistless} we consider an integrable example  in action-angle coordinates (hence all FTLEs are zero, for any time) that not only contain regions of strong shear, but also invariant tori where the ``twist condition'' breaks down (i.e. ``twistless tori''). We show that Lagrangian descriptors provide signatures of each of these features.  An advantage of the linear saddle point and linear elliptic point examples is that they allow  us to ``isolate'' hyperbolic and elliptic behaviour. In Section \ref{sec:duffing} we consider the forced  Duffing equation for three different types of time dependency: 1) no time dependency (forcing), i.e. the integrable case, 2) periodic time dependence, and 3) aperiodic time dependence (of a form that we define).  In this example hyperbolic and elliptic behaviour are ``intermingled'' in a complicated manner (that we describe), and this affords us an ideal setting to compare Lagrangian descriptors, FTLEs, and averages of velocity components along trajectories ``side-by-side'' in different regions of the phase space of the forced Duffing oscillator. In Section \ref{sec:3d} we consider a three dimensional vector field from fluid mechanics --the Hill's spherical vortex subjected to a time-aperiodic perturbation-that was studied in \cite{bw09}. For this example we show that Lagrangian descriptors provide an efficient way of discovering and visualising detailed geometrical structures in the flow. A comparison is made with FTLEs and averages of components of the vector field along trajectories, and  it is shown that these techniques may introduce ``artefacts'' that obscure the true geometric structures. In Section \ref{sec:veldata} we consider velocity fields defined as data sets. In particular, we consider a wind driven three-layer quasigeostrophic simulation in a rectangular domain in a ``double gyre'' configuration exhibiting aperiodic time dependence. This velocity field has been the subject of earlier studies (\cite{coulliette, msw04, physrep}) and therefore allows us to directly compare the performance of Lagrangian descriptors and FTLEs  with distinguished hyperbolic trajectories (DHTs) and their stable and unstable manifolds. Additionally, we compare the convergence time of Lagrangian descriptors with FTLEs to the stable and unstable subspaces of a given DHT, where we see that the Lagrangian descriptors converge to these structures more quickly. In Section \ref{sec:comp_perf} we compare  the computational effort required for Lagrangian descriptor computations with FTLE computations, and in Section \ref{sec:concl} we summarise our conclusions. In three appendices we provide some additional technical details.

\section{Lagrangian Descriptors: Definitions, Heuristics, and Analytical Results}
\label{secLd}

The concept of a Lagrangian descriptor was first  introduced in \cite{prl}. There it was shown   that this  tool
is able  to provide  a global dynamical picture of  the geometric structures for arbitrary time dependent
flows.  In particular, Lagrangian descriptors are able to detect the main ``organising centres'' in the flow --
hyperbolic trajectories and their stable and unstable manifolds and elliptic regions.  In this section we describe precisely what we mean by the term ``Lagrangian descriptor''.  We then provide a  heuristic justification for why they ``work''.  We  build on this by then considering several benchmark examples where the phase space structure and stability  characteristics are known analytically. This allows a deeper investigation of the  heuristic justification, as well as a comparison with other commonly used method for extracting phase space structures, such as finite time Lyapunov exponents (FTLEs).

We consider a general time-dependent vector field on $\mathbb{R}^n$:

\begin{eqnarray}
\frac{d{\bf x}}{dt} &=& {\bf v}({\bf x},t), \label{adv1}  \, \,{\bf x} \in \, \mathbb{R}^n, \, t \in \, \mathbb{R} \label{gds}
\end{eqnarray}

\noindent
where we assume  that ${\bf v}({\bf x},t)$ is  $C^r$ ($r \geq 1$) in ${\bf x}$ and continuous in $t$. This is a sufficient condition for the existence of  unique solutions that also allows for linearization. The time interval on which a solution exists is also an issue. However, we will proceed by assuming that the solutions exist for a sufficient time so that the integral expressions we derive have validity, and then we will verify the validity in specific examples.

We begin by describing the Lagrangian descriptor first described in \cite{prl}. In that reference it was denoted by $M$, but henceforth we will denote it by $M_1$ since we will introduce additional Lagrangian descriptors. $M_1$ is the Euclidean arc length of the curve in phase space  defined by a trajectory of \eqref{gds} that starts at ${\bf x^*}$ at time $t=t^*$  for the time interval $[t^*-\tau, t^* +\tau]$, \emph{i.e.}

\begin{equation}
M_1({\bf x^*}, t^*)_{ {\bf v}, \tau}=   \int^{t^*+\tau}_{t^*-\tau} \! \!\!\sqrt{\sum_{i=1}^n \left(\frac{d x_i(t)}{dt}\right)^2 }dt =   \int^{t^*+\tau}_{t^*-\tau} \|{\bf v}({\bf x}(t),t)\|dt,  \label{def:Mgen}
\end{equation}

\noindent
where  $(x_1(t) ,x_2(t),...,x_n(t))$ denote the components  of the  trajectory ${\bf x}(t)$ in $\mathbb{R}^n$ . Clearly, $M_1$ depends on  the initial point $\bf x^*$ and the time interval $[t^*-\tau, t^* +\tau]$ (and the vector field $\bf v$).

 \paragraph{A Heuristic Argument.}  At this point it is insightful to give a heuristic argument as to why $M_1$ is useful for revealing  the geometric structures in the phase space of \eqref{gds}.  $M_1$ measures the arc-length of trajectories on a time interval $(t^*-\tau, t^*+\tau)$. Trajectories with ``close'' initial conditions that remain ``close'' as they evolve on the time interval $(t^*-\tau, t^*+\tau)$ are expected to have  arc-lengths  that are ``close''.
 However, at the boundaries between  regions  comprising trajectories with  qualitatively different behaviour in the course of their time evolution over the  time interval $(t^*-\tau, t^*+\tau)$, we expect the arc-lengths of trajectories starting on either side of the boundary to not be ``close'' on the time interval. (Of course, this will depend on $\tau$, and the dependence on $\tau$ is an issue that we will shortly address.) Hence, these boundaries are denoted by an ``abrupt change'' in $M_1$, where ``abrupt change'' means that the derivative of $M_1$ transverse to these boundaries is discontinuous on the boundaries. This will be more precisely  discussed and defined in Section \ref{sec:hyp}. Such boundaries  correspond to the stable and unstable manifolds of hyperbolic trajectories, if we are in a  hyperbolic region, of the phase space.  Hence, in this way  $M_1$ detects stable and unstable manifolds of hyperbolic trajectories.

\paragraph{A General Construction for Lagrangian Descriptors.}
The arc length of a segment of a trajectory, defined by \eqref{def:Mgen},  is obtained by integrating the modulus of the velocity ($\|{\bf v}\|$) along a trajectory for a time that parametrizes the curve. However, consistent with the heuristic  argument given above, it  is reasonable to expect this methodology to be successful for other positive scalar valued functions that embody an intrinsic physical or geometrical property of trajectories that are integrated along trajectories over the time interval $(t^*-\tau, t^*+\tau)$.  For example,  integration along trajectories  of other
positive scalar valued quantities could have been considered,  such as  the modulus of acceleration ($\|{\bf a}\|$), the modulus of the time derivative of acceleration ($\|d{\bf a}/dt\|$), or the modulus of combinations of ${\bf v}$, ${\bf a}$ or $d{\bf a}/dt$, with the only restriction being that the integrals of these quantities along trajectories are well-defined. The heuristic argument would still apply and indicate that at the boundaries of   regions  comprising trajectories with  qualitatively different  time evolutions the  accumulated value of the chosen positive quantity will ``change abruptly'' in the sense that the derivative transverse to the boundary of the corresponding Lagrangian descriptor  will be discontinuous on the boundary, and examples of such regions would be those that are  separated by the stable and unstable manifolds of hyperbolic trajectories. For this reason abrupt changes in  a Lagrangian descriptor, which   are manifested as discontinuities on the boundary of the derivative of the Lagrangian descriptor transverse to the boundary, are seen as singular curves in contour plots of the  Lagrangian descriptor, and these are expected to be related to  invariant manifolds.  We will see specific examples that make these ideas precise in the remainder of this section.

 However, for now we note that based on this  general reasoning a  method for constructing families of Lagrangian descriptors
 for general time dependent flows would  be as follows.  Let $|\mathcal{F}({\bf x})|$ denote the  bounded, positive intrinsic physical or geometrical property of the velocity field that is of interest. Consider a trajectory ${\bf x}(t)$ satisfying ${\bf x}(t^*) = {\bf x}^*$ and defined on the time interval
 $(t^*-\tau, t^*+\tau)$ (where $\tau$ is  ``appropriately chosen'', as we will shortly discuss). Then $|\mathcal{F}({\bf x }(t) )|$  is a scalar valued function of $t$ that depends parametrically on ${\bf x}^*$ and $t^*$. Therefore for $\gamma\leq1$ we can consider its $L^\gamma$ norm:

\begin{equation}
M({\bf x^*},t^*)_{ {\bf v},\tau}=  \left( \int^{t^*+\tau}_{t^*-\tau} \vert \mathcal{F}({\bf x}(t)) \vert^\gamma \, dt \right).
\label{def:M}
\end{equation}

\noindent
Alternatively for $\gamma> 1$ the $L^\gamma$ norm is given by:

\begin{equation}
M({\bf x^*},t^*)_{ {\bf v},\tau}=  \left( \int^{t^*+\tau}_{t^*-\tau} \vert \mathcal{F}({\bf x}(t)) \vert^\gamma \, dt \right)^{\frac{1}{\gamma}}.
\label{def:Mgam}
\end{equation}

\noindent
For example, for     \eqref{def:Mgen} we considered the $L^1$ norm of the magnitude of the velocity evaluated on trajectories defined on the time interval  $(t^*-\tau, t^*+\tau)$.  Other choices are possible. For example, we could consider the $L^\gamma$ norm of the magnitude of the acceleration. For this case, for $\gamma=1$, we denote the resulting $M$ function by $M_2({\bf x^*}, t^*)_{ {\bf v}, \tau}$. We could also consider the $L^\gamma$ norm of the magnitude of the velocity or the acceleration.  In this case we denote the resulting $M$ function by $M_3({\bf x^*}, t^*)_{ {\bf v}, \tau}$, and we will consider $M_3$ for both $\gamma=2$ and $\gamma = \frac{1}{2}$.   Another possibility  is to consider the $L^1$ norm of the magnitude of the time derivative of the acceleration, and in this case we denote the $M$ function by $M_4({\bf x^*}, t^*)_{ {\bf v}, \tau}$.
We will explore more deeply the advantages  and disadvantages of these different choices in  the remainder of this section. However, we remark that the $L^\gamma$ norms, $\gamma >1$, are not useful as Lagrangian descriptors in the sense that they do not reveal significant underlying structures in the flow. We will explore the reasons for this in Section \ref{sec:linsad}.

Finally, the curvature is an intrinsic property of curves. It is a positive quantity that combines both velocity and acceleration
(\cite{kreyszig91}):
\begin{equation}
\kappa=\frac{\sqrt{({\bf v}\cdot {\bf v})({\bf a}\cdot {\bf a})-({\bf v}\cdot {\bf a})^2 } }{({\bf v}\cdot {\bf v})^{3/2}}
\label{kappa}
\end{equation}

\noindent
The value of the curvature  ranges from zero to infinity--the curvature of straight lines is zero and the curvature of fixed points is infinity.
We define a Lagrangian descriptor based on curvature as follows:

\begin{equation}
|\mathcal{F}_5({\bf x}(t))|=    \frac{1}{|\kappa({\bf x}(t))|+a}   \label{def:M5}
\end{equation}

\noindent
where singularities are avoided by introducing $a>0$ in the denominator. We have numerically verified that good choices for `$a$' lie in the
interval  $5>a>1$. This interval is determined considering that if `$a$' is very large then Eq. (\ref{def:M5}) tends to be independent of $\kappa({\bf x}(t))$. On the other hand if `$a$' is very small the expression (\ref{def:M5}) will present very sharp profiles for zero curvatures near straight lines. In the examples to follow we will present results for  $a=1$.

A key property of all of the Lagrangian descriptors is that they are  quantities  that ``accumulate'' along a trajectory, \emph{i.e.} they are  integrals of a positive quantity along a trajectory. The heuristic  argument would not be valid for the integral of a quantity that changes sign along a trajectory. We will address this issue explicitly in Section \ref{sec:ftle_ed}.
In Table \ref{LD_table} we  summarize the different Lagrangian descriptors that we have introduced in this section
\begin{table}[tpb]
\begin{center}
\begin{tabular}{|c|c|c|} \hline
Lagrangian Descriptor & Integrand &  Norm \\ \hline\hline
$M_1$  &  magnitude of velocity  &  $L^1$     \\
 \hline
$M_2$ &  magnitude of acceleration & $L^1$      \\
 \hline
$M_3$ & magnitude of acceleration or velocity  &   $L^{\frac{1}{2}}$ or $L^2$  \\
 \hline
$M_4$ &  magnitude of the time derivative of the acceleration &   $L^1$   \\
 \hline
$M_5$ & positive quantity related to curvature ( Eq. \eqref{def:M5})  &  $L^1$   \\
\hline
\end{tabular}
\caption{The left hand column denotes the Lagrangian descriptors used, the middle column describes the positive quantities used as the integrand for the Lagrangian descriptors, the right hand column describes the norm that is used. }
\end{center}
\label{LD_table}
\end{table}
and  Appendix A describes in detail issues associated with the numerical computation of Lagrangian descriptors.

\medskip
\paragraph{Hyperbolic and Elliptic Regions.} The phrases ``hyperbolic regions'' and ``elliptic regions'' occur throughout the literature. Here we want to consider their  meaning more carefully.

Traditionally, the words ``elliptic'' and ``hyperbolic'' have referred to the stability characteristics of {\em individual trajectories}.  Trajectories are said to be hyperbolic if none of its Lyapunov exponents are zero (with the exception of the Lyapunov exponent tangent to the direction of motion of the trajectory, \emph{i.e.} the Lyapunov exponent in the direction of the velocity). Exponential dichotomies are also, equivalently, used to characterise hyperbolicity, see
\cite{c,dieci1,dieci3}. These characterisations of hyperbolicity are independent of the nature of the time dependence.  The definition of ``elliptic'' for general time dependent trajectories has not received the same level of attention, at least in the applied areas. Certainly, the notion of ``elliptic stability type'' for  time periodic trajectories is well established. A time periodic trajectory is elliptic if all of its Floquet multipliers lie on the unit circle (there are some issues when a multiplier is $1$ or $-1$, but that level of technicality is not important for our discussion).

However, there is a significant body of literature that does address the notion of ``elliptic stability'' for trajectories in vector fields having arbitrary (e.g. aperiodic) time dependence. It would be very interesting to explore the implications of these ideas in applications, in much the same way that Lyapunov exponents have been explored and exploited in applications. However, here we are considering what is meant by the phrases ``hyperbolic region'' and ``elliptic region''.

With the notion of hyperbolic and elliptic stability of individual trajectories in hand, we can now address the question of what is meant by the phrases ``hyperbolic region'' and ``elliptic region''. Initially, one might guess that these are regions (e.g. open sets) having the property that all trajectories  in such regions are either elliptic or hyperbolic. However, there is an issue with this particular definition that requires careful consideration. Namely, traditionally notions of ``elliptic'' or ``hyperbolic'' stability type are infinite time characterisations of  trajectories. Hence, in order to establish that all the trajectories in a region are of a particular stability type, one would need the trajectories to remain in that region for all time. In other words, one would need the region to be invariant (at least in forward time). This requirement could  (naturally) be met for linear or integrable Hamiltonian systems, but not for general dynamical systems. There is an additional complication that occurs for ``generic'' dynamical systems (at least for dynamical systems defined by time periodic velocity fields). General theoretical results imply that it is unreasonable to assume that we can ``cleanly'' separate dynamical behaviour into ``hyperbolic'' and ``elliptic''. In particular, conservative generalisations of the ``Newhouse phenomena'' \cite{newhouse77, duarte99, gon00}  imply that tangencies between the stable and unstable manifolds of hyperbolic trajectories are ``persistent'' and will give rise to elliptic ``islands of stability''. If one ``perturbs away'' such tangencies, then new ones will be created elsewhere. We emphasis that, to date, such results have only been proven in the time-periodic setting.

Despite ambiguities, it has nevertheless proved useful in  the fluid mechanical situation to attempt to partition the flow into ``elliptic'' and ``hyperbolic'' regions.   Early attempts in the fluids community to achieve such a decomposition of the flow led to the Okubo-Weiss criterion (\cite{okubo70,weiss91}), which is essentially Eulerian in nature.  Efforts to generalize the idea to a Lagrangian setting appeared in \cite{haller01} (but see also \cite{Duc08,brawigg}). This is not a problem that we will explicitly address in this paper.  Nevertheless, we will use the phrases ``hyperbolic region'' and ``elliptic region'', but for most usages we will be considering linear and integrable systems, where the difficulties mentioned above are not an issue.

\subsection{Lagrangian Descriptors in  Hyperbolic Regions}
\label{sec:hyp}

Hyperbolic regions  contain the ``seeds'' of change, uncertainty, and chaos in flows.  In particular, they contain hyperbolic trajectories and their stable and unstable manifolds.  In this section we will explore the performance of the Lagrangian descriptors defined in the previous section in familiar examples that exhibit hyperbolic behaviour. We will also relate their performance to more familiar diagnostics for ``diagnosing'' phase space structure, such as  finite time Lyapunov exponents (FTLEs) and the ergodic decomposition.

\subsubsection{The Linear Saddle Point}
\label{sec:linsad}

The first example that we analyze is the linear saddle. Initially, one might consider this example as ``too trivial''. However, we will show that it  yields some interesting insights. Moreover, it lends itself to either exact, or accurate approximate, solutions for several interesting quantities in Section \ref{sec:ftle_ed}.

The velocity field  that we consider is given by:

\begin{eqnarray}
\dot{x} & = & \lambda x, \nonumber\\
\dot{y} & = & -\lambda y, \quad \lambda >0,
\label{linex}
\end{eqnarray}

\noindent
and the flow generated by this velocity field is given by:

\begin{eqnarray}
x(t, x_0)  & = & x_0 e^{\lambda t},  \nonumber \\
y(t, y_0)  & = & y_0 e^{-\lambda t}, \quad \lambda >0,
\label{linex_flow}
\end{eqnarray}

\noindent
The origin, $(x, y) =(0, 0)$ is a hyperbolic fixed point with stable and unstable manifolds given by:

\begin{eqnarray}
W^s \left(0,0 \right) & = & \{ (x, y)  \in \mathbb{R}^2 \vert x=0, \, y \ne 0 \},  \label{linex_stab}\\
W^u \left(0,0 \right) & = & \{ (x, y)  \in \mathbb{R}^2 \vert y=0, \, x \ne 0 \}. \label{linex_unstab}
\end{eqnarray}

\begin{figure}
a)\includegraphics[width=7cm]{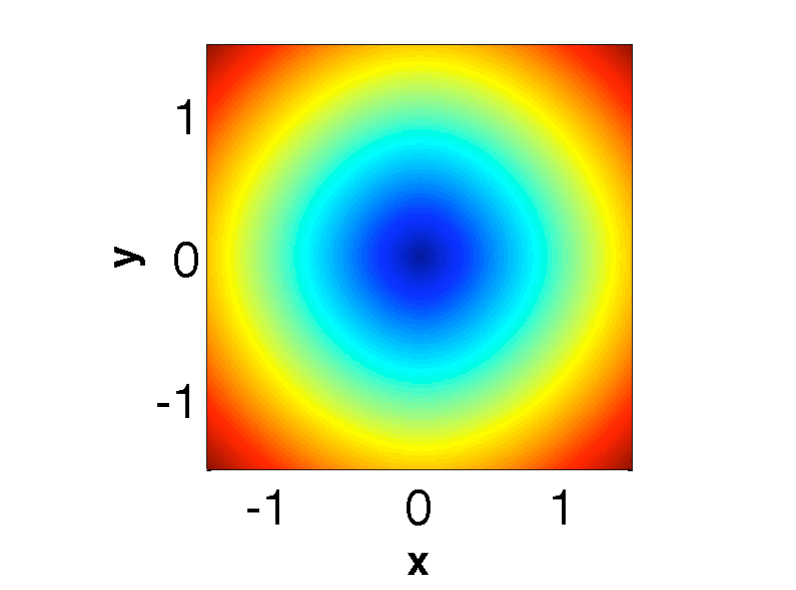}
b)\includegraphics[width=7cm]{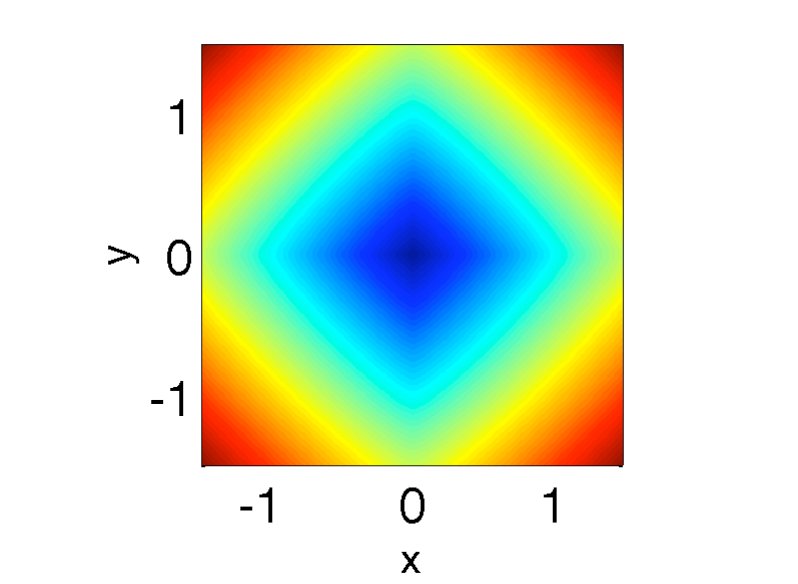}
c)\includegraphics[width=7cm]{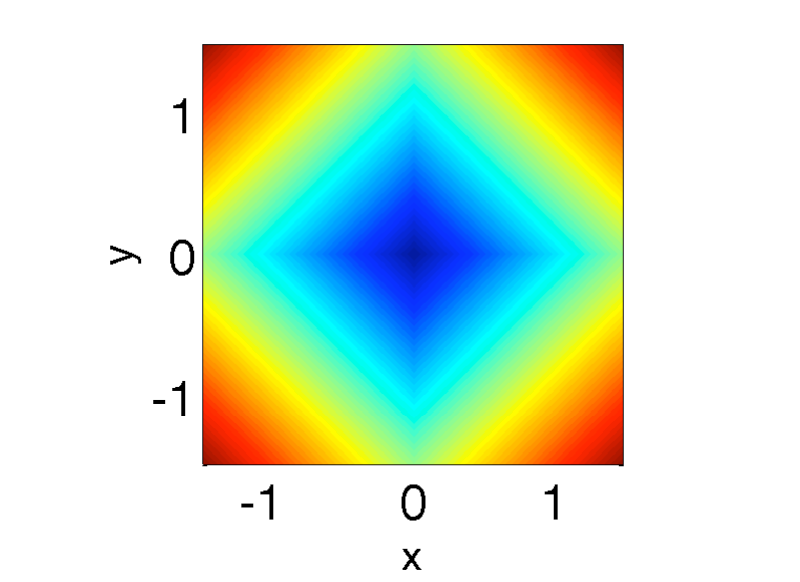}
d)\includegraphics[width=7cm]{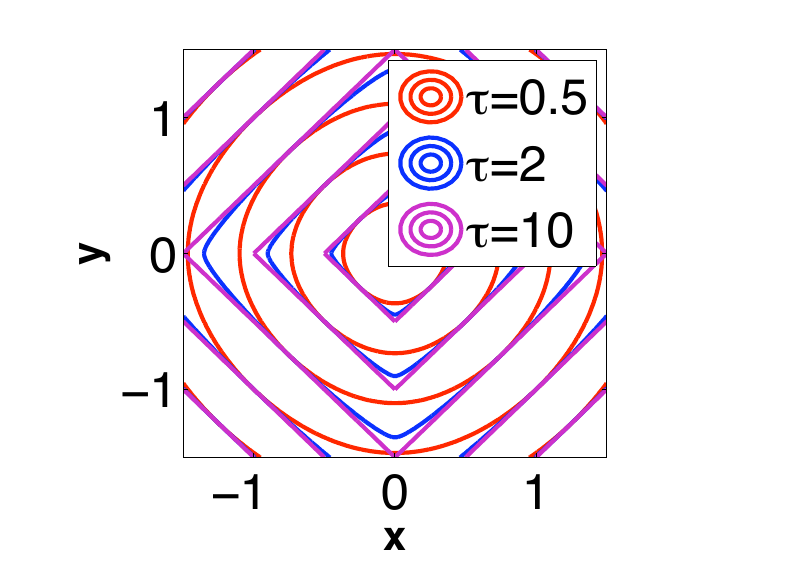}
\caption{\label{contours_M1} Evolution of the contours as $\tau$ increases. a) $\tau=0.5$, b) $\tau=2$, c) $\tau=10$, d) isolated contours showing the evolution for $\tau=0.5, \,2,\, 10$. }
\end{figure}

In Figure \ref{contours_M1} we  show the evolution of the contours of $M_1$ for $\tau=0.5, 2,   \, 10$.  The first thing to observe is that the patterns of the contours displayed depend on $\tau$. For $\tau$ small the structure of $M_1$ is smooth
and  for increasing $\tau$ the contour patterns converge towards a structure that displays the manifolds by means of discontinuity in the derivatives.
An analytical argument concerning the convergence time is given in Section \ref{sec:asymp}.

\begin{figure}
a)\includegraphics[width=7cm]{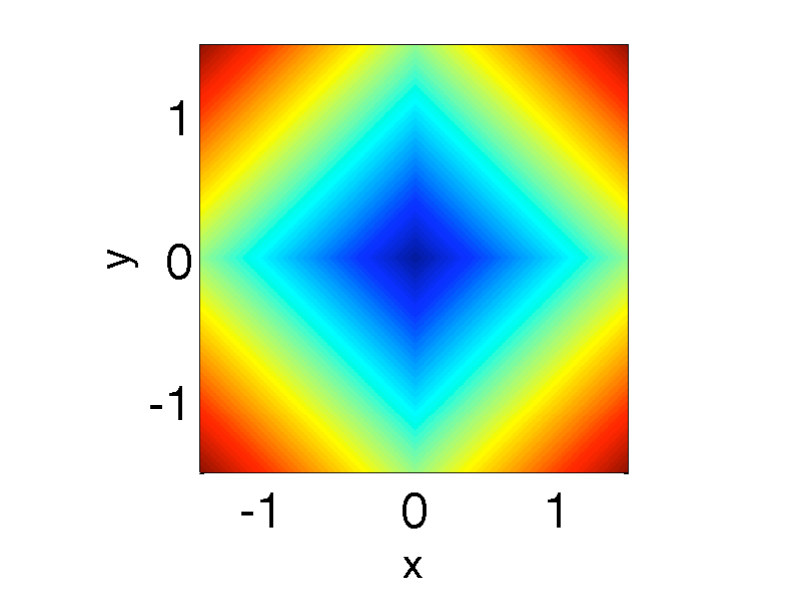}
b)\includegraphics[width=7cm]{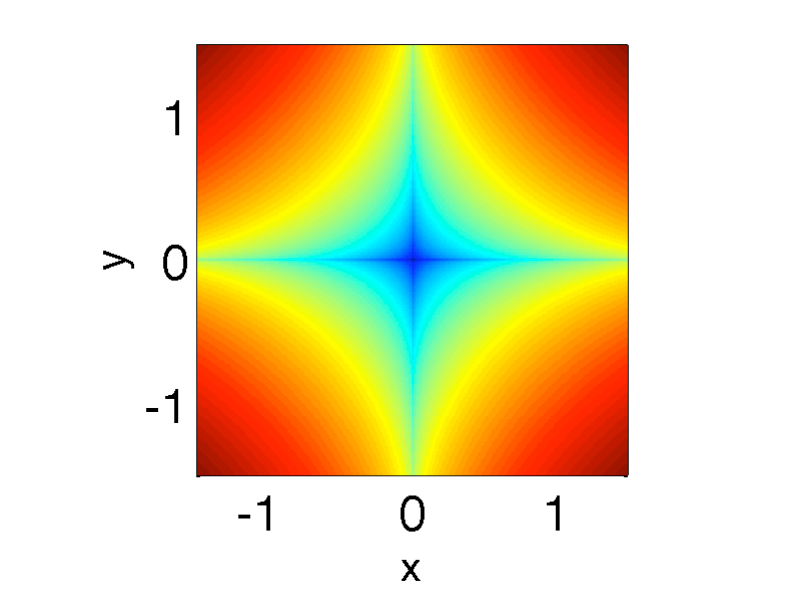}
c)\includegraphics[width=7cm]{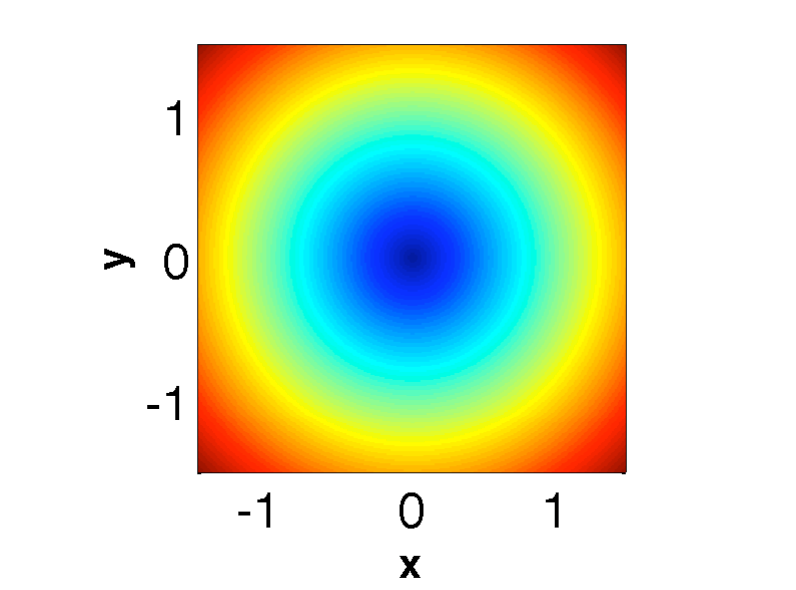}
d)\includegraphics[width=7cm]{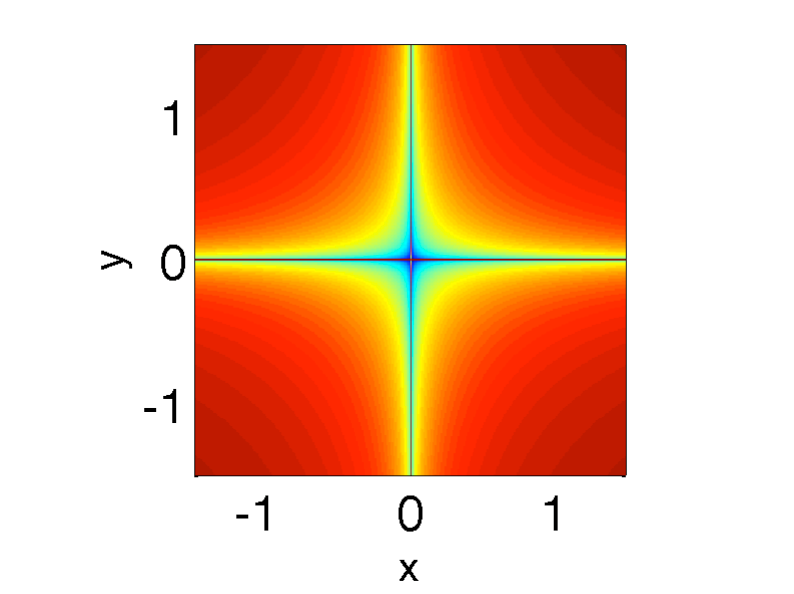}
\caption{\label{dufflin} Contour plots for a selection of Lagrangian  descriptors for \eqref{linex}. a) $M_1$; b) $M_3$, using the $L^{\frac{1}{2}}$ norm;
c) $M_3$ using the $L^2$ norm; d) $M_5$.}
\end{figure}

In Figure \ref{dufflin} we show several different Lagrangian descriptors, where each is computed for $\tau =10$.
Figure  \ref{dufflin}a) shows contours of the Lagrangian descriptor $M_1$ for $\lambda =1$. We see that the stable and unstable manifolds of the hyperbolic fixed point at the origin are easily identified.  For $\lambda =1$ it is easy to see from the simple form of the velocity field \eqref{linex} and its flow given in \eqref{linex_flow} that $M_1 =M_2 = M_4$, and therefore the latter two Lagrangian descriptors are not shown.

Figure \ref{dufflin}b) shows the contours of $M_3$ for $\gamma = 0.5$--the $L^{\frac{1}{2}}$ norm of the modulus of the acceleration--for which the stable and unstable manifolds of the hyperbolic fixed point at the origin are readily apparent. However, figure \ref{dufflin}c) shows the contours of $M_3$ for $\gamma = 2$--the $L^{2}$ norm of the modulus of the acceleration, and in this case the invariant manifolds structure is completely lost. We explain this behaviour in Section \ref{sec:asymp}. In Figure  \ref{dufflin}d)  we show contours of the Lagrangian descriptor $M_5$ (for $a=1$). In this case the manifolds are clearly shown. This is not surprising since the stable and unstable manifolds of the origin are the only trajectories having zero curvature. In fact, in this case we would expect that the contours of $M_5$ would reveal the manifolds for relatively small $\tau$. In particular, we have verified numerically that essentially the same contours are obtained for $\tau =2$.

Finally, we note that the Lyapunov exponents can be computed analytically for every trajectory in this example. Since the system is linear they are the same for every trajectory--$\pm \lambda$ (except for the exception case of trajectories starting on either the stable or unstable manifold of the origin, in which case one of the Lyapunov exponents will be zero). Hence the contours of constant Lyapunov exponents reveal no structure in this case (as noted in \cite{brawigg}). This will be explored in more detail in Section \ref{sec:ftle_ed}.

\subsubsection{Analytical Justification}
\label{sec:asymp}

In this section, for the specific example (\ref{linex}), we give analytical justification  underlying the heuristic argument as to why contours of the Lagrangian descriptors having the property that the derivative of the Lagrangian descriptor transverse to the contours correspond to material curves. This also provides insight as to why the $L^\gamma$ norms, $\gamma > 1$, do not reveal this singular property in the contours for  Lagrangian descriptors that use such norms.

For \eqref{linex}, using \eqref{def:Mgen}, the $M_1$ function is given by:

\begin{equation}
M_1(x_0, y_0 ;  t_0, \tau) = \int_{t_0-\tau }^{t_0+\tau} \sqrt{\dot{x}(t, x_0) ^2+ \dot{y}(t, y_0) ^2} dt =
 \int_{t_0-\tau }^{t_0+\tau} \sqrt{x_0^2 \lambda^2 e^{2 \lambda t} + y_0^2  \lambda^2 e^{-2 \lambda t}} dt
\label{M_1}
\end{equation}

\noindent
This function is clearly nonnegative since it is a measure of arc length of a trajectory.
It is clear by inspecting \eqref{def:Mgen} that  $M_1(0, 0 ;  t_0, \tau) =0$. This makes sense since the arc length of a point is zero.  We want to compute the integral $M_1$. For simplicity we assume without loss of generality that $t_0=0$ (this is always possible for an autonomous system  after a time shift $t'=t-t_0$ and by redefining the initial conditions $x_0$ and $y_0$).

As already noted the patterns of the contours displayed in Figure \ref{contours_M1} depend on $\tau$. For $\tau$ small the contours
are smooth but for increasing $\tau$ they develop singular features along the unstable and stable manifolds.  The nature of what we mean by the phrase ``singular features'' will be explained in the following where we provide analytical justification that underlies this behaviour.

\medskip
\paragraph{Behavior for Small $\tau$.}

We now want to argue that for $\tau$ small, \eqref{M_1} does not display ``sharp features'' that are related to the stable and unstable manifolds of the hyperbolic fixed point of  \eqref{linex}.   Towards this end, we consider an expansion of the  expression   \eqref{M_1} for fixed $y_0$   that is valid for $x_0$ small (\emph{i.e.} near the stable manifold) and for  small $\tau$. The expansion has the following form:

\begin{eqnarray}
M_1(x_0, y_0 ;  t_0, \tau) &=&   \int_{-\tau }^{\tau} \sqrt{(x_0 \lambda )^2 e^{2 \lambda t} + (y_0 \lambda)^2 e^{-2 \lambda t}} dt \nonumber\\
&=& \int_{-\tau }^{\tau} \left( e^{-\lambda t} \lambda |y_0| + \frac{e^{3\lambda t} \lambda |y_0|}{2 y_0^2} x_0^2 \right) dt + O(x_0^4).
\label{M_1_2}
\end{eqnarray}

\noindent
It is important to notice that in the integrand  $x_0$ is always multiplied by $e^{2 \lambda t} $, thus if we want the ``error terms'' in the series expansion around $x_0=0$ to be small,   then $e^{2 \lambda t} $ should not be large, which means  that  $\tau$ should not be large, as we are assuming.
The leading order terms in \eqref{M_1_2} can be explicitly computed, and are found to be:

\begin{eqnarray}
M_{1, {\rm lo}}(x_0, y_0 ;  t_0, \tau) & = &  \int_{-\tau }^{\tau} \left( e^{-\lambda t} \lambda |y_0| + \frac{e^{3\lambda t} \lambda |y_0|}{2 y_0^2} x_0^2 \right) dt, \nonumber
\\
& =&   {\left( e^{\lambda \tau} - e^{-\lambda \tau} \right)}  |y_0| + \frac{\left( e^{3\lambda \tau}-e^{-3\lambda \tau} \right)|y_0|}{6 y_0^2} x_0^2.
\label{M_1_2_lo}
\end{eqnarray}

\begin{figure}
 \centering
\includegraphics[width=8cm]{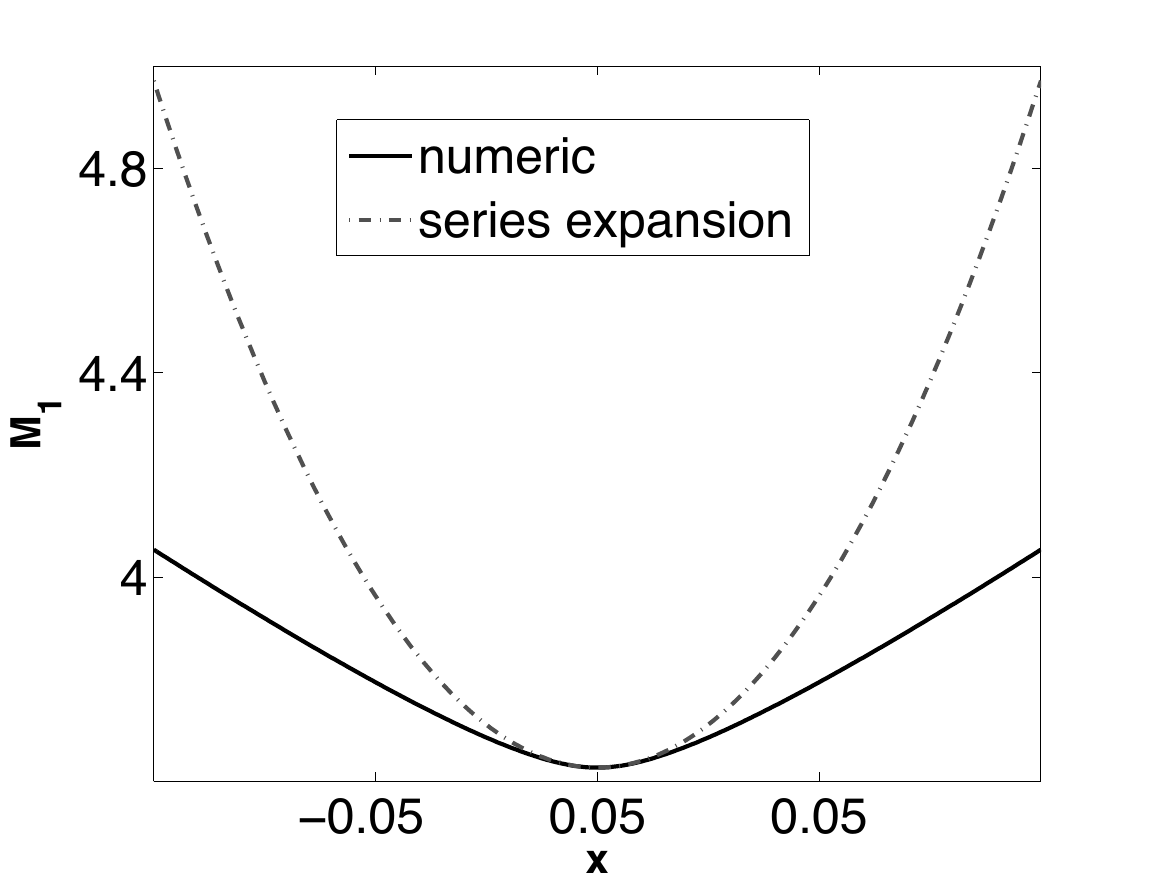}
\caption{Solid line, the numerical evaluation of $M_1$ along $x_0$ for $y_0=0.5$, $\tau=2$, $\lambda=1$. Dashed line, the evaluation of the leading order terms  \eqref{M_1_2_lo}.}
\label{comp_M1_linsad_small}
\end{figure}

\noindent
For $y_0$ fixed, \eqref{M_1_2_lo} is a smooth function of $x_0$.
Figure \ref{comp_M1_linsad_small} shows a comparison of the accuracy of \eqref{M_1_2} with \eqref{M_1_2_lo} for $\tau = 2$ along a horizontal line perpendicular to the stable manifold. More precisely, we fix $y_0 =0.5$ and plot \eqref{M_1_2} (computed numerically, and shown with a solid line) and \eqref{M_1_2_lo} (shown with a dashed line).
It is clear from the figure that leading order terms \eqref{M_1_2_lo}  accurately approximate \eqref{M_1_2} in the interval $x_0 \in [-0.01, 0.01]$.

\medskip
\paragraph{Behavior for Large $\tau$.}

Now we consider the behaviour of \eqref{M_1} when $\tau$ is large. Towards this end, we re-write \eqref{M_1} as follows:

\begin{equation}
M_1(x_0, y_0 ;  t_0, \tau) =   \int_{0 }^{\tau} \sqrt{X_0^2 + Y_0^2} dt + \int_{-\tau }^{0}  \sqrt{ X_0^2 + Y_0^2 } dt
\label{M_1_3}
\end{equation}

\noindent
where we have defined $X_0^2= x_0^2 \lambda^2 e^{2 \lambda t} $ and $Y_0^2= y_0^2 \lambda^2 e^{-2 \lambda t} $.

In the first integral term in \eqref{M_1_3},  since $t$ is positive if $\tau$ is large $X_0$ may assume large values for $t$ values approaching the upper limit of the integral. By similar reasoning, $Y_0$ will become small for $t$ values approaching the upper limit of integration.
For the second integral term in \eqref{M_1_3}, since the integration range is negative $Y_0$ assumes large values for $t$ approaching the lower limit of the integral. By similar reasoning,  $X_0$ will become small for $t$ values approaching the lower limit of integration.
These facts suggest isolating the large parts of both integral terms by choosing constants $\tau_p, \, \tau_n > 0$ (as we will see, precise values for these constants are not important for the argument) and re-writing \eqref{M_1_3} as follows:

\begin{eqnarray}
M_1(x_0, y_0 ;  t_0, \tau) &=&   \int_{\tau_p }^{\tau} \sqrt{X_0^2 + Y_0^2} dt +  \int_{0 }^{\tau_p} \sqrt{X_0^2 + Y_0^2} dt \nonumber \\&+& \int_{-\tau_n }^{0}  \sqrt{ X_0^2 + Y_0^2 } dt+  \int_{-\tau }^{-\tau_n}  \sqrt{ X_0^2 + Y_0^2 } dt
\label{M_1_3b}
\end{eqnarray}

We expand the first term in \eqref{M_1_3b} in $\frac{X_0}{Y_0}$  and the second term in \eqref{M_1_3b} in $\frac{Y_0}{X_0}$  and rearrange the terms to obtain the following:

 \begin{equation}
M_1(x_0, y_0 ;  t_0, \tau) =   \int_{\tau_p }^{\tau} | X_0|   dt + \int_{-\tau }^{-\tau_n} |Y_0| dt  +O(1/X_0) +O(1/Y_0)+B,
\label{M_1_4}
\end{equation}

\noindent
where $B$ is defined as:

\begin{equation}
B =    \int_{0 }^{\tau_p} \sqrt{X_0^2 + Y_0^2} dt + \int_{-\tau_n }^{0}  \sqrt{ X_0^2 + Y_0^2 } dt.
\label{B}
\end{equation}

\noindent
Using the fact that for $\tau$ large   $ |x_0| e^{\lambda \tau} >>|x_0| e^{\lambda \tau_p}$ and $|y_0| e^{\lambda \tau} >>|y_0| e^{\lambda \tau_n}$ it is straightforward to determine that the leading order terms of \eqref{M_1_4} are given by:

 \begin{equation}
M_{1, {\rm lo}}(x_0, y_0 ;  t_0, \tau) =   |x_0| e^{\lambda \tau} + |y_0| e^{\lambda \tau}.
\label{M_1_4_lo}
\end{equation}

In the above expressions it is important to note that the estimates  are valid even if $x_0$ or $y_0$ are small as far as  the
 products   $x_0 e^{\lambda t}$, $y_0 e^{-\lambda t}$ are large. This occurs
 very close to the origin,  $x_0\approx 0$ or $y_0\approx 0$,   if $\tau$ is large, and this corresponds to a large integration time in the definition of the Lagrangian descriptor.
 In this case discontinuities in the derivatives (or even on the functions themselves in other possible examples) observed along the stable and unstable manifolds rise because the asymptotic expansions at both sides of the manifolds, do not match well across these lines. In practice for any finite $\tau$ the series expansion  \eqref{M_1_2} is still valid, thus strictly speaking the function $M$ is regular, but it is accurate only in an extremely narrow gap  around $x_0=0$, exponentially decreasing  for increasing $\tau$.

\begin{figure}
 \centering
\includegraphics[width=8cm]{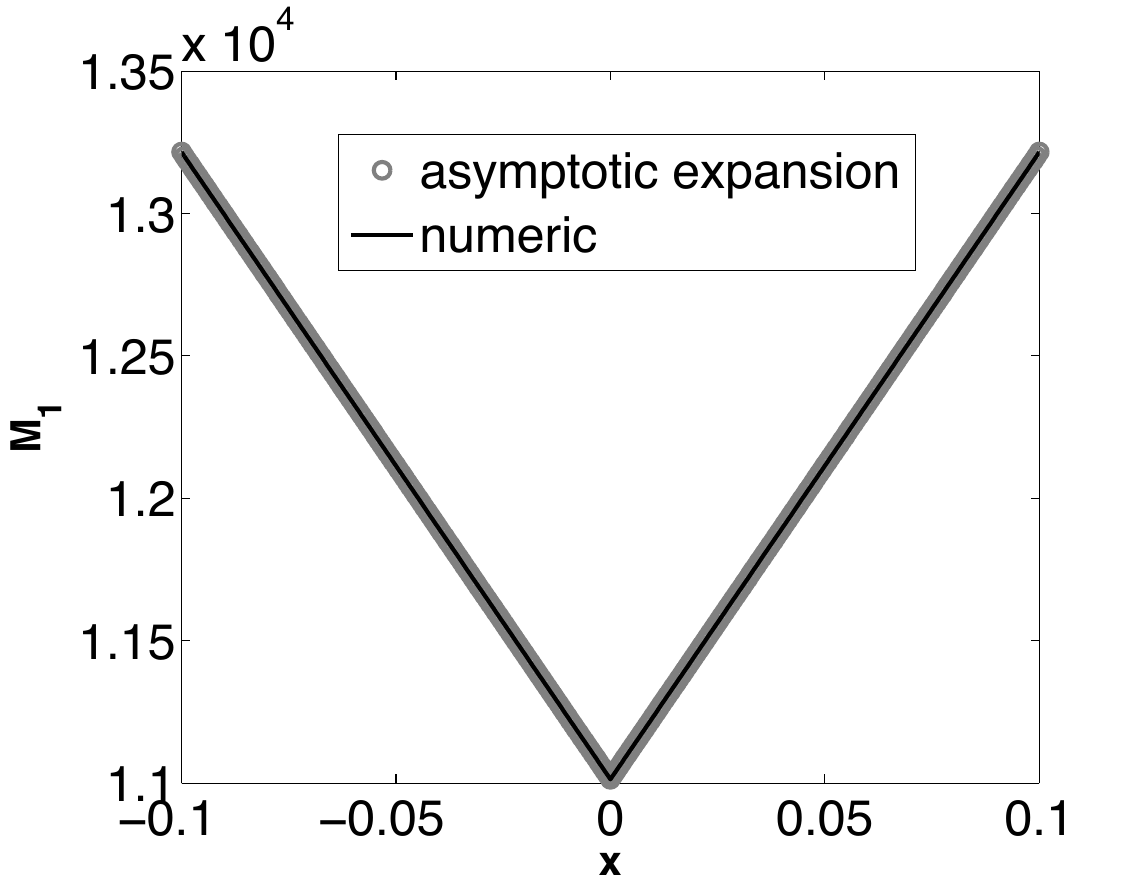}
\caption{ In black the numerical evaluation of \eqref{M_1_3} along $x_0$ for $y_0=0.5$, $\tau=10$, $\lambda=1$. In gray the evaluation of expression  \eqref{M_1_4_lo}.}
\label{comp_M1_linsad_large}
\end{figure}

Figure \ref{comp_M1_linsad_large} shows a comparison of the accuracy of \eqref{M_1_3} with \eqref{M_1_4_lo} for $\tau = 10$  along a horizontal line perpendicular to the stable manifold. More precisely, we fix $y_0=0.5$, $\tau=10$, $\lambda=1$  and plot \eqref{M_1_3} (computed numerically, and shown in black) and \eqref{M_1_4_lo} (shown in gray). It is clear from the figure that leading order terms \eqref{M_1_4_lo}  accurately approximate \eqref{M_1_3} in the interval $x_0 \in [-0.1, 0.1]$ and that the derivative of both functions appears discontinuous on the stable manifold.

\medskip
\paragraph{The $L^\gamma$ norm, $\gamma >1$.}

In Section \ref{secLd} we stated that the contours of the  $L^\gamma$ norm, $\gamma >1$, of  a bounded, positive intrinsic physical or geometrical property of the velocity field integrated along trajectories do not yield interesting structures in the flow. Here we give an argument why this is the case in the context of the example \eqref{linex}.

 We first extend the previous results for the $L^\gamma$ norm, with $\gamma \leq 1$.  
The essential feature to consider is the exponent on the integrand, which we will take as the magnitude of velocity for this example:

\begin{equation}
M(x_0, y_0 ;  t_0, \tau) =   \int_{-\tau }^{\tau} \left( X_0^2 + Y_0^2 \right)^{\delta} dt.
\label{M_1_5}
\end{equation}

\noindent
Following the same argument as above, in this case the asymptotic expansion is:

\begin{eqnarray}
M(x_0, y_0 ;  t_0, \tau) &\sim&   \int_{\tau_p }^{\tau}  \left( X_0^{2 \delta}+\delta Y_0 X_0^{2(-1+\delta)}  \right) dt + \int_{-\tau }^{-\tau_n}    \left(Y_0^{2 \delta}+\delta X_0 Y_0^{2(-1+\delta)}  \right)  dt \nonumber\\&+&O(1/X_0^{4-2\delta})+O(1/Y_0^{4-2\delta})
\label{M_1_5}
\end{eqnarray}

\noindent
Following an argument similar to that given above, for  $0<\delta<1$ the leading order  terms in the expansion are:

\begin{equation}
M(x_0, y_0 ;  t_0, \tau) \sim   \frac{|x_0|^{\gamma}}{\gamma} e^{\gamma \lambda \tau} +   \frac{|y_0|^{\gamma}}{\gamma } e^{\gamma \lambda \tau}
\label{M_1_6}
\end{equation}

\noindent
where $\gamma=2 \delta$. If $\gamma<1$ the derivative across   the stable manifold is  singular, which explains the enhancement of the features displayed in Figure \ref{dufflin}b)   for this choice of exponent. The case $\gamma=1$ corresponds to \eqref{def:Mgen}.
On the other hand, if $\gamma>1$ the expression above should be raised  to the power $1/\gamma$.  In this case the derivative across   the stable manifold is continuous. We see this in Figure  \ref{dufflin}c)   ($\gamma=2$) for which the contours of $M$ are smooth.

\subsection{Lagrangian Descriptors in Elliptic Regions}
\label{sec:ell}

Now we will consider the behaviour of Lagrangian descriptors in ``elliptic regions''.  In the same spirit as in  the discussion in Section \ref{sec:linsad}, we will begin by discussing the familiar linear elliptic fixed point.

\subsubsection{The Linear  Elliptic Point }
\label{sec:linep}

We consider the velocity field:

\begin{eqnarray}
\dot{x} & = & y,  \nonumber \\
\dot{y} & = & -x,
\label{eq:linellip}
\end{eqnarray}

\noindent
and the flow generated by this velocity field is given by:

\begin{eqnarray}
x(t, x_0)  & = & x_0 \cos t + y_0 \sin t,  \nonumber \\
y(t, y_0)  & = & -x_0 \sin t + y_0 \cos t.
\label{eq:flowellip}
\end{eqnarray}

\noindent
The origin, $(x, y) =(0, 0)$ is an elliptic  fixed point. In this case the
$M_1$ function has the form:

\begin{eqnarray}
M_1(x_0, y_0 ;  t_0, \tau) &=& \int_{t_0-\tau }^{t_0+\tau} \sqrt{\dot{x}(t, x_0) ^2+ \dot{y}(t, y_0) ^2} dt \nonumber\\
&=&  \int_{t_0-\tau }^{t_0+\tau} \sqrt{ (-x_0\sin t + y_0 \cos t)^2   +     (-x_0\cos t - y_0 \sin t)^2} \,\,dt,
\label{M_ell}
\end{eqnarray}

\noindent
which is easily computed to give:

\begin{equation}
M_1(x_0, y_0 ;  t_0, \tau) = \int_{t_0-\tau }^{t_0+\tau} \sqrt{ x_0^2+y_0^2} dt = 2\tau \sqrt{ x_0^2+y_0^2}.
\label{M_ell_2}
\end{equation}

\noindent
which is a smooth function for all $\tau$ except at the origin. The contours of $M_1$, for any fixed $\tau$, are smooth circles surrounding the origin. Also, note that for this example it is easily verified that $M_1 = M_2$.

\subsubsection{An Integrable Example: Shear and Resonance}
\label{sec:twistless}

In this section we consider an example that, in some sense, is the ``simplest'' nonlinear problem that has entirely elliptic behaviour--a one degree-of-freedom integrable Hamiltonian system in action--angle variables. Nevertheless, this system provides some useful insights that raises questions for more complex systems.

The Hamiltonian is given by:

\begin{equation}
H(I) = \frac{I^4}{4}-\frac{I^2}{2},
\label{twistless_ham}
\end{equation}

\noindent
and the associated Hamiltonian vector field is given by:

\begin{eqnarray}
\dot{\theta} = \frac{\partial H}{\partial I} & = &  I^3 -I, \nonumber \\
\dot{I} = -\frac{\partial H}{\partial \theta} & = &  0. \qquad \qquad (I, \theta) \in \mathbb{R} \times S^1,
\label{eq:hamvf5}
\end{eqnarray}

 \begin{figure}
a)\includegraphics[width=4.6cm]{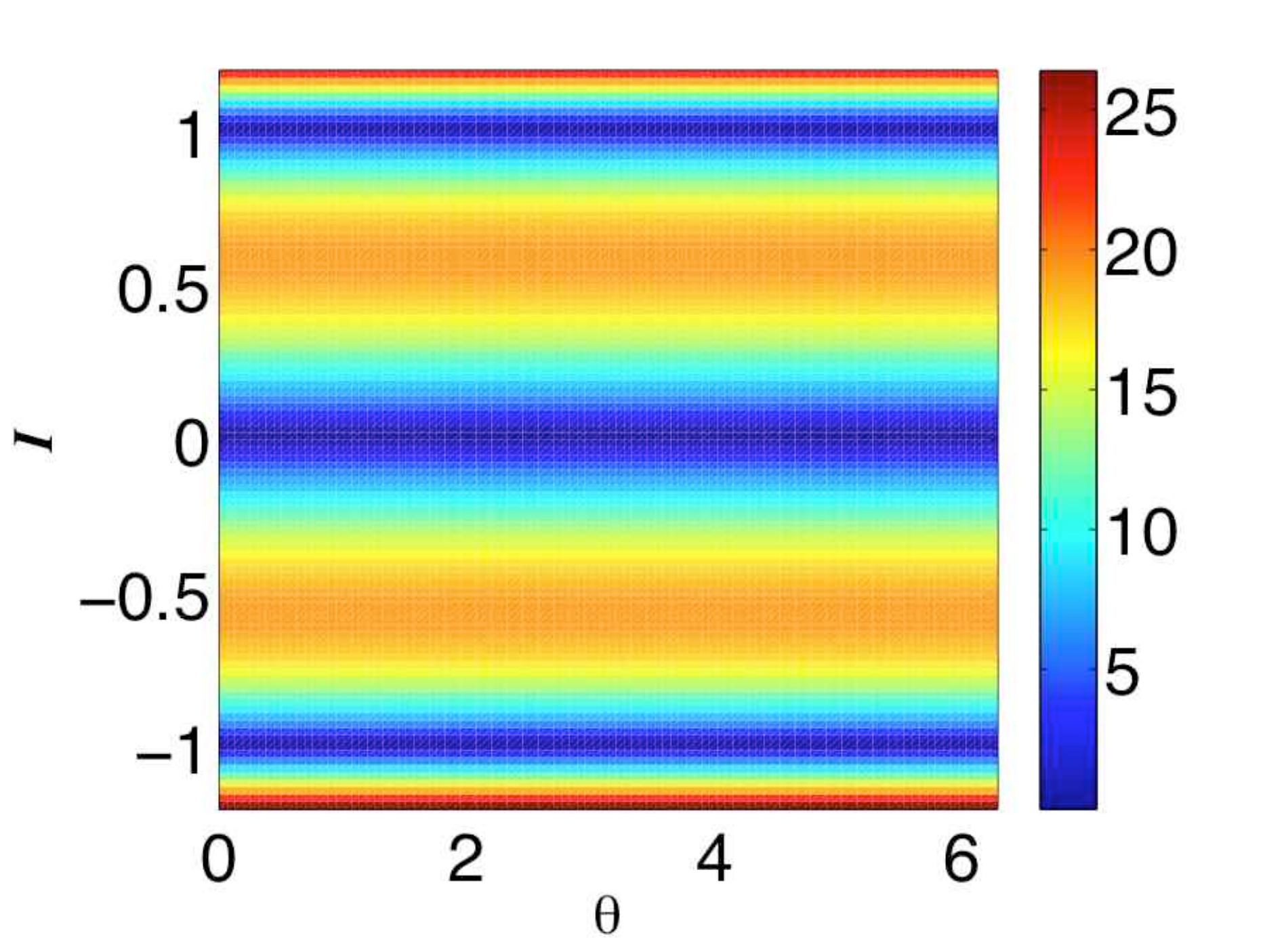}
b)\includegraphics[width=4.6cm]{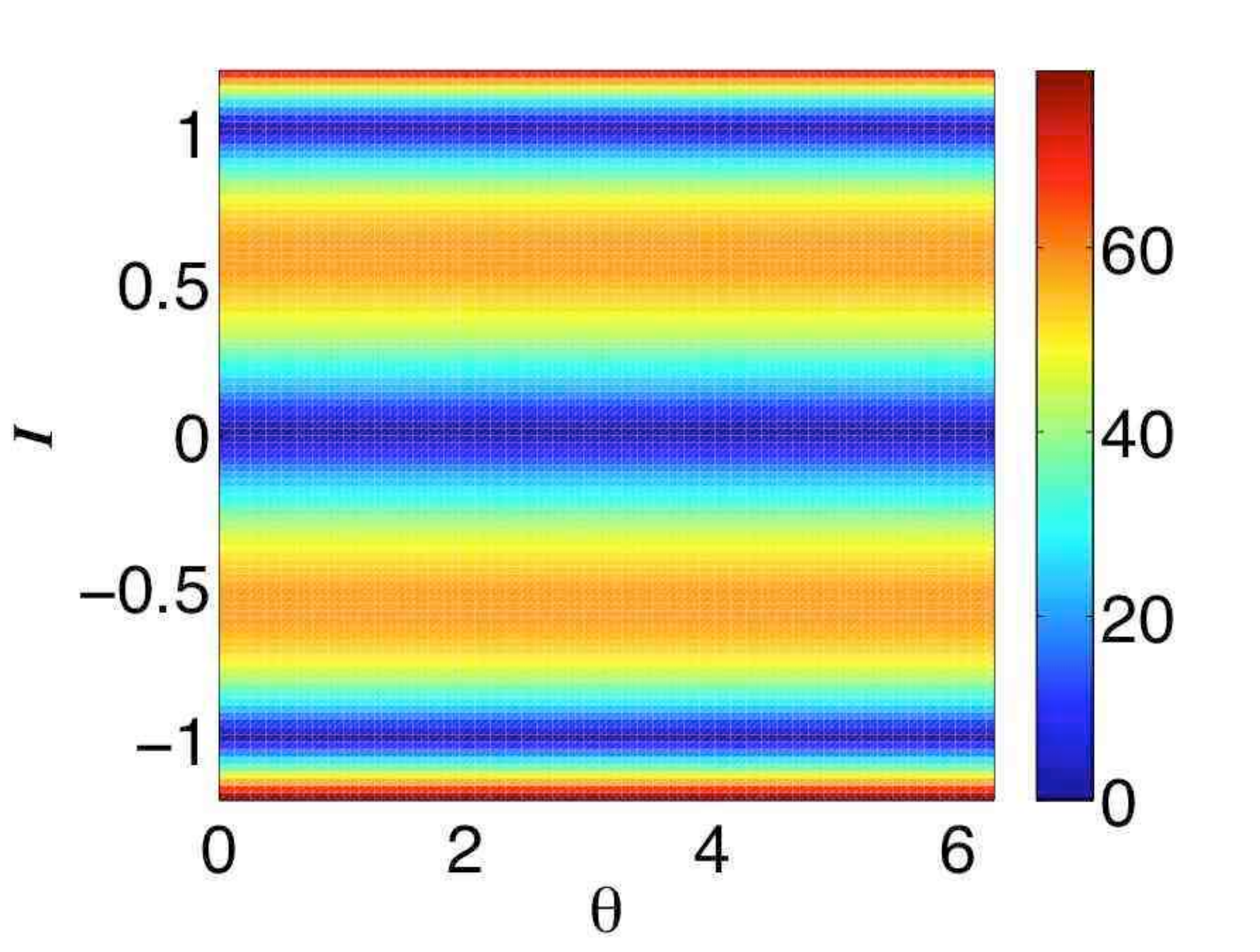}
c)\includegraphics[width=4.6cm]{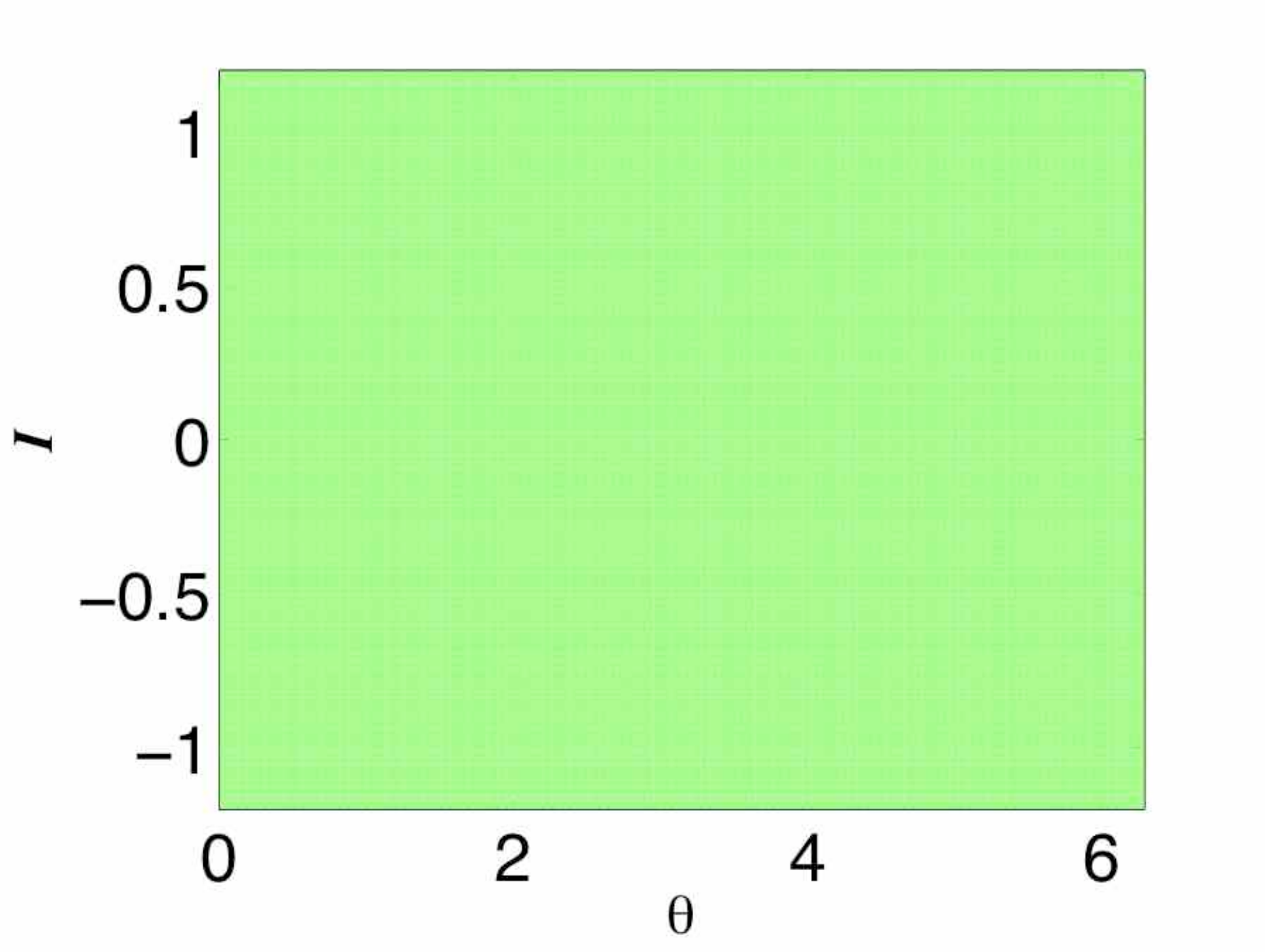}
\caption{\label{twtori} Evaluation of  Lagrangian Descriptors for the system (\ref{eq:hamvf5}). a) $M_1$ for $\tau=25$; b)$M_1$ for $\tau=75$ ;c) $M_3$, using $||{\bf a}||$  in the integrand with the $L^{\frac{1}{2}}$ norm.}
\end{figure}

\noindent
It is evident from \eqref{eq:hamvf5} that $I$ is constant, and each value of $I$ corresponds to an invariant circle, or ``invariant one torus''  (since $\theta$ is periodic).  The frequency of each torus is given by $\frac{\partial H}{\partial I} = I^3 - I$ and the shear, or ``twist'' associated with each torus is $\frac{\partial^2 H}{\partial I^2} = 3 I^2 -1$.  The typical behaviour for each $I = $ constant invariant circle is that it has nonzero frequency and nonzero shear. However, there are five exceptional tori that we note.

\begin{description}

\item[Resonance.] The three invariant circles $I=0$, $I= \pm 1$ correspond to circles of fixed points since the frequency is identically zero on each of these circles.

\item[Zero shear.]  On the invariant circles $I = \pm \frac{1}{\sqrt{3}}$ the frequencies are $ \mp \frac{2}{3 \sqrt{3}}$ but the shear, or ``twist'' is identically zero on each of these invariant circles. In the dynamical systems literature these are referred to as ``twistless'' invariant circles. The choice of ``twist less'' tori will allow us to compare with recent work in \cite{beron-vera-twistless}  when we consider FTLEs in Section \ref{sec:ftle_ed}.

\end{description}

Figure \ref{twtori}a) illustrates how $M_1$  detects the three circles of fixed points (dark blue) as well as the high shear ``twistless tori'' (the reddish colours), for $\tau = 25$. Figure \ref{twtori}b) shows the contours of $M$ for $\tau = 75$, which contain essentially the same information as for $\tau=25$.

The function $M_3$, based on $||{\bf a}||$ and the $L^{\frac{1}{2}}$, gives a different perspective on the structure of the flow. Since in this system all trajectories move with constant velocity, their acceleration is zero, from which it follows that  $M_3$ is identically zero. In this case Figure \ref{twtori}c) confirms a flat structure which is obtained for all $\tau>0$. That is, there is no convergence threshold in $\tau$.  We will explore this  example again when we discuss FTLEs  in Section \ref{sec:ftle_ed}.

\subsection{Remarks on the Application of Finite Time Lyapunov Exponents (FTLEs) and Time Averages for These Examples}
\label{sec:ftle_ed}

In this section we want to compare two other techniques that are used to visualise phase space structures--finite time Lyapunov exponents (FTLEs) and the ergodic decomposition--with the method of Lagrangian descriptors.

\subsubsection{Finite Time Lyapunov Exponents (FTLEs)}
\label{sec:ftle}

FTLEs have proven to be a useful tool for obtaining a ``global picture'' of phase space structure. The fundamental theorem concerning Lyapunov exponents (\emph{i.e.} for infinite time) is the Oseledets multiplicative ergodic theorem (\cite{oseledets}).  This theorem gives sufficient conditions for  finite time Lyapunov exponents to converge to their infinite time values. However, it does not provide information how accurately FTLEs, for a specified finite time, approximate their asymptotic values, nor does it relate Lyapunov exponents (either finite or infinite time) to invariant manifolds (e.g. ``material''  curves for two-dimensional, time dependent velocity fields). Discussions of  Oseledets theorem in the context of applications can be found in \cite{legraslv, lapeyre}.

FTLEs have been used successfully as ``proxies''  for invariant manifolds (\cite{haller, shad}), in the following sense. An initial time is chosen and the spatial domain is decomposed into a discrete grid. These grid points are then integrated forward (resp.,  backward) in time, for a fixed time (each grid point is integrated for the same time).  The maximal Lyapunov exponent for each trajectory is computed for this time (the ``FTLE'') and a value is given to the initial point corresponding to this maximal FTLE. This is done for each grid point and the result is a forward (resp., backward) FTLE field at the chosen initial time. Then the level curves of the FTLE are displayed--different colours for different values. Locally maximal level curves of the forward (reps. backward) FTLE field are typically identified with the stable (resp.,  unstable) manifolds of hyperbolic trajectories.  In the ``topographical map'' of the level curves of the FTLE field the locally maximal level curves have the property of a ``ridge'', \emph{i.e.}, moving in a direction that is not tangent to the locally maximal level curve results in moving in regions of smaller FTLEs. The quantification of this notion of a ``ridge curve'' is carried out in \cite{shad}, and we mention this again below. For now we note that the identification of  locally maximal level curves of the FTLE field has been noted to provide good approximations to material curves in certain examples. Nevertheless, there are numerous issues , the full details of which are sometimes not reported in papers, that use FTLEs for specific physical applications. In particular, we point out three common issues.

\begin{description}

\item[The choice of finite time.] This is a problematic issue for which there is no theoretical guidance. The Oseledets theorem gives sufficient conditions for the FTLE to converge. However, for any finite time, it is not clear how ``close'' the FTLEs computed at that time are to their asymptotic values. The limited theoretical results that are available predict a ``relatively slow'' convergence time (e.g. logarithmic in $\tau$, in our notation, see, e.g., \cite{goldhirsch}). Moreover, it is observed in the course of  time evolution trajectories that are ``finite time hyperbolic'' or ``finite time elliptic'' can change their ``finite time stability type'' (see, e.g. \cite{brawigg, jpo}). There are numerous issues related to this topic that demand further theoretical investigations.

\item[Discretization issues.] Once the FTLE fields are computed the goal is then to look at the level curves of the FTLE field. However,
for any realistic system, FTLEs must be computed numerically, requiring discretization in space (and possibly time). Setting the issue of the choice of finite time aside, the computation of FTLE fields {\em may} be somewhat noisy, and this means that the ``raw output'' of FTLE computations does not yield smooth level curves. In general, some form of filtering and smoothing is  required in order to obtain a FTLE field that yields smooth level curves. This can be a subjective process.  Examples of the effects this can yield, in the context of known ``benchmark'' examples can be found in \cite{brawigg}.

\item[The connection to invariant manifolds (``material curves'').] Setting aside the crucial issue of the smoothness of the level curves of the FTLE field, it is often stated  in the literature that the level curves that locally maximal level curves of the FTLE field are ``material curves''. In general, this is not true, although they may be ``close'' to material curves and provide a good approximation for methods that compute material curves via particle advection. This issue was carefully considered in \cite{shad}, who introduced the notion of a {\em ridge curve} in the FTLE field as an approximation to a material curve. They showed that ridge curves tend to be a good approximation to a material curve by deriving an expression for the flux across the ridge curve, which may be small, but generally nonzero. There is a great deal of ambiguity in the literature concerning the notion of a ``ridge curve'' or ``ridges of the FTLE field''.  In many papers the notion of a ``ridge'' is taken to by synonymous with locally maximal level curves of the FTLE field. However, this is not the case, and ridge curves are precisely defined in \cite{shad}.

\end{description}

Of course,  a goal of this article is not to present a critical discussion of FTLEs and the range of their applicability (this is dealt with, to varying degrees, in \cite{physrep, brawigg}). Rather, we will compute FTLEs, for different times, for the simple examples that we have previously considered and compare them with the results obtained from Lagrangian descriptors. While these examples are certainly far simpler than a typical ``realistic''  geophysical fluid dynamics application, they have the advantage that the FTLEs, hyperbolic trajectory (if it exists) and their stable and unstable manifolds are known analytically and, in that sense, they may serve as benchmarks for testing various methods.

\begin{figure}
a)\includegraphics[width=8cm]{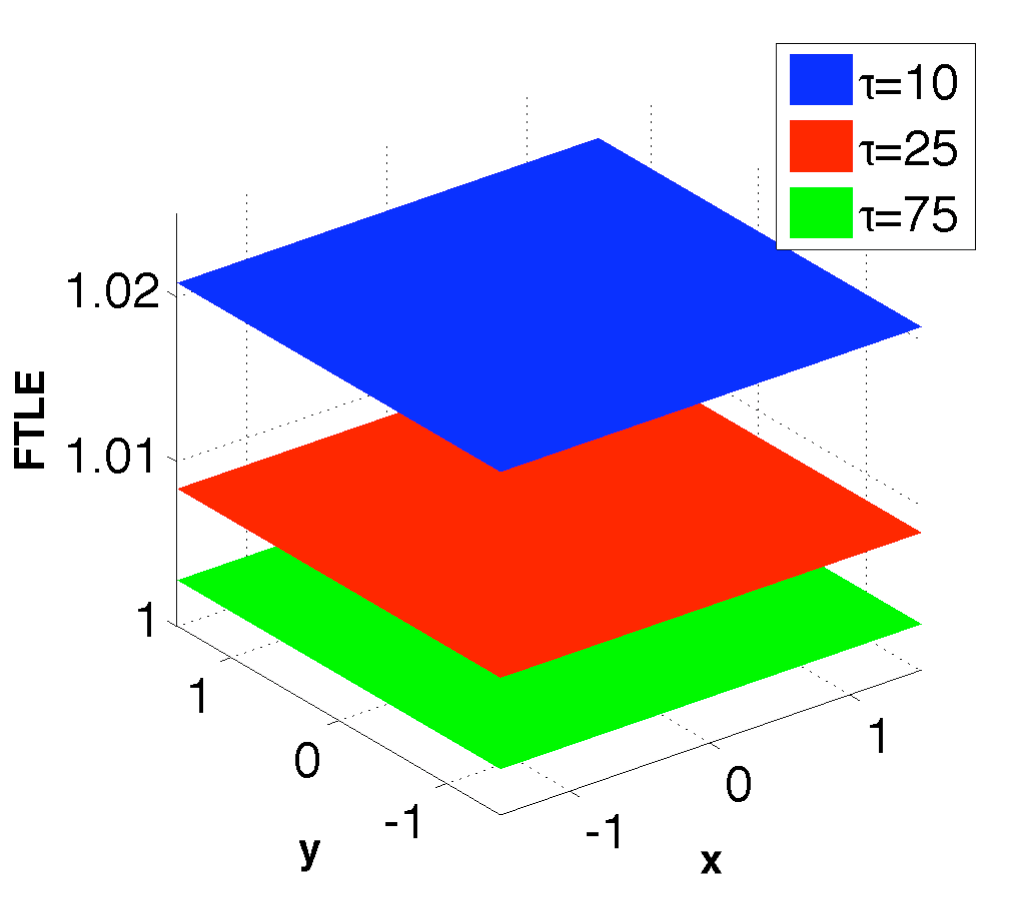}
b)\includegraphics[width=8cm]{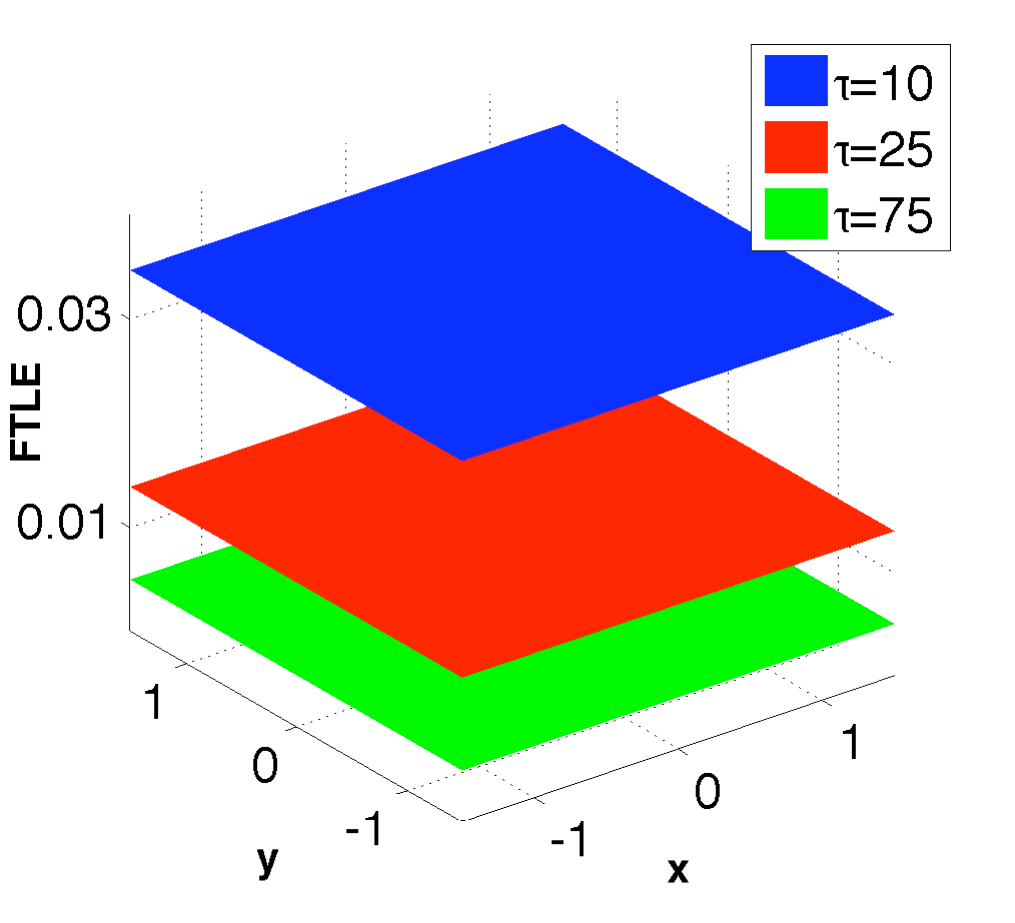}
\caption{\label{linear_ftle} a) FTLE for the linear saddle point, \eqref{linex},  at different $\tau$.; b) FTLE for the linear elliptic point, \eqref{eq:linellip}, at different $\tau$.}
\end{figure}

For linear, time independent velocity fields the (infinite time) Lyapunov exponents are simple to compute. They are the real parts of the eigenvalues of the matrix associated with the linear velocity field. Therefore, for the linear saddle point, \eqref{linex},   the Lyapunov exponents are $\pm \lambda$ (with the maximal Lyapunov exponent being $ \lambda >0$) and for the linear elliptic point, \eqref{eq:linellip}, the Lyapunov exponents are both zero.
The FTLE fields for each of these examples is shown in Figure \ref{linear_ftle} for $\tau =10, \, 25, \, 75$. For each $\tau$ a ``flat field'' is shown, which is to be expected for linear velocity fields since the Lyapunov exponents are the same for all trajectories. Hence,  the FTLEs reveal no structure in the flows, in contrast to the Lagrangian descriptors.

\begin{figure}
a)\includegraphics[width=8cm]{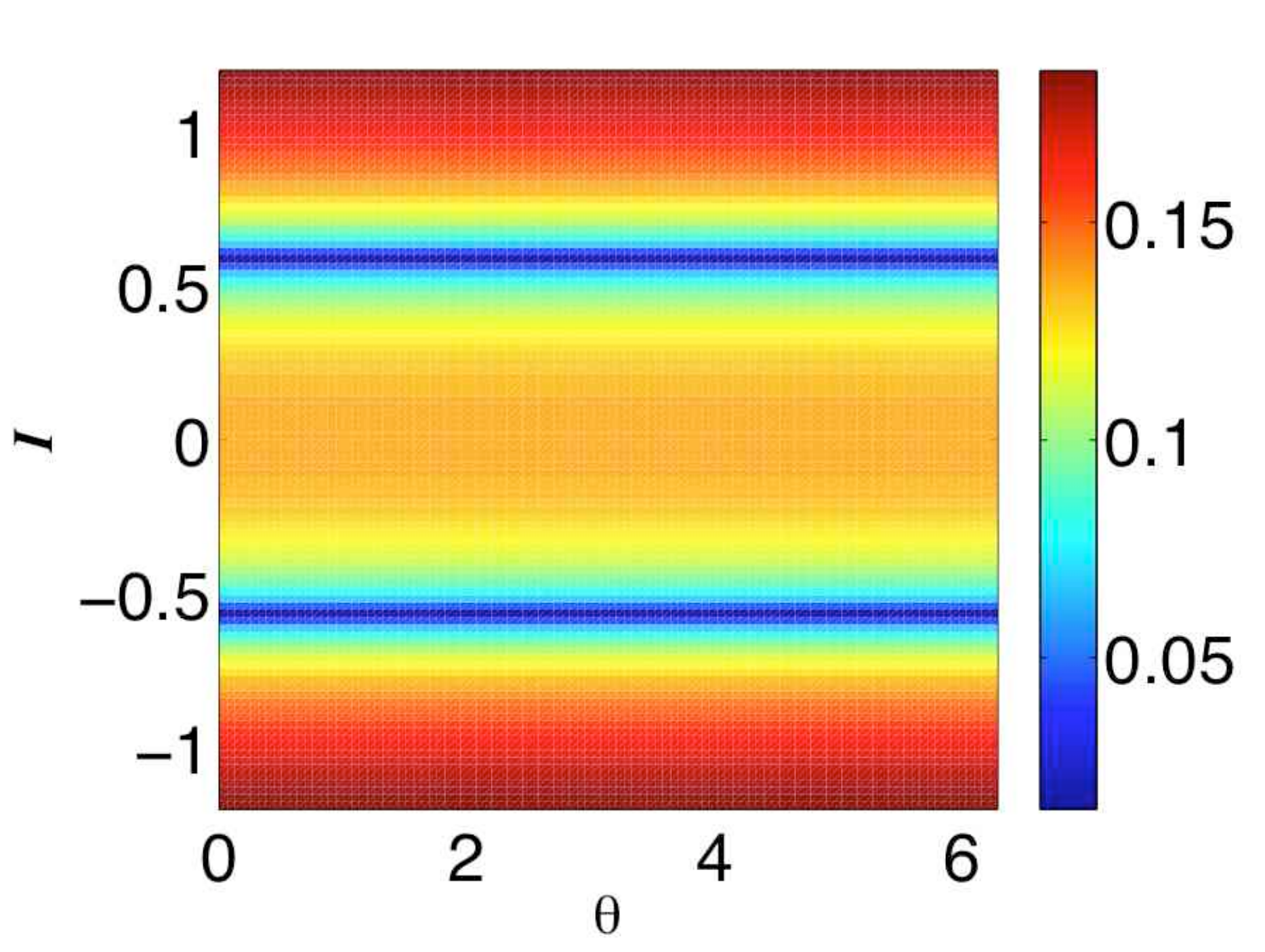}
b)\includegraphics[width=8cm]{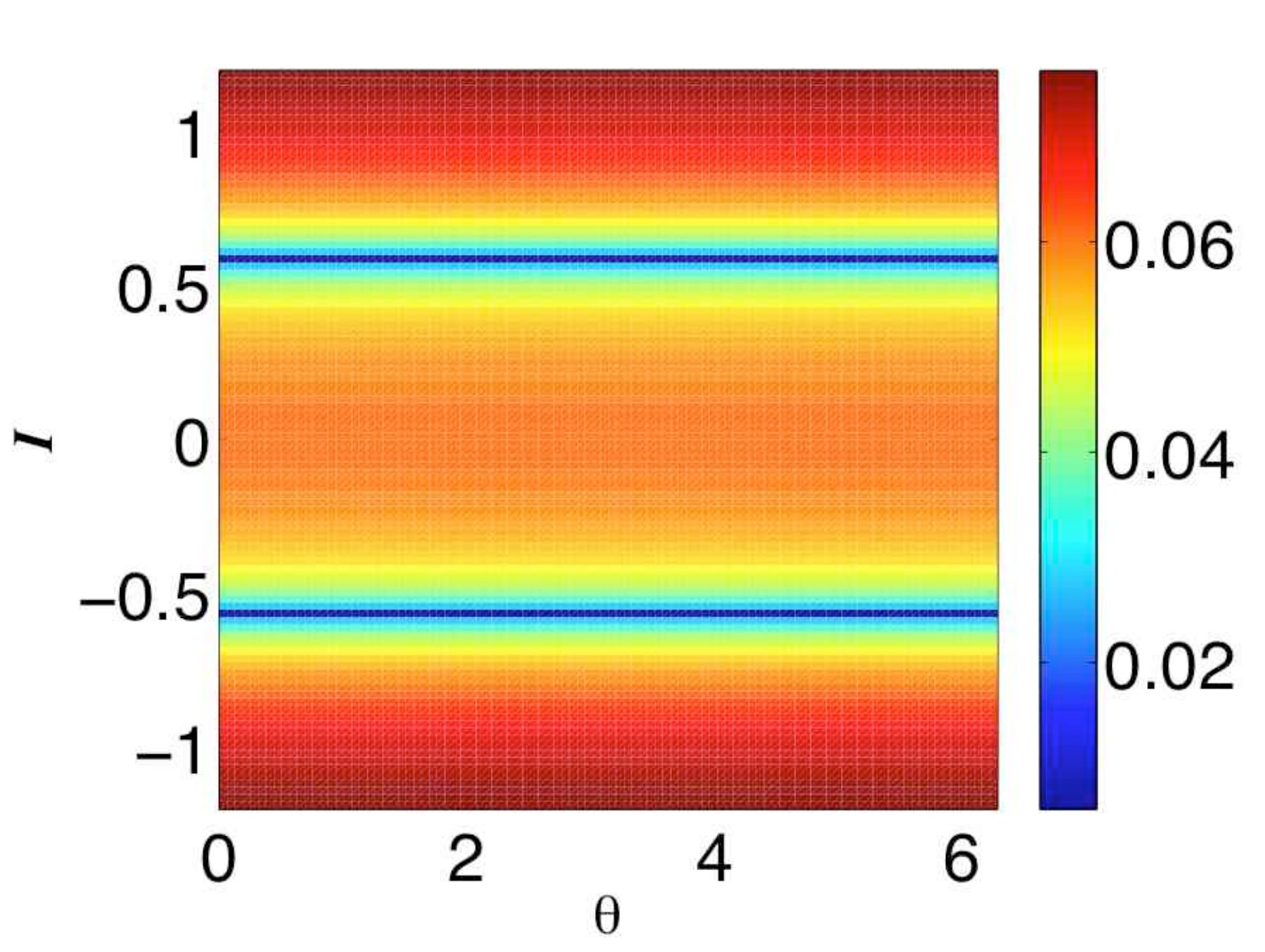}
\caption{\label{shear_ftle} FTLE for \eqref{eq:hamvf5}.  a) $\tau=25$; b) $\tau=75$.}
\end{figure}

In Figure \ref{shear_ftle} we compute the FTLEs for \eqref{eq:hamvf5} for $\tau = 25$ and $\tau = 75$. The velocity field is integrable and expressed in action-angle coordinates. Therefore we know that the Lyapunov exponents are identically zero. However, since the velocity field is nonlinear, the approach to the limit is ``nonuniform'' in the sense that different trajectories show approach the limit at different rates. Consequently, Figure \ref{shear_ftle} shows ``structure'' in the FTLE fields (in contrast to our example linear velocity fields), but this structure will disappear in the infinite time limit.  It is clear from the colour bar in Figure \ref{shear_ftle} that the FTLE are decreasing with $\tau$. Note that the smallest values for the FTLE appear near the invariant circle $I = \pm \frac{1}{\sqrt{3}}$,  which correspond to the ``twistless'' invariant tori.  It was shown in \cite{beron-vera-twistless} that twistless tori in integrable systems expressed in action-angle variables will have zero FTLE, for any time.

\subsubsection{Time Averages}
\label{sec:ergod_decomp}

In this section we will describe another approach for discovering phase space structure in dynamical systems that utilises time averages of certain functions along trajectories. Relating time averages to phase space structure is reminiscent of notions of ergodicity, and a general framework that has been used in the context of fluid flows (among other applications) is the {\em ergodic decomposition}. This approach was developed in
\cite{mezic2,mezic1,MW99}, and  is based on fundamental work of  \cite{rokhlin}. Superficially, this approach appears to have some similarity to the method of Lagrangian descriptors.  We will explain that this is not the case, and consider the application for the linear saddle \eqref{linex}, where this approach fails in several aspects.

We begin by giving a very brief description of the method. The setting is that of smooth ergodic theory. We have a compact domain, $A$, and a differentiable dynamical system defined on $A$ possessing an invariant measure. The dynamical system can be either discrete time, \emph{i.e.} a map, or continuous time, \emph{i.e.} a flow, and by ``flow'' we mean the precise mathematical definition as a one-parameter group of transformations of $A$. We denote by $L^1 (A)$ the space of scalar valued functions on $A$ having the property that the integral of its absolute value over $A$ exists.  It is essential that the Birkhoff ergodic theorem is applicable. In particular, we require that the time average of functions $ f \in L^1(A)$ along trajectories of the dynamical system exist for almost all (with respect to the invariant measure) initial conditions in $A$, and we  point out that by ``time average'' we mean infinite time average (as in the statement of the Birkhoff ergodic theorem).

Suppose we choose $f \in L^1 (A)$, and consider the function on $A$ defined by the time average of $f$ along all possible trajectories in $A$. It can be shown that the level sets of this function are invariant sets with respect to the dynamics. Now suppose we have an inner product on $A$ and we are able to choose a set of mutually orthogonal functions, $\{ f_i \}, \, i \in \mathbb{N}$, having the property that the set of all possible linear combinations of the $\{f_i \}$ are dense in $L^1 (A)$. For each $f_i$, we consider the function of $A$  given by the time average of trajectories through initial points in $A$, which we denote by $f^*$. Finally we consider the ``joint level sets''  that are constructed  by considering the  intersections of all level sets of all of the $f^*$.  This collection of ``joint level sets'' defines a partition of $A$, and the dynamical of each element of the partition is ergodic (the ``ergodic partition'').

An essential tool used  in this procedure is the Birkhoff ergodic theorem which requires compactness of the phase space as an essential ingredient.  Unfortunately, this theorem has not been proven for time dependent systems with general, aperiodic, time dependence.  The theorem holds for maps and for time periodic and quasi periodic velocity fields. In the latter case the dimension of the system can be increased by considering time  as an additional  dependent variable in the  definition of the velocity field. Since the velocity field is quasi periodic (of which periodicity is a special case) time is treated by including the independent (but finite) angles associated with each frequency component of the quasi periodic time dependence. In this way compactness is preserved (each included angular variable is compact) and the  velocity field, in this higher dimensional setting, is autonomous and generates a flow on a compact space.  For general, aperiodic time dependence transforming the problem to an autonomous setting in which compactness is preserved is generally not possible (see, e.g. \cite{bal10} for a recent review of the relevant issues for aperiodic time dependence).

Even though the Birkhoff ergodic theorem has not been proven for general aperiodic time dependence, the general approach suggested by the ergodic  decomposition has proven fruitful, especially when the $f_i$ under consideration have a physical meaning (and this is not unlike the approach for constructing Lagrangian descriptors).
In particular,   the finite time average of the horizontal component of the velocity field has been shown to provide insight in the flow in numerous different situations by \cite{mezic, mezic1, mezic2}. For the general velocity field, \eqref{gds},  let $\mathbf{x} (t, t^*, \tau)$ denote the trajectory satisfying $x (\mathbf{x} (t^*, t^*, \tau) = \mathbf{x}_0$ and let $v_x$ denote the $x$ component of the velocity field. Then the finite time average of $v_x$ along this trajectory is denoted by:

\begin{equation}
\mathbf{v}_x^*(\mathbf{x}_0, t^*, \tau) = \frac{1}{\tau} \int_{t^*}^{t^* + \tau} \mathbf{v}_x (\mathbf{x} (t, t^*, \tau), t ) dt
\label{eq:vxav}
\end{equation}
\noindent
Since the integral is only over a finite length (in time) of the trajectory, conditions for its existence are relatively mild.  However, the ``meaning'' of such results for such finite time averages is not a priori clear. The lack of a rigorous theoretical framework is similar to the situation for FTLEs.  The typical approach is to compute the relevant quantities (e.g., \eqref{eq:vxav}),  ``see'' if the computations yield something ``interesting'', and then to investigate  further.

The linear velocity field \eqref{linex} is defined on a non compact domain, thus the Birkhoff ergodic theorem does not apply to it,  in the sense that it does not allow us to conclude that time averages along trajectories exist. Nevertheless, we can still compute the finite time averages along trajectories of physically interesting quantities in the spirit of the patchiness work described above.
For this example the finite time average of
the horizontal component of the velocity field can be explicitly computed. Taking $t_*=0, \, \lambda=1$, and denoting the finite time average of $v_x$ by $v_x^*$,  we have:

\begin{equation}
v_x^*=\frac{x_0}{\tau} (exp(\tau)-1).
\label{linear_fta}
\end{equation}

\begin{figure}
 \centering
\includegraphics[width=8cm]{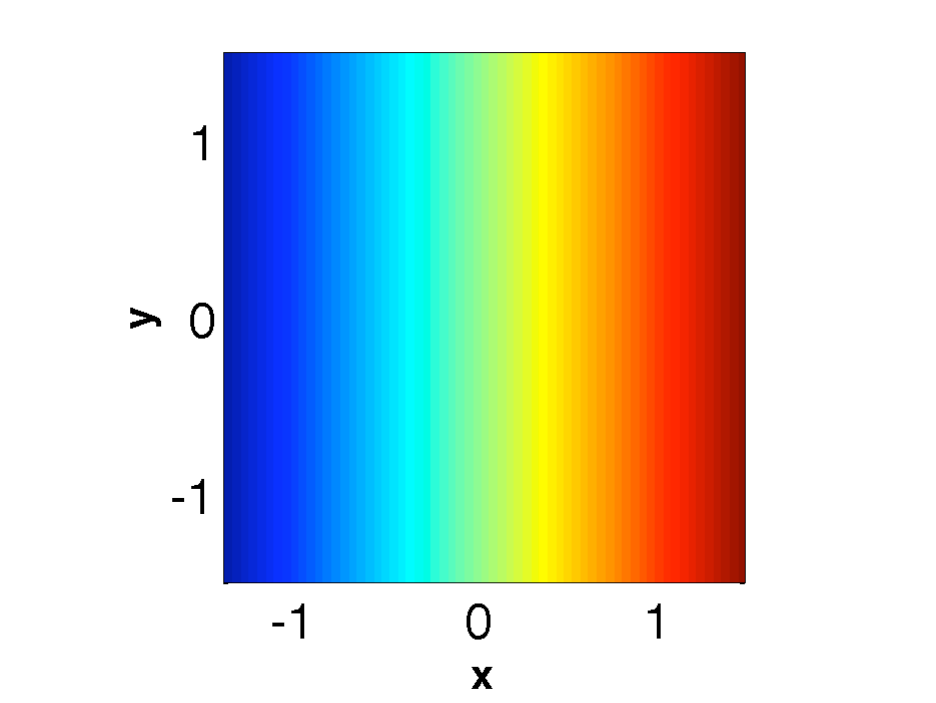}
\caption{\label{meziclinear} Integral of the horizontal component of the velocity along trajectories of  the  linear saddle point \eqref{linex}  for $\tau=10$.}
\end{figure}

\noindent
Clearly, the limit does not exist as $\tau \rightarrow \infty$.  In figure \ref{meziclinear} we plot the contours of \eqref{linear_fta}  for $\tau=10$. It is also clear that the contour plot does not reveal the presence of the stable manifold nor invariant sets, despite the system, which  is Hamiltonian, contains invariant sets defined by $H(x,y)=\lambda x y=c$ . The stable manifold of the hyperbolic trajectory at the origin is given by $x_0=0$. We see from \eqref{linear_fta} that the finite time average changes sign at $x_0 =0$ (and this is true for any $\tau$). However, the derivative of \eqref{linear_fta} with respect to $x_0$ is smooth across $x_0=0$. Hence, the finite time average of the horizontal velocity component has no singular features that would indicate the presence of material curves. A reason for this behaviour is that general finite time averages do not require that the integrand is non-negative along trajectories, which is an essential requirement for Lagrangian descriptors, and motivation for this was given in the heuristic argument. Typically integrals of oscillating quantities along trajectories lead to a distorted output which makes interpretation of the results, in terms for phase space structure, difficult.  In certain cases it was noted in \cite{mezic1}  that for increasing averaging time  oscillations could lead to  average velocities that approached zero everywhere,
and as a consequence, in this limit any spatial structure is completely lost.

\subsection{Application to the Forced Duffing Oscillator}
\label{sec:duffing}

In this section we will discuss further the ability of Lagrangian descriptors to highlight manifolds. Also  we will compare and contrast the behavior of Lagrangian descriptors with FTLEs and time averages of velocity components along trajectories  using a familiar, but much more complicated nonlinear system --the forced Duffing equation:

\begin{eqnarray}
\dot{x}&=&y,\nonumber \\
\dot{y}&=&x-x^3+\varepsilon f (t),\label{eq:duffing}
\end{eqnarray}

\noindent
We will consider three cases.

\begin{description}

\item[Integrable case.] In this case we take $\epsilon=0$ in \eqref{eq:duffing}. The phase space structure of the resulting system is well-known. The system has a hyperbolic fixed point at the origin connected by a symmetric pair of homoclinic orbits.  The region inside and outside the homoclinic orbits consist of continuous families of periodic orbits, and correspond to elliptic regions.

\item[Periodically time dependent case.] For this case we set  $\varepsilon=0.1$  and  $f(t)=\sin(t)$ \eqref{eq:duffing}. The system has
a hyperbolic periodic trajectory near the origin, referred to as a {\em distinguished hyperbolic trajectory}, or DHT  (see \cite{physd,chaos,kayo}).

\item[Aperiodically time dependent case.]  The aperiodically time dependent forcing, $f(t)$, has the form shown in  figure \ref{forcing}, which his obtained from a  trajectory of the Duffing equation in a chaotic regime. We have chosen $\varepsilon=0.15$, which results in an amplitude of the time dependence  that is, roughly, and order of magnitude smaller that the time periodic case considered above.  In this case the system also has a hyperbolic trajectory near the origin.

 \begin{figure}
 \centering
\includegraphics[width=10cm]{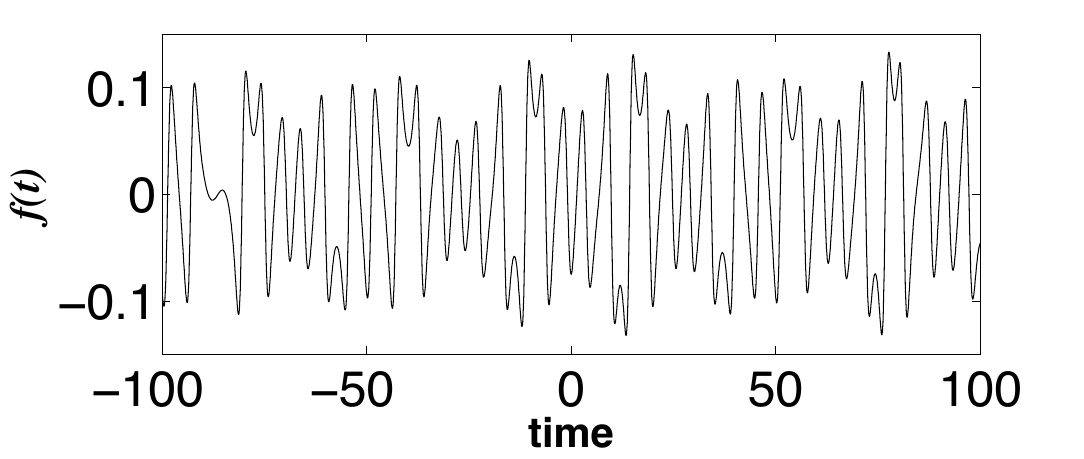}
\caption{\label{forcing} The time series of the function $f(t)$ corresponding to aperiodic time dependence  for the Duffing Equation.}
\end{figure}

\end{description}

\medskip
\paragraph{Singular Features of the Contour Plots of Lagrangian Descriptors.}

\begin{figure}
a)\includegraphics[width=0.5\linewidth]{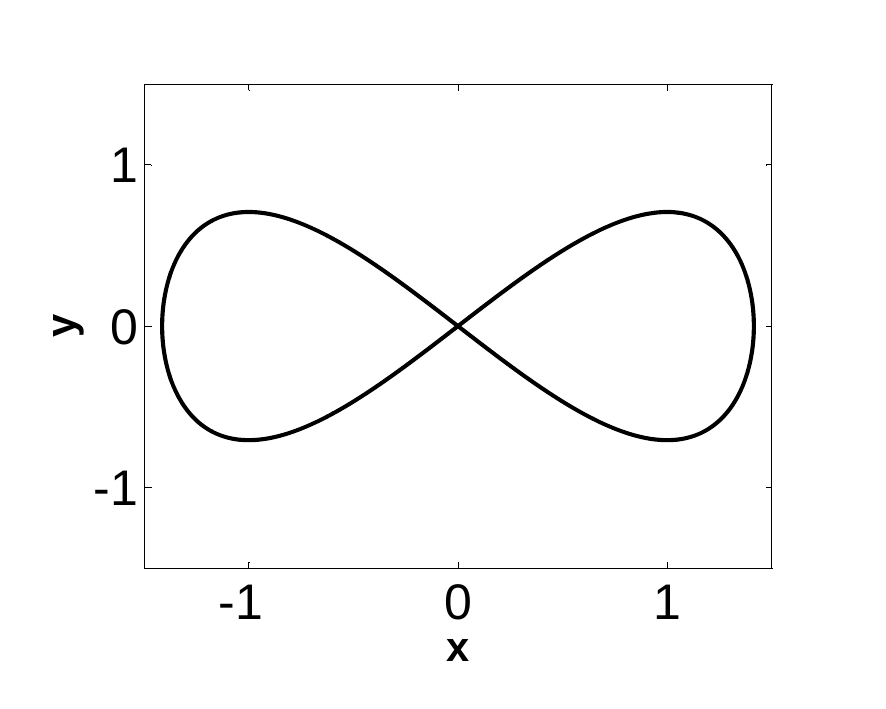}
d)\includegraphics[width=0.5\linewidth]{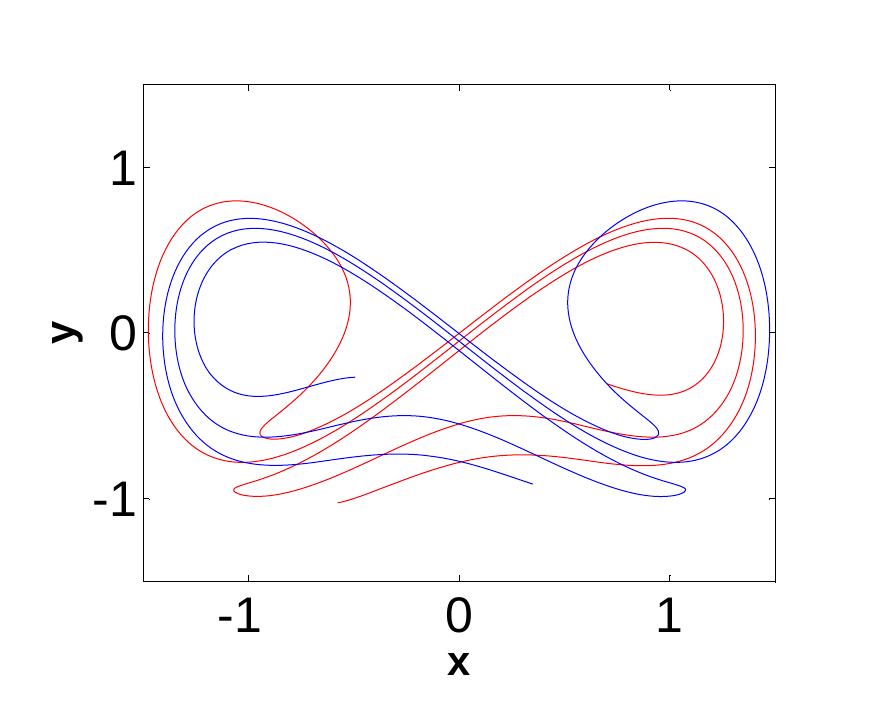} \\
b)\includegraphics[width=0.5\linewidth]{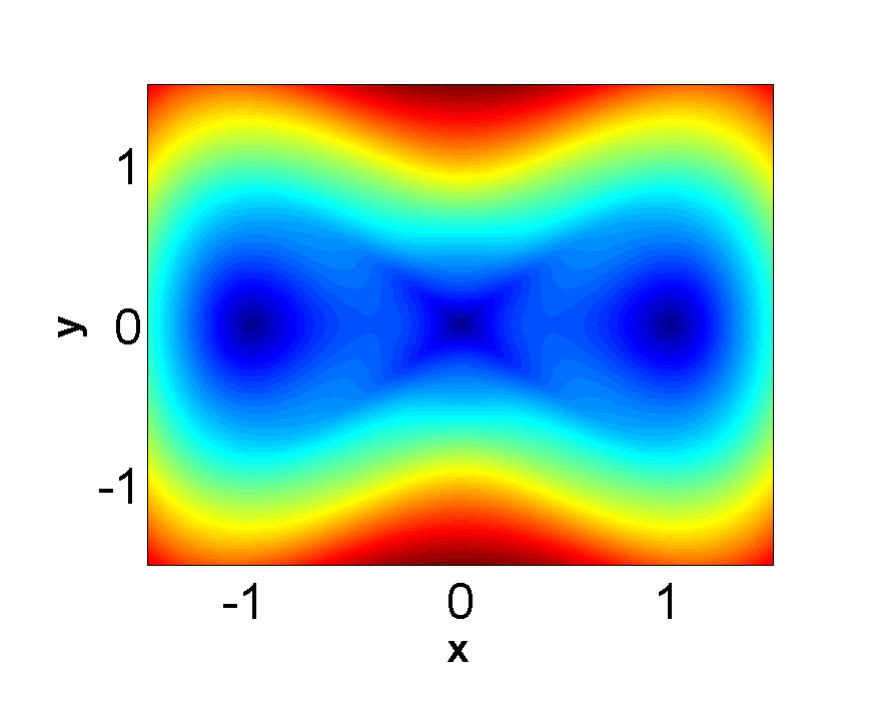} 
e)\includegraphics[width=0.5\linewidth]{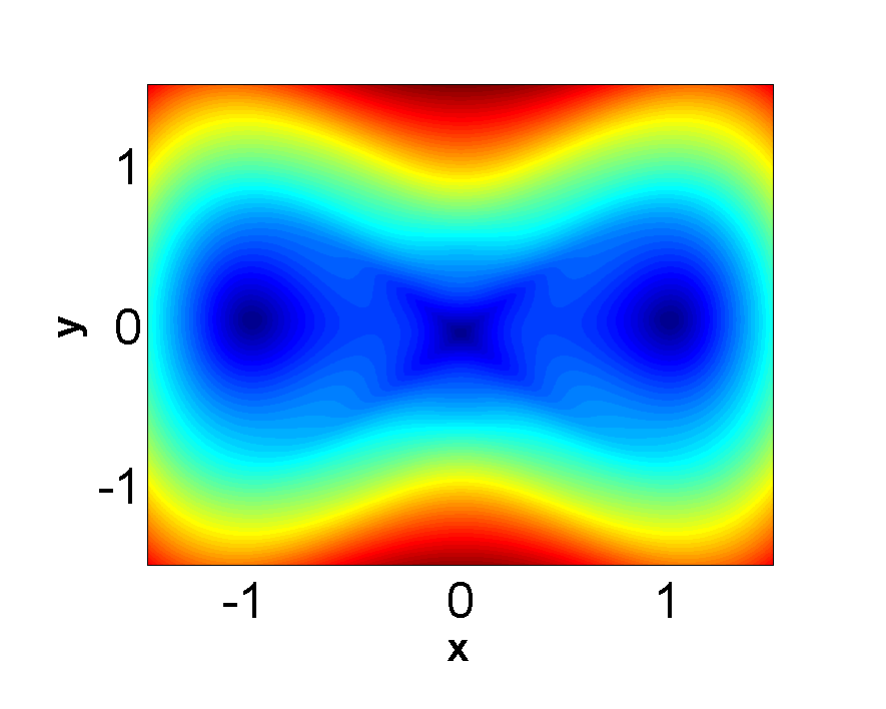}\\
c)\includegraphics[width=0.5\linewidth]{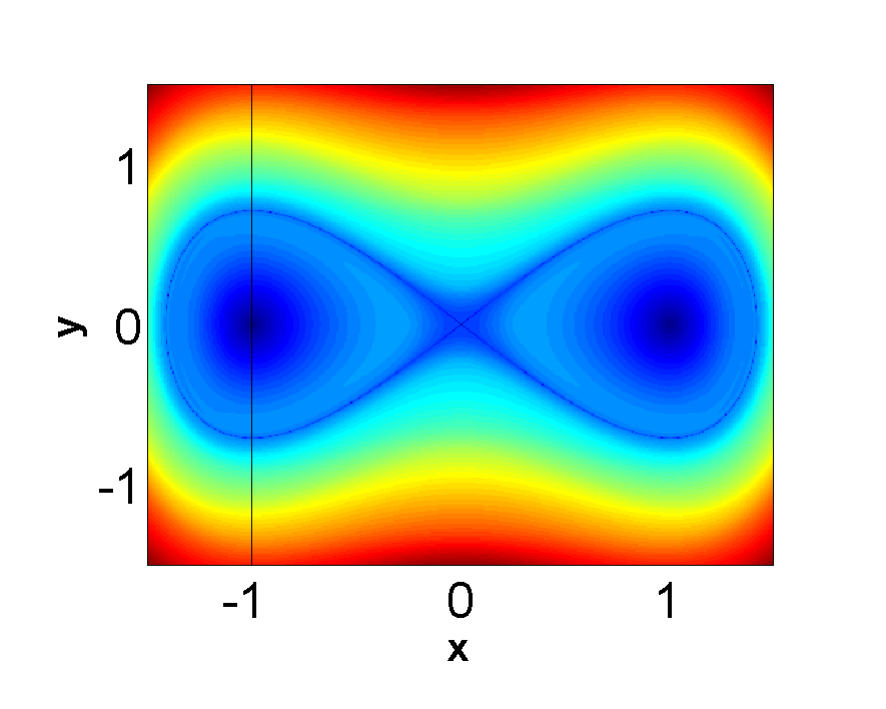}
f)\includegraphics[width=0.5\linewidth]{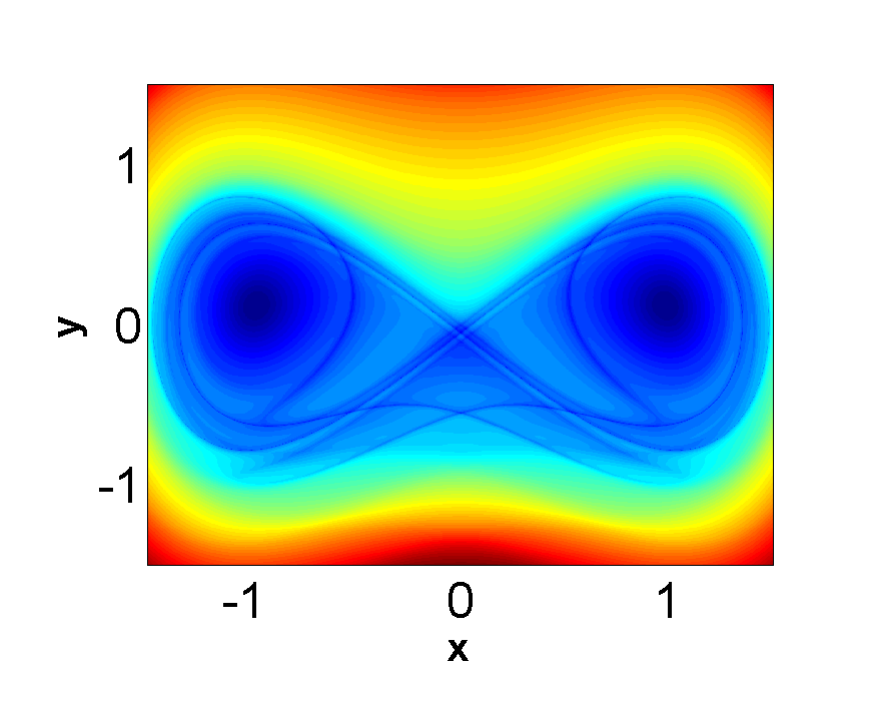}
\caption{\label{duffing2} For the integrable case: a) the stable and unstable manifolds of the hyperbolic fixed point; 
b) contours of $M_1$ for $\tau=2$; c) contours of $M_1$  for $\tau=10$.
For the periodically forced Duffing equation: d) Segments of the stable and unstable manifolds of the hyperbolic trajectory near the origin ; e) contours of $M_1$ for $\tau=2$; f) contours of $M_1$ for $\tau=10$.}
\end{figure}

In our study of the linear saddle in Section \ref{sec:linsad} we argued that Lagrangian descriptors had two important properties. 1) They depend on the time
of integration $\tau$.  2) After a certain convergence time they display ``singular contours'' that correspond to the stable and unstable manifolds of hyperbolic trajectories.
By  ``singular contours'', we mean that the Lagrangian descriptor is a function lacking regularity, either because the Lagrangian descriptor is itself discontinuous across the manifolds  (we will see this is the case in Section 3) or because   its derivative transverse to the manifolds is discontinuous.
Here we will show that these two properties also hold for the cases of the forced Duffing equation that we have considered.  In figure \ref{duffing2} we show the stable and unstable manifolds of the hyperbolic trajectory and contours of the function $M_1$ for $\tau=2$ and $\tau=10$.
For both cases we see that the contours of $M_1$ converge to the stable and unstable manifolds.

\begin{figure}
\centering
\includegraphics[width=8cm]{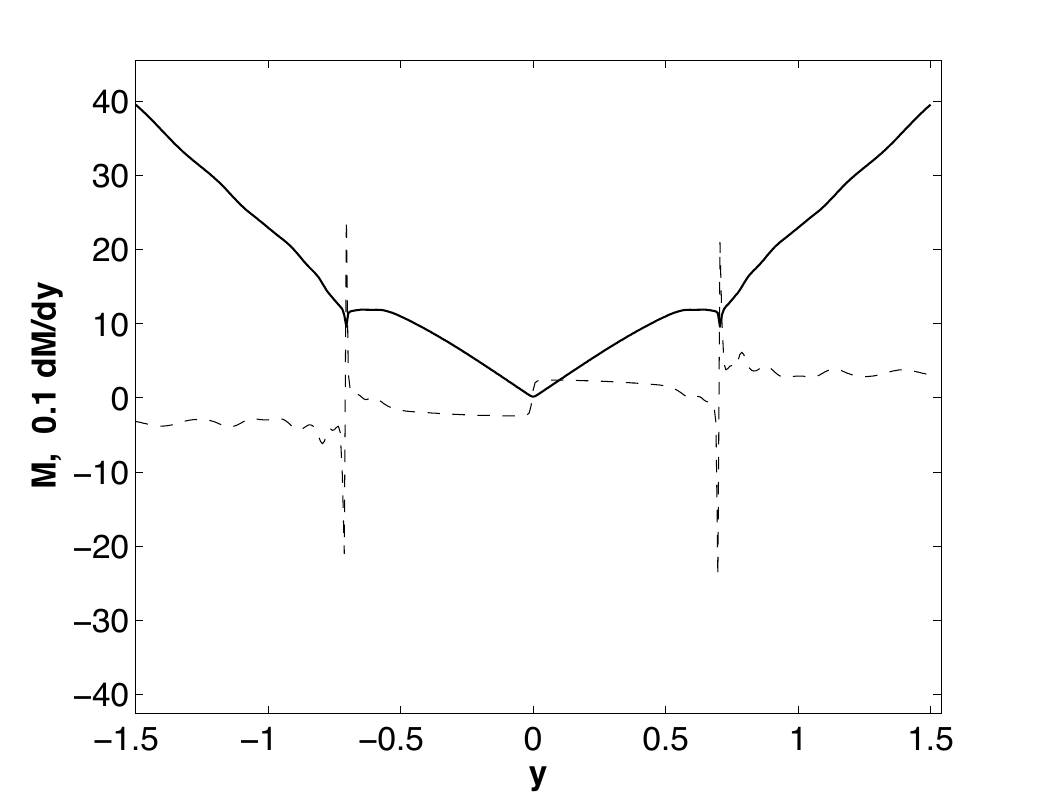}
\caption{\label{duffingsharp} The black solid line represents the function $M_1$ vs $y$
at a fixed $x$. $M_1$  along the vertical  black line shown in Figure \ref{duffing}c). Abrupt changes in $M_1$, occurring at the location of the manifolds, correspond to
discontinuities of the derivative of $M_1$. The dashed line represents  0.1 times the derivative of $M_1$ with respect to $y$.}
\end{figure}

Figure \ref{duffingsharp} further illustrates how singular features of the contours of $M_1$ correspond to stable and unstable manifolds of hyperbolic trajectories. The figure shows values of $M_1$ along the vertical line shown in Figure \ref{duffing2}c). We see that the value of $M_1$  ``change abruptly'' at the locations of the manifolds. Discontinuities in the derivative of $M_1$ quantify the notion  of ``abrupt change''  of $M_1$. The latter are illustrated  in the dashed line in the figure.

\begin{figure}
a)\includegraphics[width=7cm]{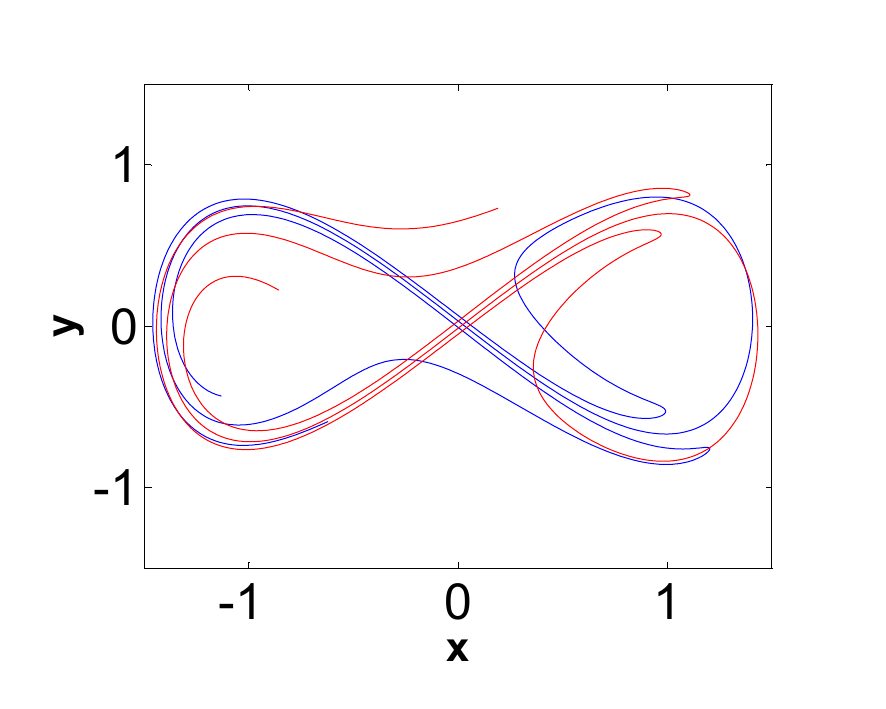}
b)\includegraphics[width=7cm]{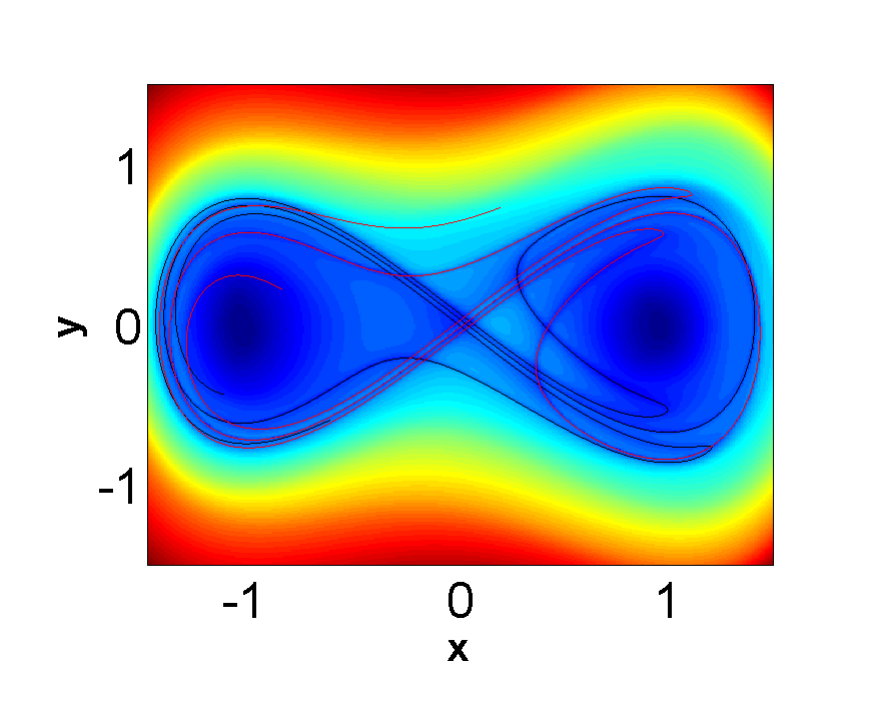}\hfill
\caption{\label{overlay} The aperiodically time dependent case. a) The stable and unstable manifolds of the hyperbolic  trajectory near the origin. b) An overlay of the stable and unstable manifolds of the hyperbolic trajectory near the origin with the contours of the $M_1$. The manifolds and $M_1$ are computed for $\tau=10$.  }
\end{figure}

In figure \ref{overlay} a) we show segments of the stable and unstable manifolds of the hyperbolic trajectory near the origin. In  figure \ref{overlay} b) we show the same manifolds overlaid with the contours of $M_1$ computed for $\tau =10$. The manifolds align well with the ``singular contours'' of $M_1$.

\medskip
\paragraph{Lagrangian Descriptors vs FTLE.}

\begin{figure}
\centering
a)\includegraphics[width=0.3\linewidth]{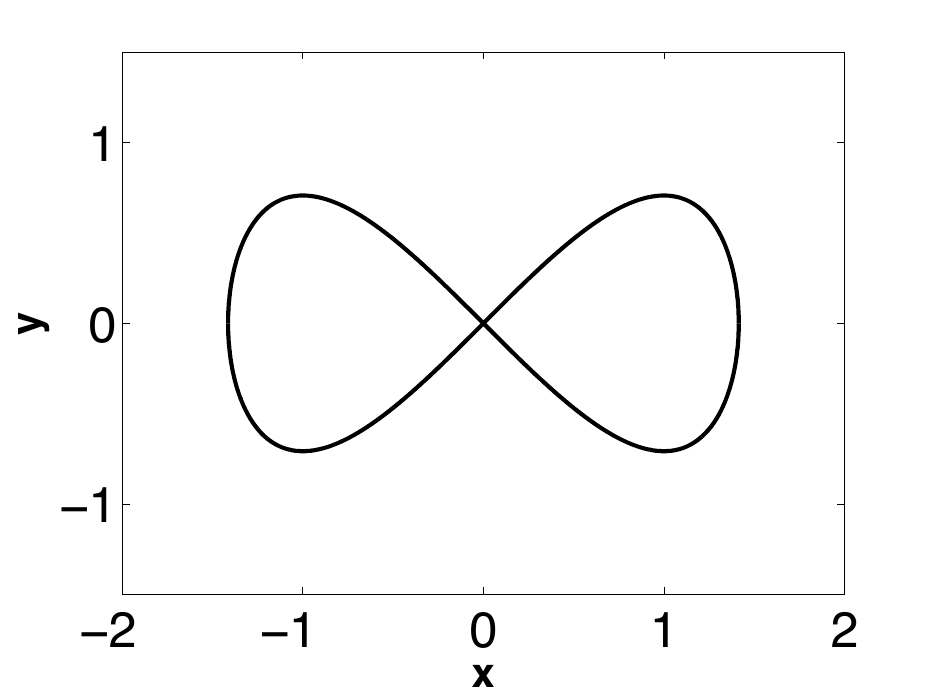}
b)\includegraphics[width=0.3\linewidth]{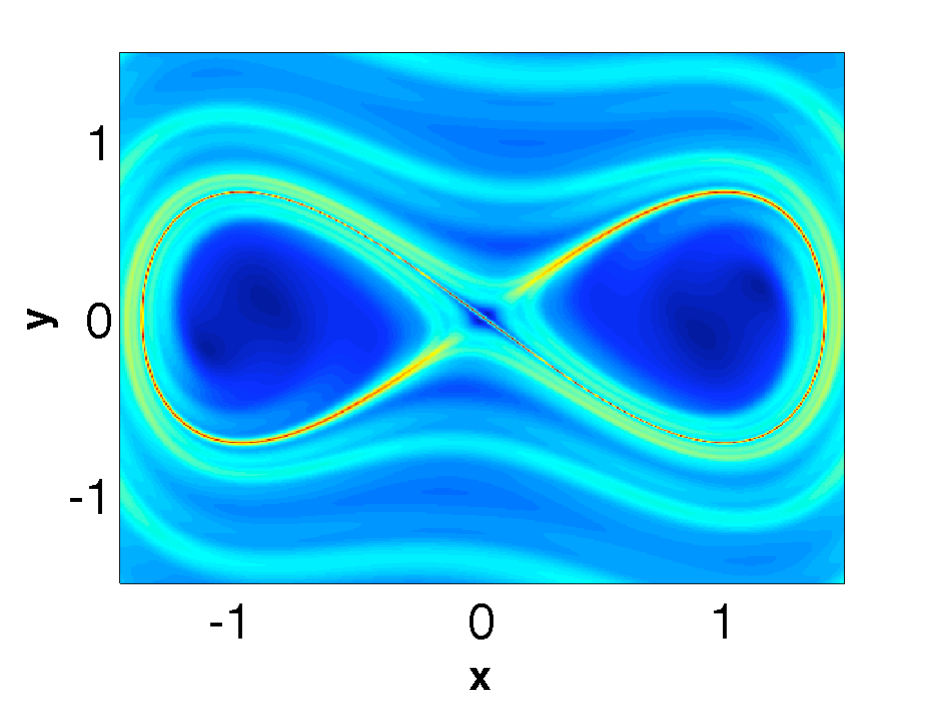}
c)\includegraphics[width=0.3\linewidth]{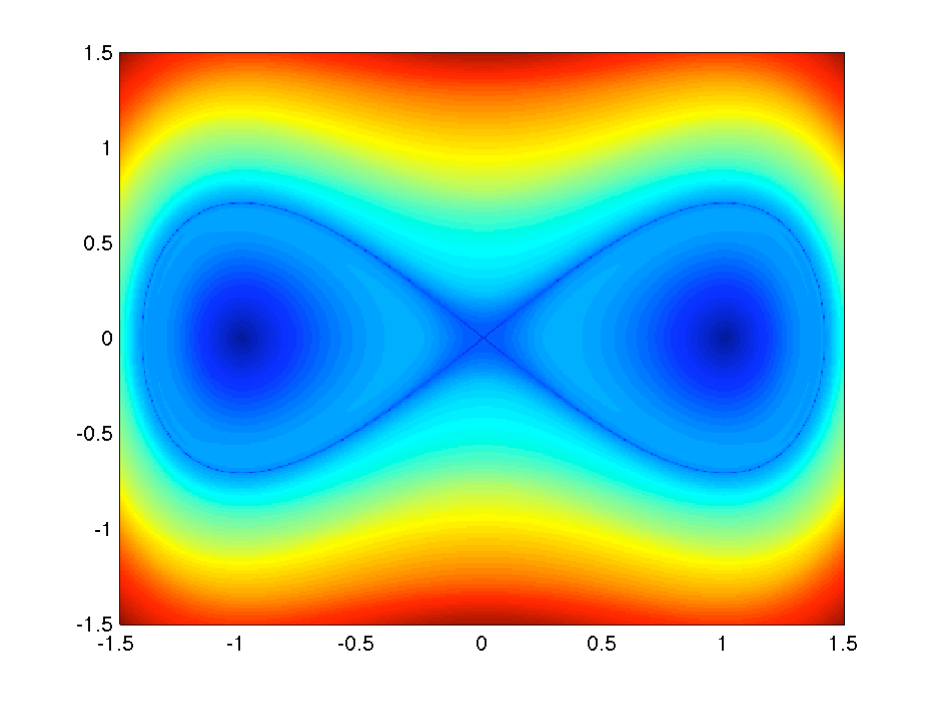} \hfill

\medskip
d)\includegraphics[width=0.3\linewidth]{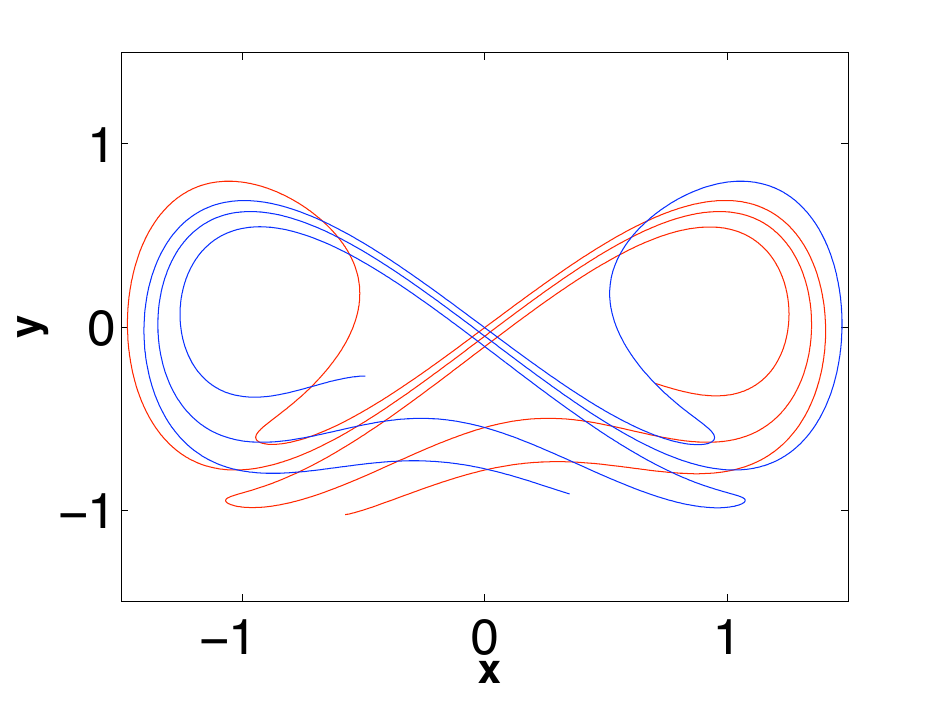}
e)\includegraphics[width=0.3\linewidth]{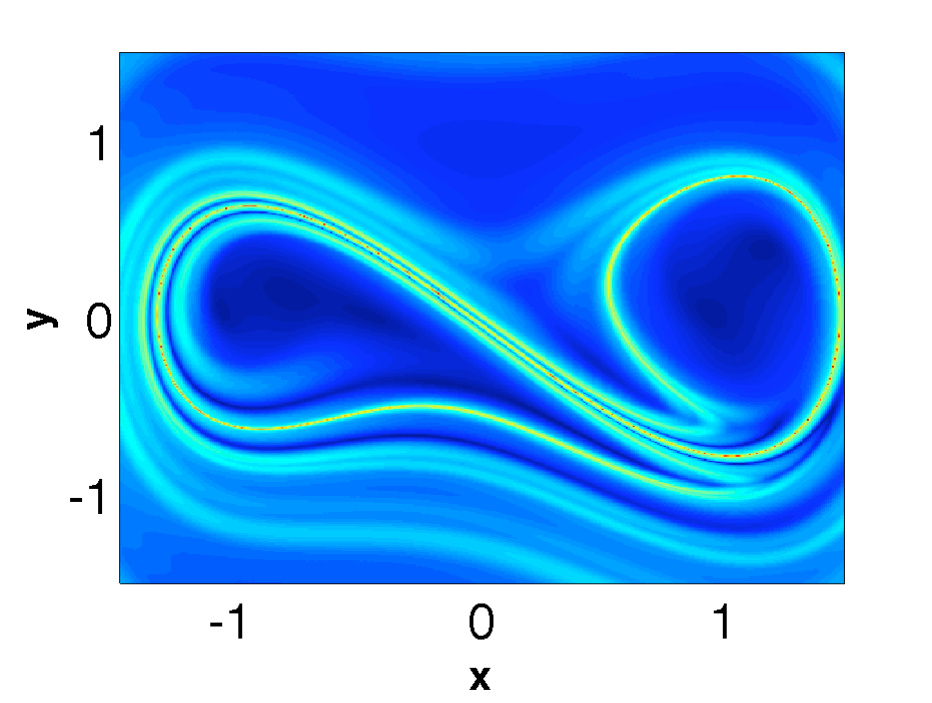}
f)\includegraphics[width=0.3\linewidth]{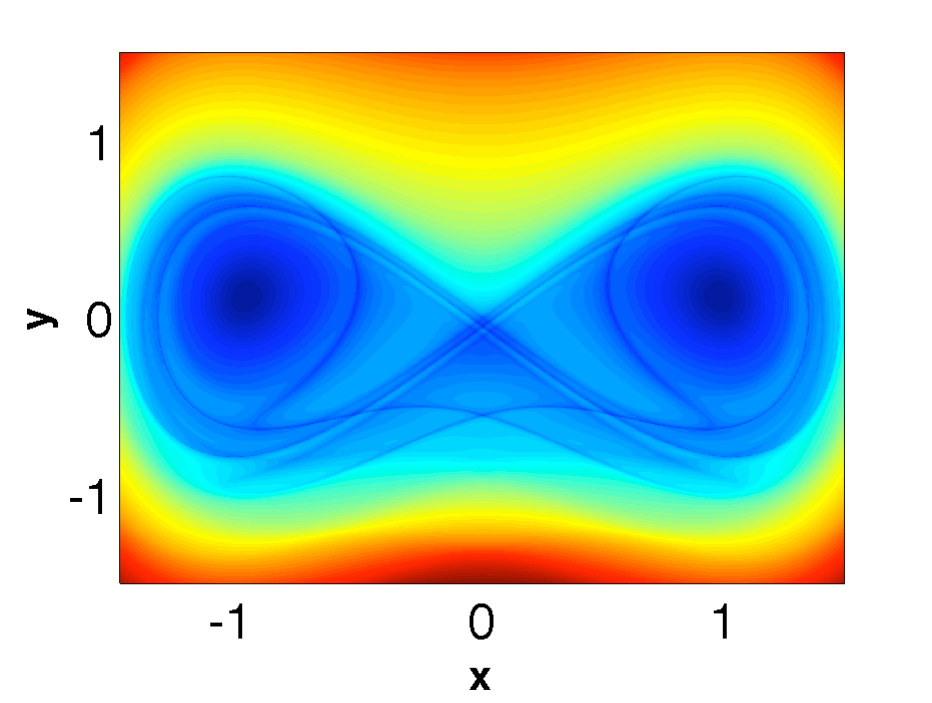} \hfill

\medskip
g)\includegraphics[width=0.3\linewidth]{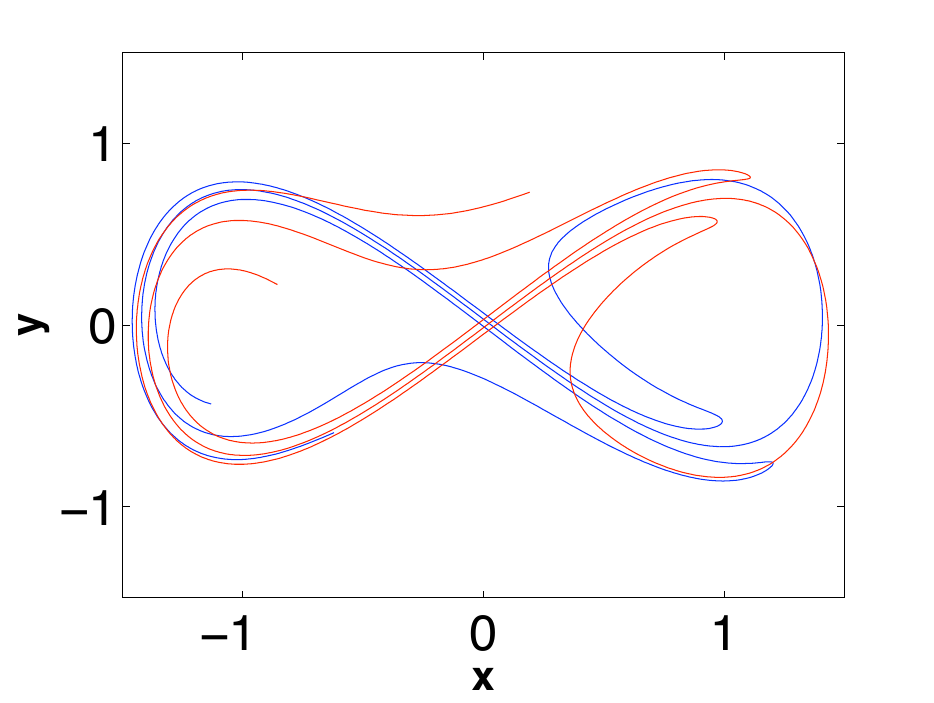}
h)\includegraphics[width=0.3\linewidth]{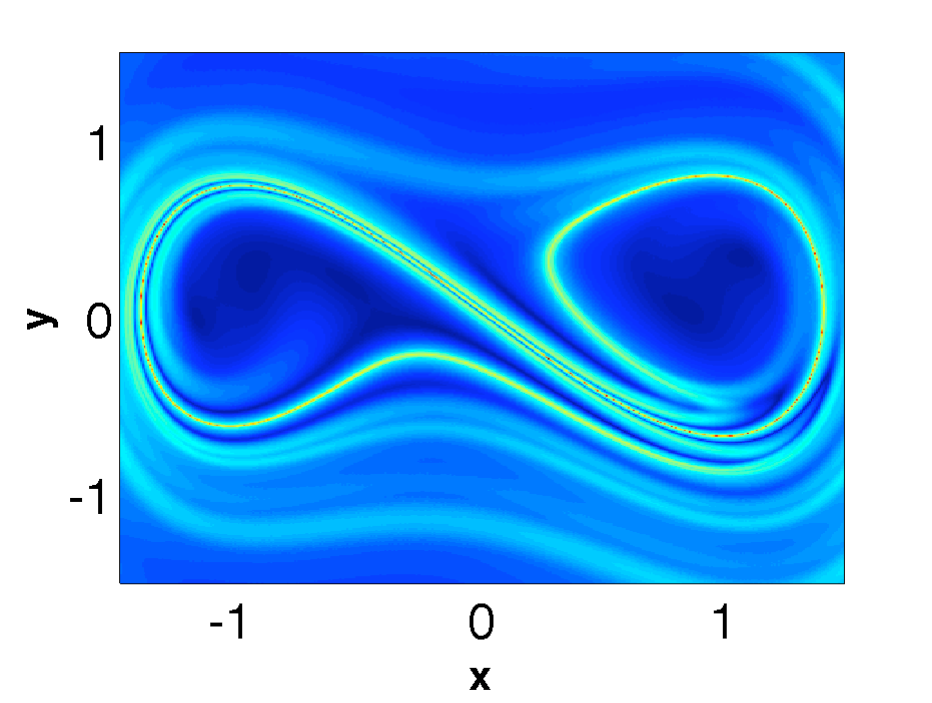}
i)\includegraphics[width=0.3\linewidth]{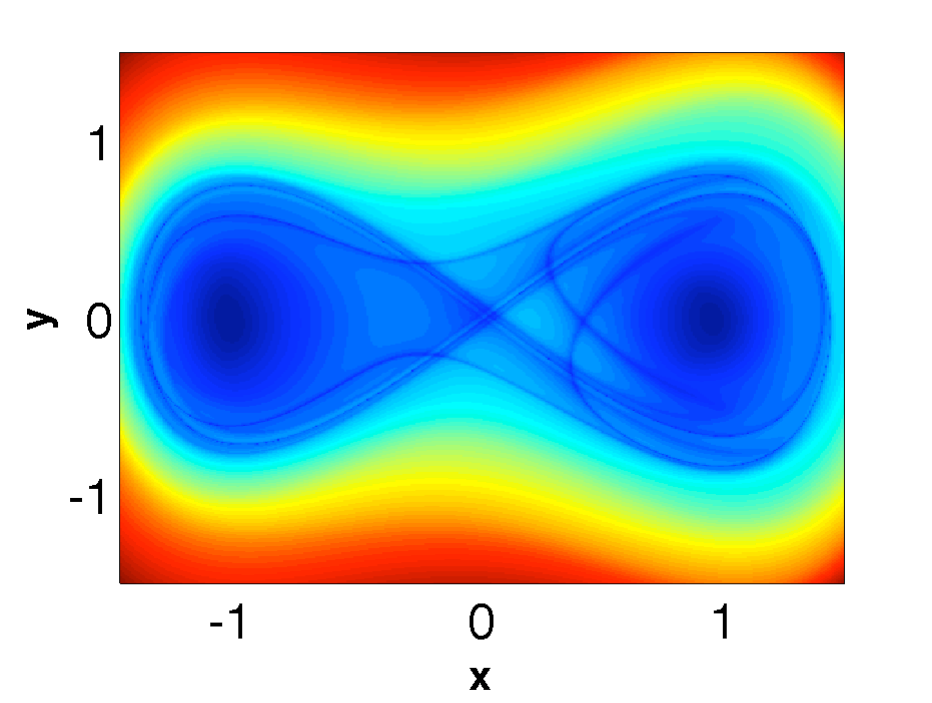} \hfill
\caption{\label{duffing} For the integrable Duffing equation: a) the stable and unstable manifolds of the hyperbolic fixed point; b) forward FTLE for $\tau =10$; c) contours of  $M_1$ for $\tau=10$.
For the periodically forced Duffing equation: d) segments of the stable and unstable manifolds of the hyperbolic trajectory near the origin  computed for $\tau =10$ (and displayed at $t=0$); e) forward FTLE for $\tau =10$; f) contours of  $M_1$ for $\tau=10$.
For the aperiodically forced Duffing equation: g) segments of the stable and unstable manifolds of the hyperbolic trajectory near the origin  computed for $\tau =10$ (and displayed at $t=0$); h) forward FTLE for $\tau =10$; i) contours of  $M_1$ for $\tau=10$.}
\end{figure}

In figure \ref{duffing} we show the stable and unstable manifolds of the hyperbolic trajectory near the origin (left most column), the forward time FTLEs (middle column) and the contours of the Lagrangian descriptor $M_1$ for the three cases, and all for $\tau=10$.
In all cases the ``singular'' contours of $M_1$ approximate the manifolds much better than the contours of the forward FTLE, since although the forward FTLE only captures the stable manifold of the hyperbolic trajectory near the origin,  it displays ridges which are artifacts in the FTLE field that have no Lagrangian interpretation, and that are not present on the contour plots of $M_1$.

\medskip
\paragraph{Lagrangian Descriptors in Elliptic Regions.}

Next we consider the behaviour of several Lagrangian descriptors  in ``elliptic regions'' (cf. the discussion of elliptic and hyperbolic regions in Section \ref{secLd}) for both the periodic and aperiodically time dependent cases. The elliptic region that we consider consists of the region enclosed by the left hand homoclinic orbit in the $\epsilon = 0$ case (but we consider the behaviour of the Lagrangian descriptors in this region for the periodic and aperiodically time dependent cases).

\begin{figure}
a)\includegraphics[width=7.5cm]{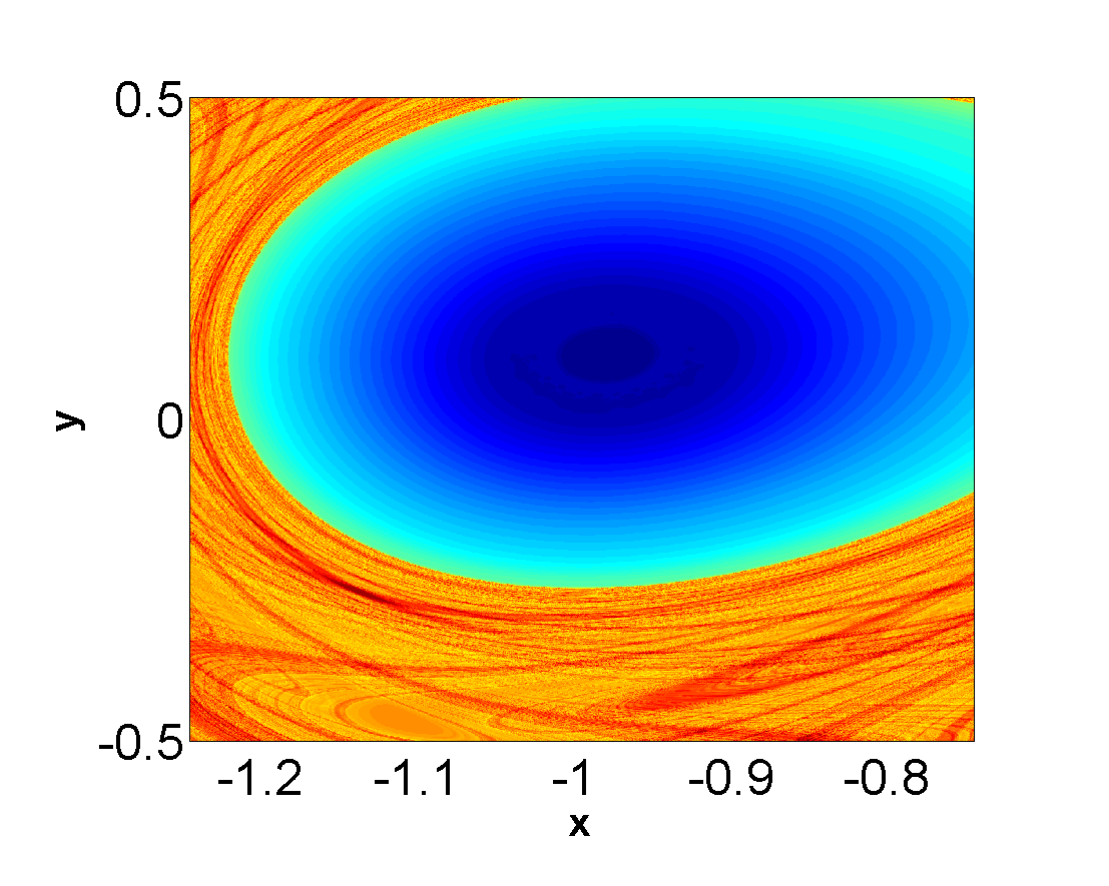}
b)\includegraphics[width=7.5cm]{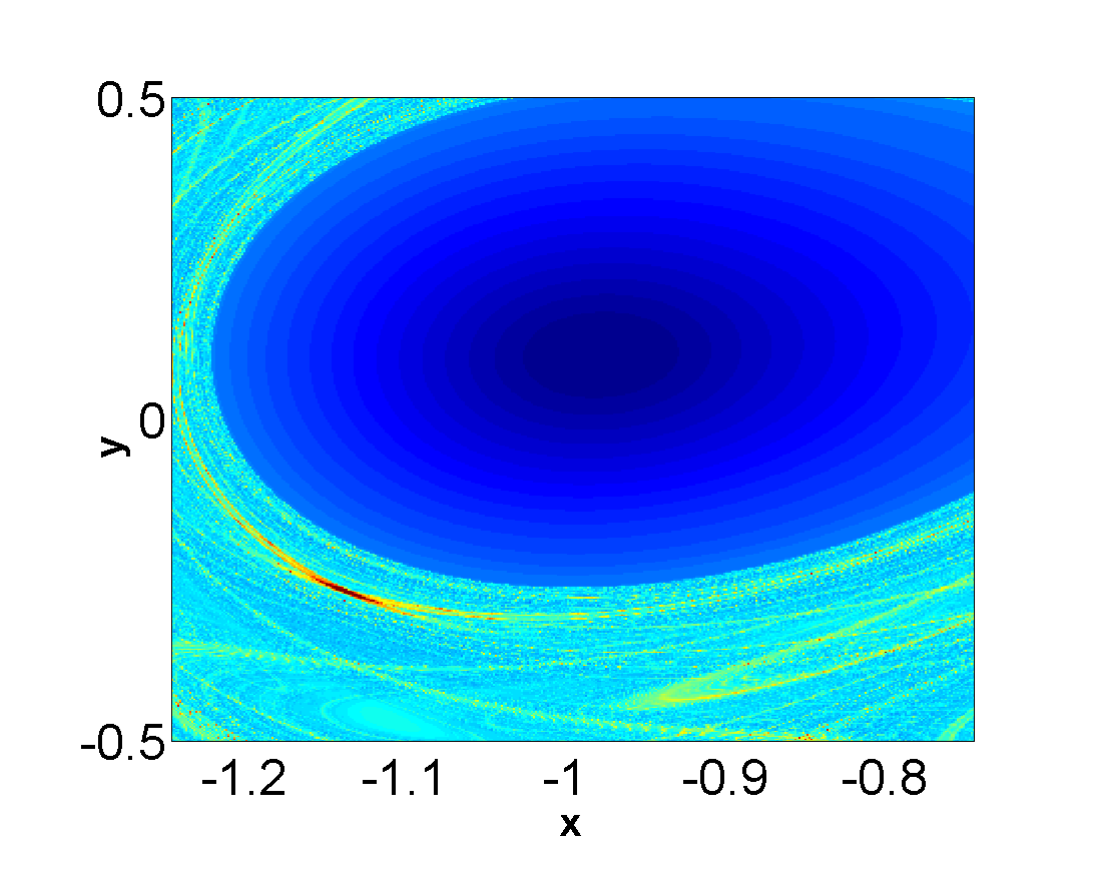}\hfill
\medskip
c)\includegraphics[width=7.5cm]{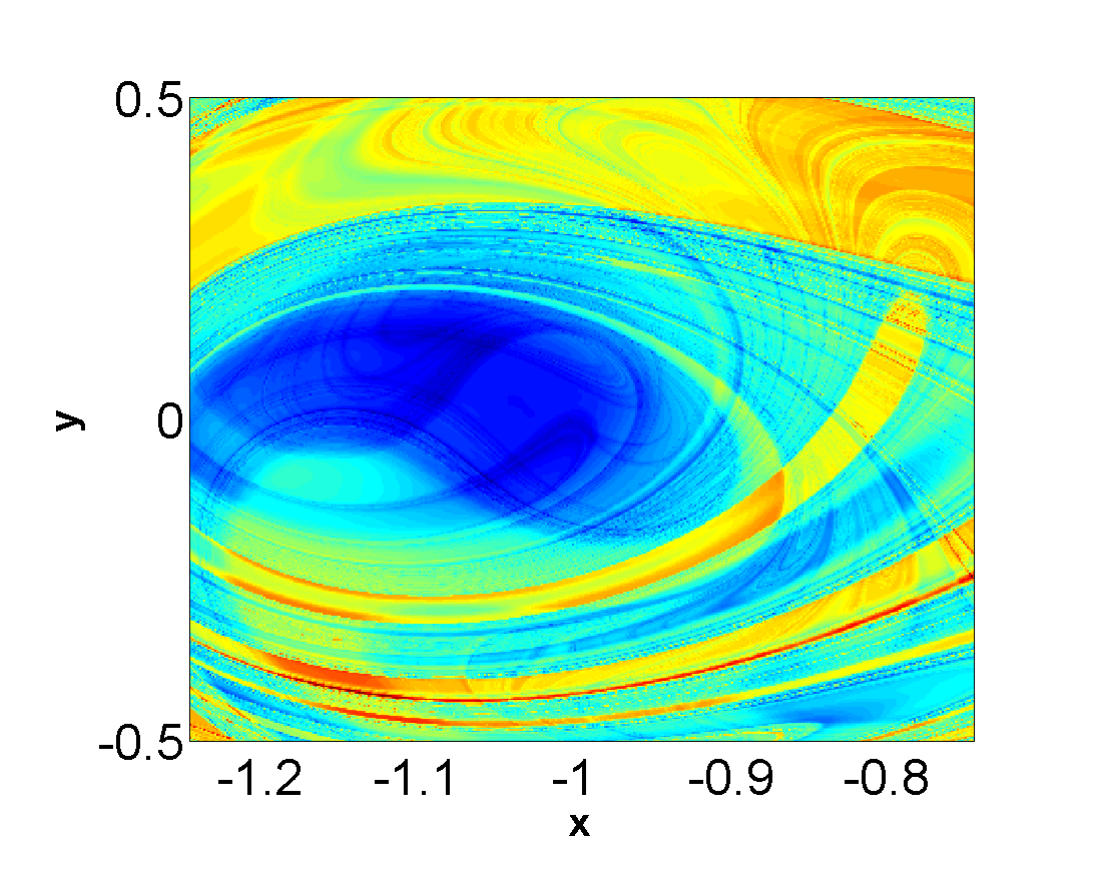}
d)\includegraphics[width=7.5cm]{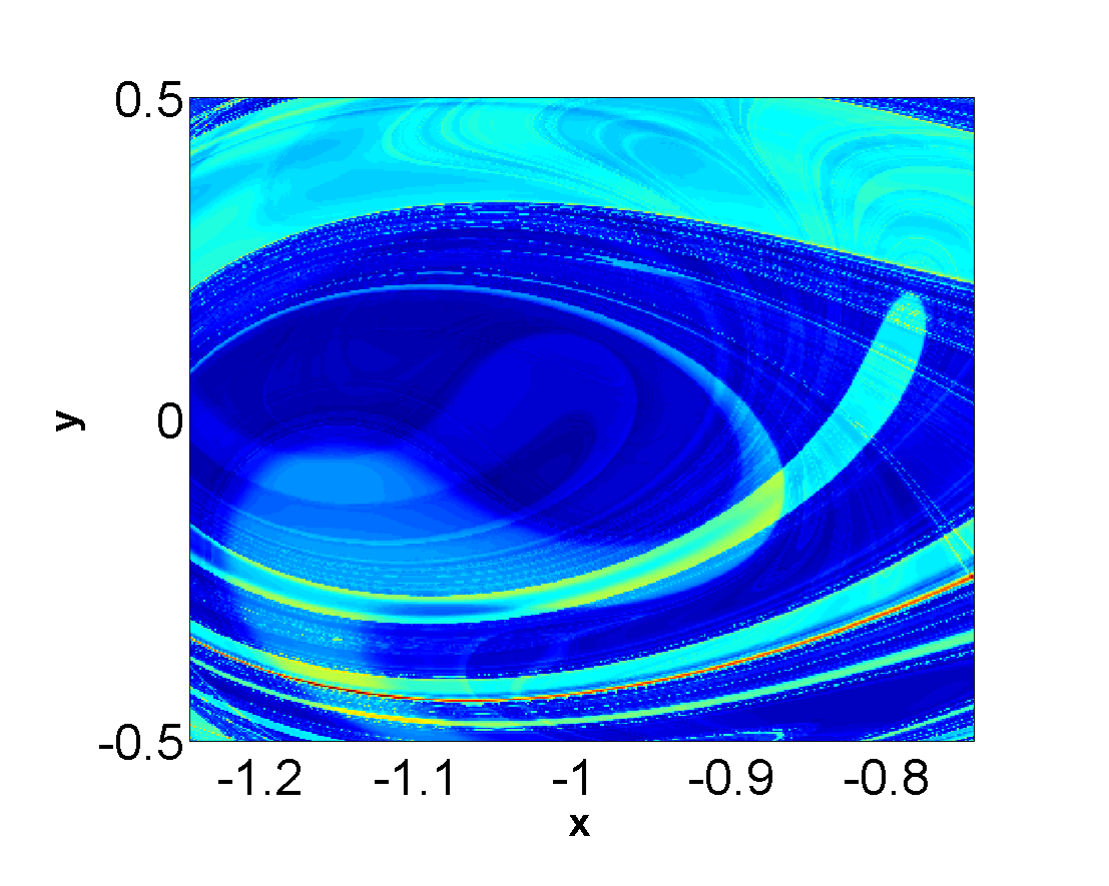}\\
\caption{\label{duffedd} Contour plots of the of Lagrangian descriptors in the region corresponding to that  enclosed by the left hand homoclinic orbits for $\epsilon =0$. Panels  a) $M_5$; b) $M_2$ correspond to the periodically time dependent case. Panels  c) $M_3$; d) $M_4$ correspond to the aperiodically time dependent case. The Lagrangian descriptors are computed for $\tau=70$. }
\end{figure}

We first consider the Lagrangian descriptors $M_5$ and $M_2$  in the periodically time dependent case.
 Figure \ref{duffedd} a) and b)  shows the contours of these Lagrangian descriptors computed for $\tau=70$. The contours of each Lagrangian descriptor show a smooth structure near the centre of the elliptic region. This is not surprising since  for a ``weak'' time-periodic perturbation of an integrable system we expect ``most'' of the closed trajectories to be preserved as (two-frequency) Kolmogorov-Arnold-Moser tori. This  would imply that trajectories starting ``close'' to each other would have a similar past and future on the time interval $(t-\tau, t+\tau)$.
Outside this smooth region the contours of $M_5$ and $M_2$ reveal a more complex structure. This ``complex structure'' appears to be associated with strong mixing. This suggests that $\tau$ may be a useful parameter for quantifying the notion of ``finite time mixing'', but making this notion precise requires further research. Practically, the notion of ``finite time mixing'' is very important in applications, and lies outside the scope of notions of mixing in traditional ergodic theory.

Next we  consider  the aperiodically time dependent case in the same region. In this case the isolated region of  ``trapped'' trajectories shown in Figure \ref{duffedd} a) and b)  is  ``broken'' as it is ``invaded'' by segments of the stable and unstable manifolds of the hyperbolic trajectory near the origin, as shown in Figure \ref{duffedd} c) and d) for the Lagrangian descriptors  $M_3$ and $M_4$ (more details are evident from the contours of $M_3$).
Although not shown, we note that the Lagrangian descriptors $M_1$, $M_2$ and $M_5$ give results similar to those of $M_4$ and $M_3$ in this case.

\medskip
\paragraph{Time Averages of Velocity Components.}

We now return to the issues raised in Section \ref{sec:ergod_decomp}. In particular, we compute the finite time average of the horizontal component of the Duffing equation, following \eqref{eq:vxav}, for the three different time dependencies that we are considering.

  \begin{figure}
a)\includegraphics[width=7.5cm]{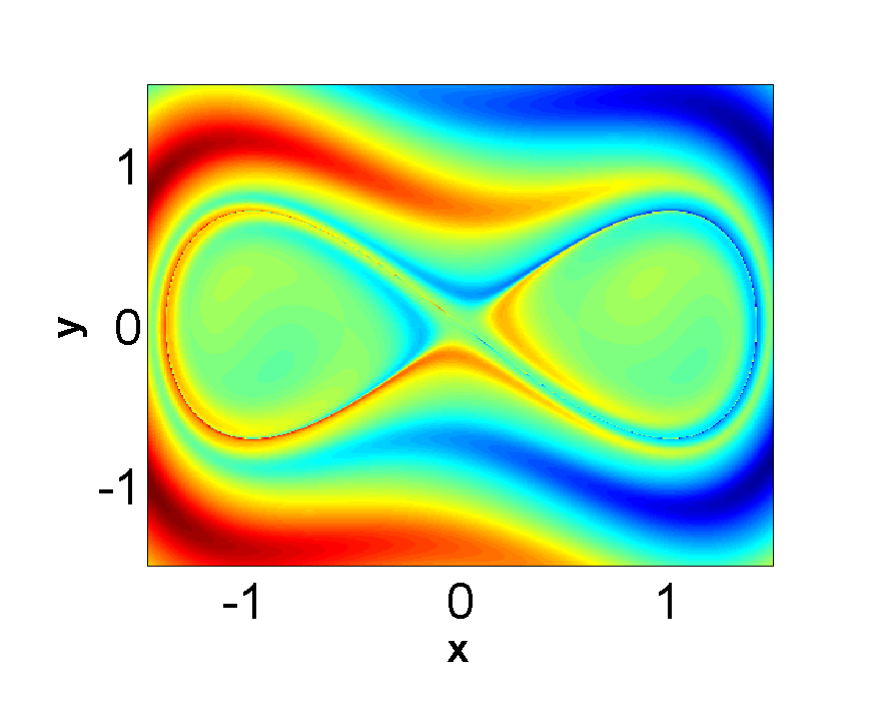}
b)\includegraphics[width=7.5cm]{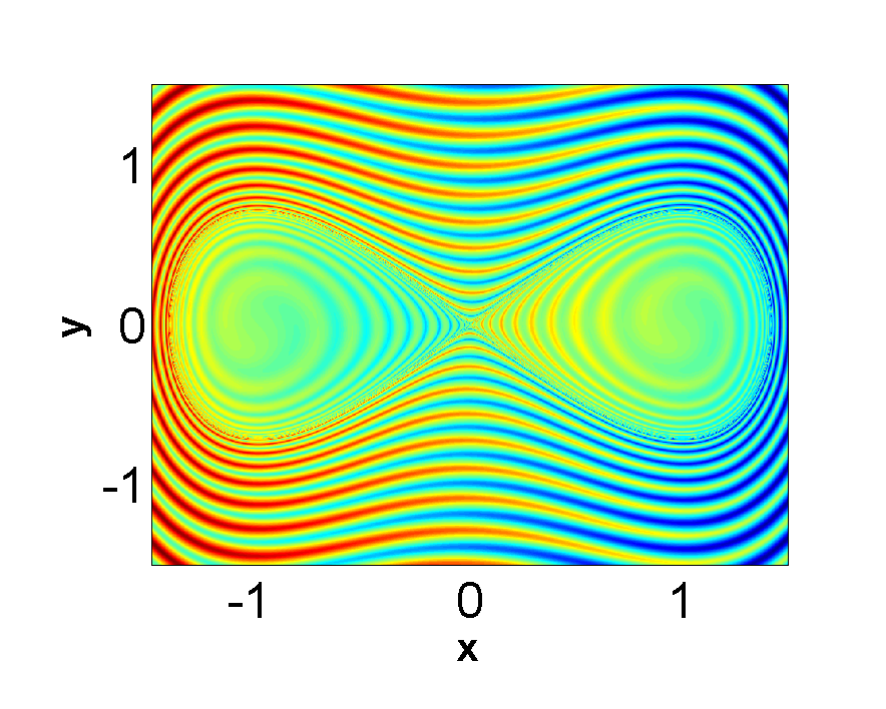} \hfill
\medskip
c)\includegraphics[width=7.5cm]{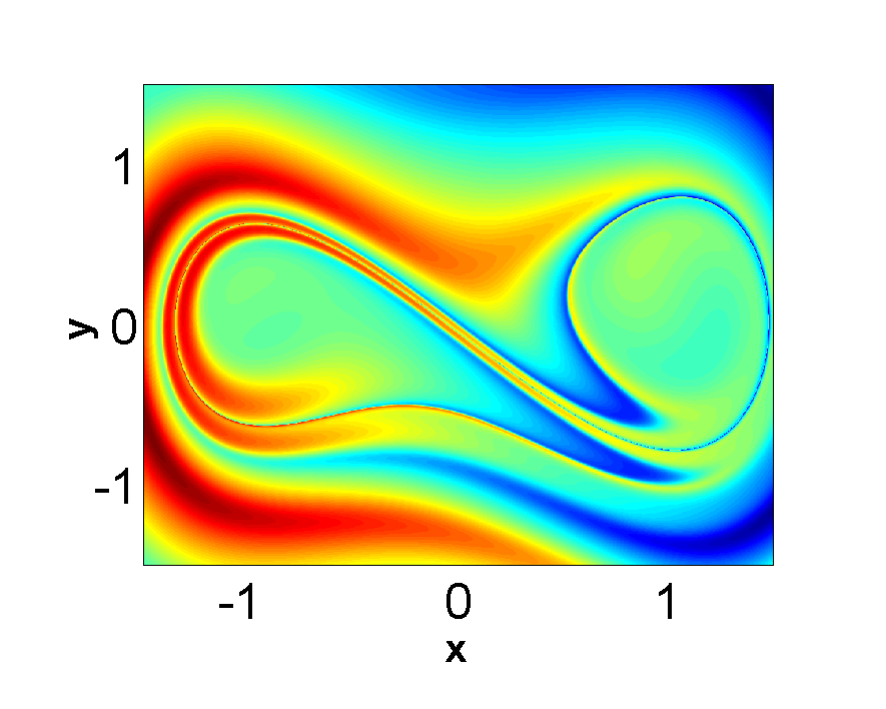} \hfill
c)\includegraphics[width=7.5cm]{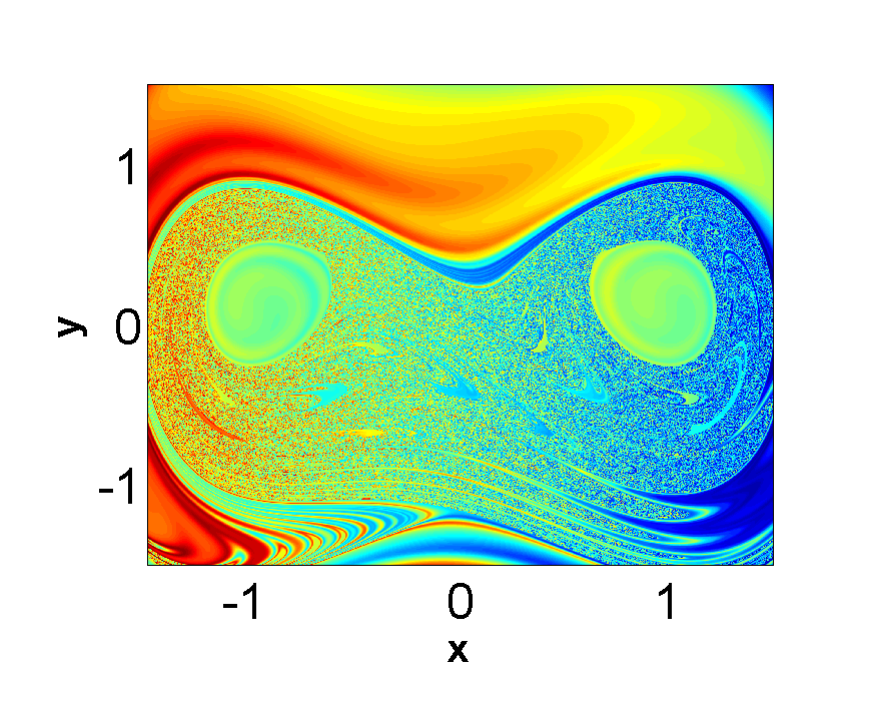} \hfill
\medskip
e)\includegraphics[width=7.5cm]{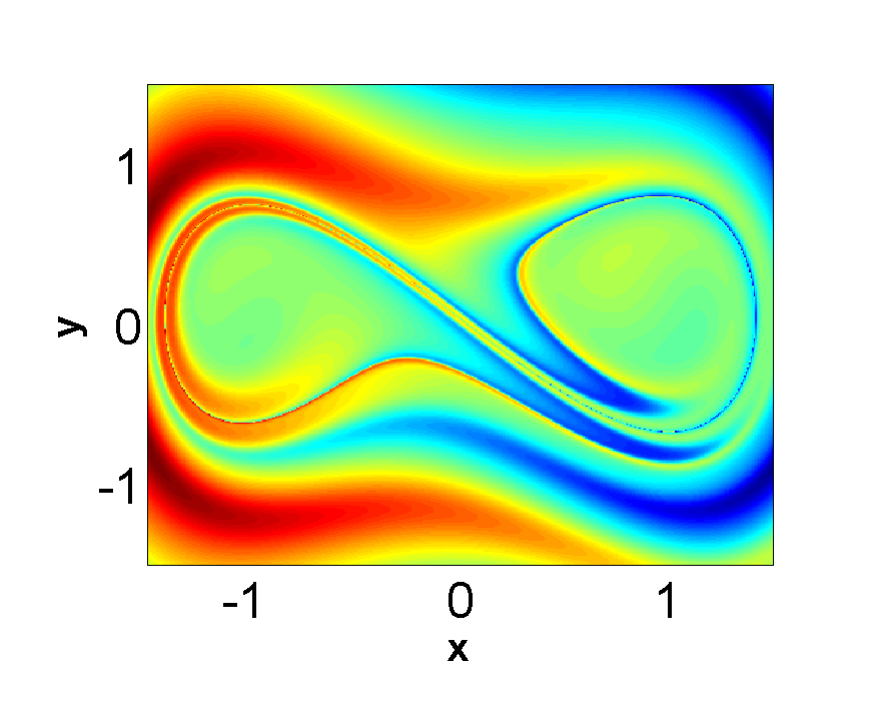}
f)\includegraphics[width=7.5cm]{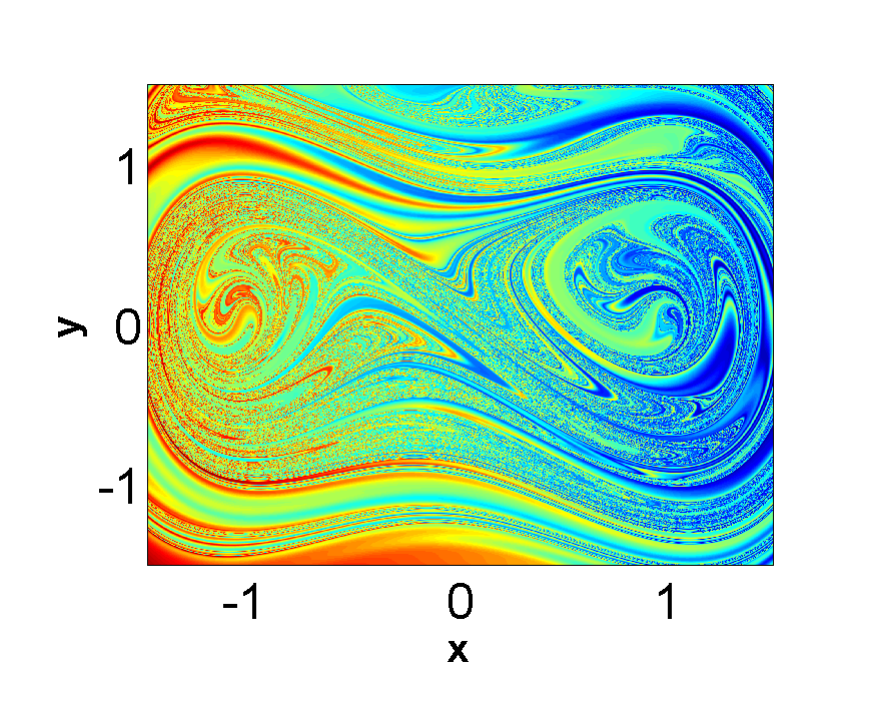} \hfill
\caption{\label{mezicduffeps0} Contours of the finite time average of the horizontal component of  the Duffing Equation along trajectories;
  The integrable case ($\epsilon =0$) for a) $\tau=10$; b)  $\tau=70$; the periodic case for c) $\tau =10$ d) $\tau=70$; the aperiodic case for e)  $\tau =10$; f) $\tau = 70$. }
\end{figure}

The results are shown in figure \ref{mezicduffeps0}, where the time averages are computed in each case for $\tau = 10$ and $\tau =70$.  For each time dependence we see similar behavior. For the smaller time we see ``a hint'' of structure that bears some resemblance to ``fattened'' versions of short segments of manifolds of the unstable manifold of the hyperbolic trajectory. For the longer time the contours of the finite time average of the horizontal component of velocity develop a more complex spatial structure that %completely
 obscures the underlying unstable manifold structure of the hyperbolic trajectory.

 \section{Applications to time dependent 3D flows}
 \label{sec:3d}

 In this section we show that Lagrangian descriptors can also provide accurate information on the stable and unstable manifolds of hyperbolic trajectories in three dimensional (3D) time dependent flows. Computation  of the stable and unstable manifolds of hyperbolic trajectories in aperiodically time dependent flows was discussed in \cite{bw09}, where an algorithm for their calculation was developed and several ``benchmark'' examples were considered. The particular example that we will consider is the perturbed Hill's spherical vortex, which we will take as a benchmark for the performance of our methods in 3D.  We give a brief description of the velocity field. More details on the background of Hill's spherical vortex can be found in \cite{bw09}.

The velocity field, ${\bf v}$, has the general form:

$$ {\bf v}=H(x,y,z)+S(x,y,z,t)$$

\noindent
where $H(x,y,z)$ is given by:

\begin{eqnarray}
H_x&=&(u_r\sin\Theta+u_\Theta \cos\Theta) \cos \Phi,\\
H_y&=&(u_r\sin\Theta + u_\Theta \cos\Theta) \sin \Phi,\\
H_z&=&(u_r \cos\Theta -u_\Theta \sin \Theta),
\end{eqnarray}

Here:

$$ r=\sqrt{x^2+y^2+z^2}, \quad \Theta = \arccos (z/r), \quad \Phi= \arccos (x/\sqrt{x^2+y^2})$$

\begin{equation}
u_r=\left\{  \begin{array}{ll} U(1-a^3/r^3)\cos\Theta,& {\rm if } \,\, r\geq a,\\ -\frac{3}{2} U(1-r^2/a^2)\cos\Theta, & {\rm if }  \,\, r<a, \end{array} \right.\\
u_\Theta=\left\{\begin{array}{ll} -U(1+a^3/(2r^3))\sin\Theta,& {\rm if }  \,\, r\geq a,\\ \frac{3}{2} U(1-2r^2/a^2)\sin\Theta, & {\rm if }  \,\, r<a, \end{array} \right.
\end{equation}

\noindent
The matrix $S$ is given by:

$$ S=\mathcal{A}(t)\cdot \delta^{-1}\left[\begin{array}{ccc} \alpha(t) & 0 & 0 \\ 0 & \beta(t) &0 \\ 0 & 0 & \gamma(t)\end{array}\right]\cdot \delta \cdot \left[\begin{array}{c}x\\ y\\ z \end{array}\right]$$

\noindent
where  $\delta$ is the orthogonal matrix:

$$ \delta=\left[ \begin{array}{ccc} cos \psi \cos \varphi - \sin\psi \cos \theta \sin \varphi & \cos\psi\sin\varphi+\sin\psi\cos\theta \cos \varphi & \sin\psi \sin \theta \\ -\sin \psi \cos\varphi -\cos \psi \cos \theta \sin \varphi & -\sin \psi \sin \varphi+ \cos \psi \cos \theta \cos \varphi & \cos \psi \sin \theta \\ \sin \theta \sin \varphi & -\sin \theta \cos \varphi &  \cos \theta \end{array}\right],$$

\noindent
and $A(t)$ is the time dependent function:

\begin{equation}
\mathcal{A}(t)=(0.3+0.27\sin(3.3t))\exp(-(t-6)^2/2.5^2).
\label{def:A(t)}
\end{equation}

\noindent
Additionally:

$$\theta=0.5+0.05\sin(2t), \; \varphi=5t, \; \psi=0 $$

\noindent
and

$$\alpha(t)= -0.5, \; \beta(t)=-0.5 \,\, { \rm and }  \,\, \gamma(t)=1.$$

\begin{figure}
\includegraphics[width=6cm]{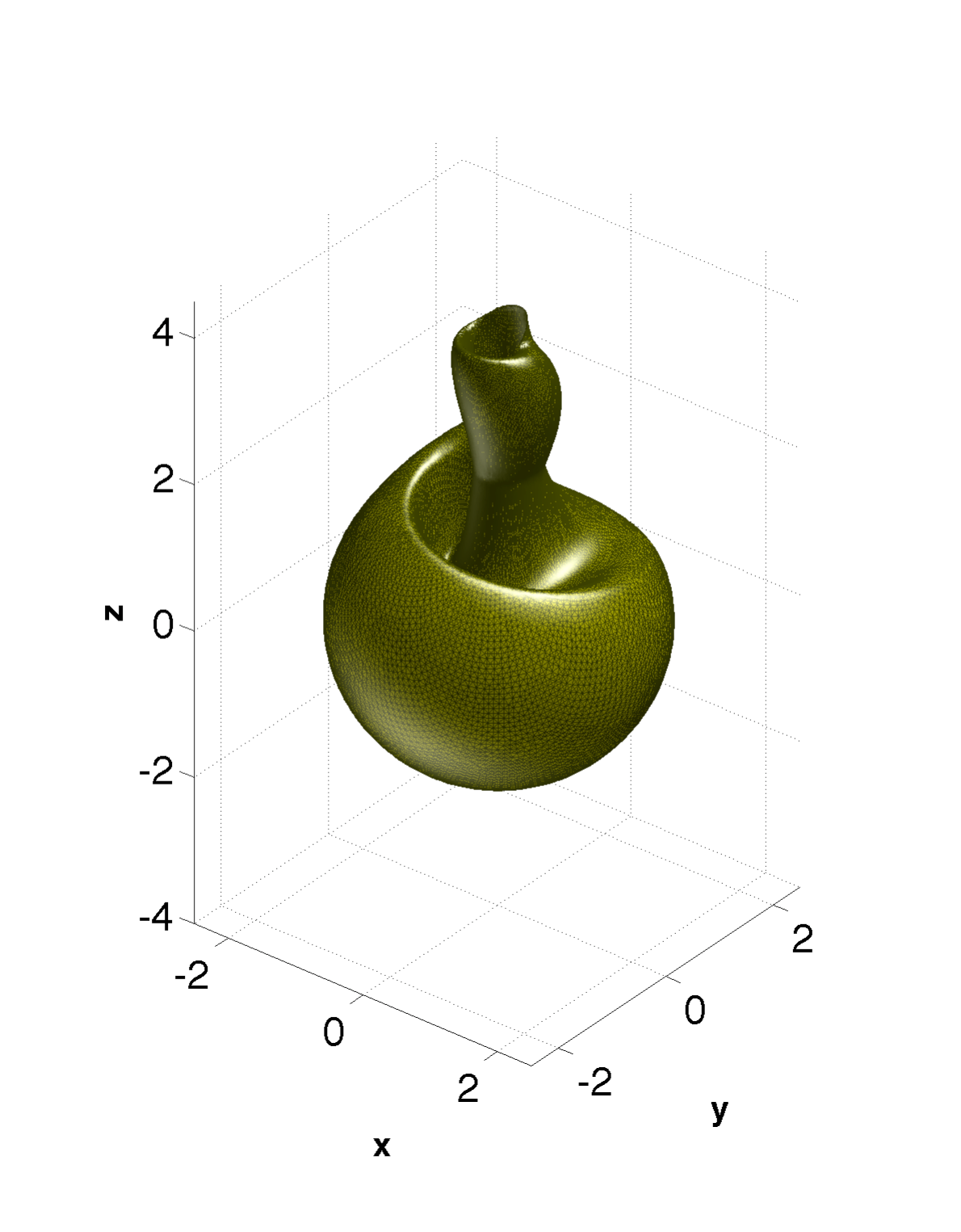}\includegraphics[width=6cm]{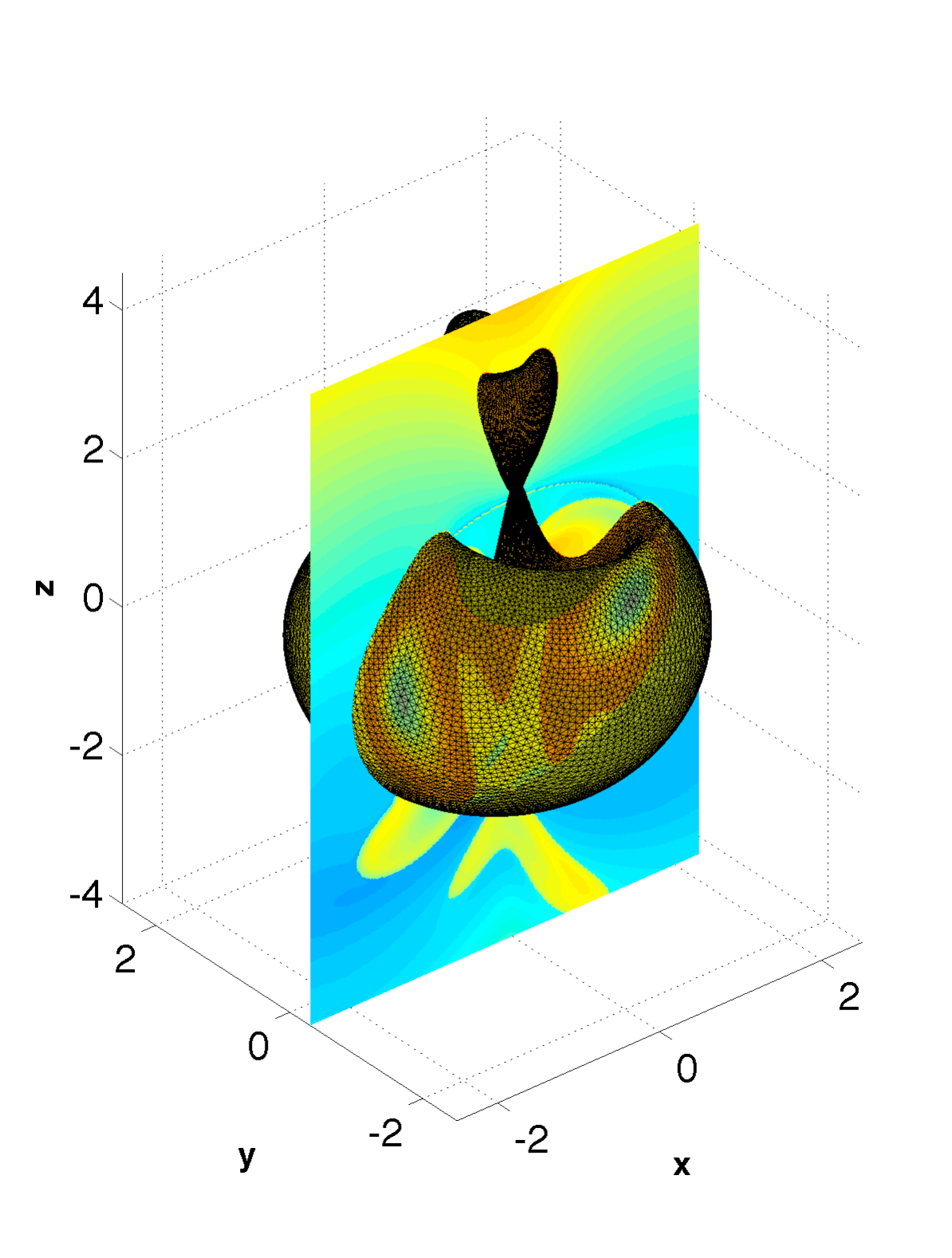}\includegraphics[width=6cm]{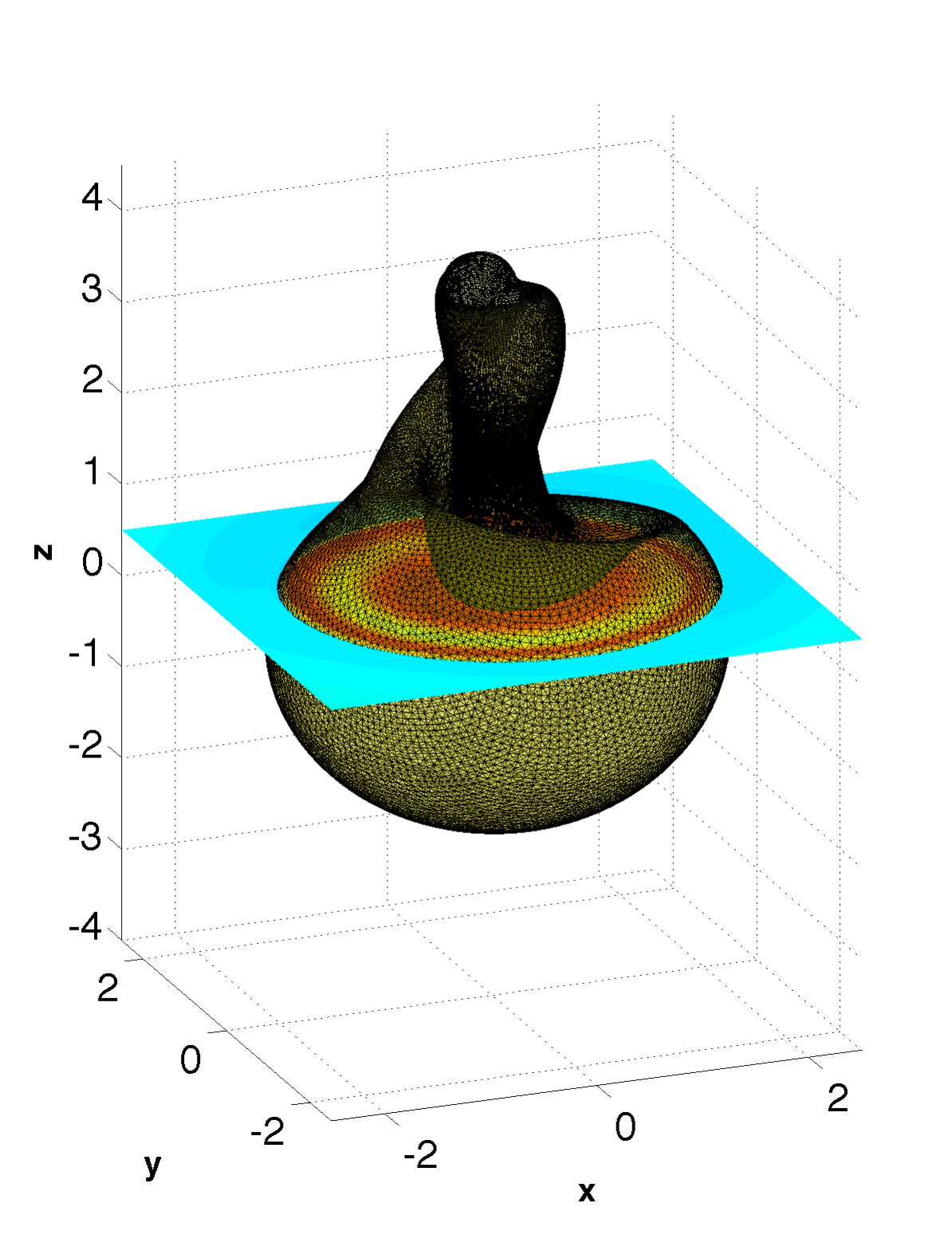}
\caption{a) The two dimensional unstable manifold of $\gamma_1$ is shown at $t=5.3$ for $U=a=2$ (image courtesy of M. Branicki); b) the   intersection of the  unstable manifold with a vertical cross section of $M_2$ obtained at $y=-0.3$ and $t=5.3$ for $\tau=10$; c) the   intersection of the  unstable manifold with a horizontal cross section of $M_2$ obtained at $z=0$ and $t=5.3$ for $\tau=10$}
\label{BW_manifolds}
\end{figure}

\noindent
For $\mathcal{A}(t) =0$ the flow is steady and has hyperbolic stagnation points at:

\begin{eqnarray}
h_1 & = & (0, 0, -a), \nonumber \\
h_2 & = & (0, 0, a),
\end{eqnarray}

\noindent
where $h_1$ has a  one dimensional stable manifold and a  two dimensional unstable manifold, and $h_2$ has a two  dimensional stable manifold and a one dimensional  dimensional manifold. In \cite{bw09} numerical algorithms are used to show that when $\mathcal{A}(t)$ increases from $0$ to the expression in  \eqref{def:A(t)}, $h_1$ and $h_2$ ``continue'' to be hyperbolic (time-dependent) trajectories, denoted $\gamma_1$ and $\gamma_2$, respectively, and at each instant of time $\gamma_1$ has a one dimensional stable manifold and a two dimensional unstable manifolds, and $\gamma_2$ has a two dimensional stable manifold and a one dimensional stable manifold. The unstable manifold at  $t=5.3$ with $U=2$ and $a=2$ computed as in \cite{bw09} is displayed in  Figure \ref{BW_manifolds}a).

We now illustrate the performance of  Lagrangian descriptors for this three dimensional, aperiodically time dependent example.
Figure \ref{BW_manifolds}b) and c) show  respectively  the   intersection of the  unstable manifold with vertical (at $y=-0.3$) and horizontal (at $z=0$) cross sections of $M_2$ obtained under the same conditions. The agreement is excellent.

\begin{figure}
a)\includegraphics[width=4.5cm]{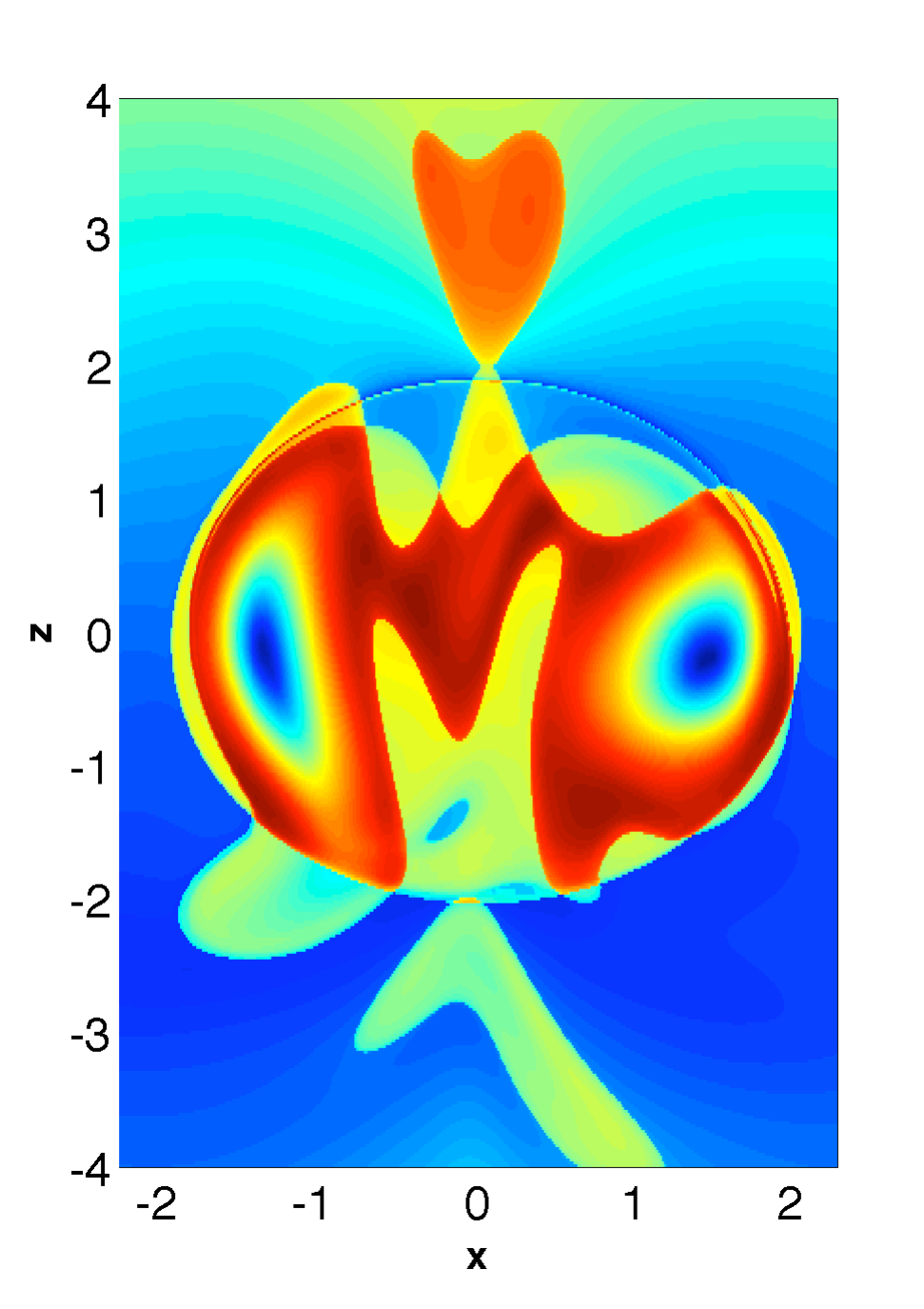}c)\includegraphics[width=4.5cm]{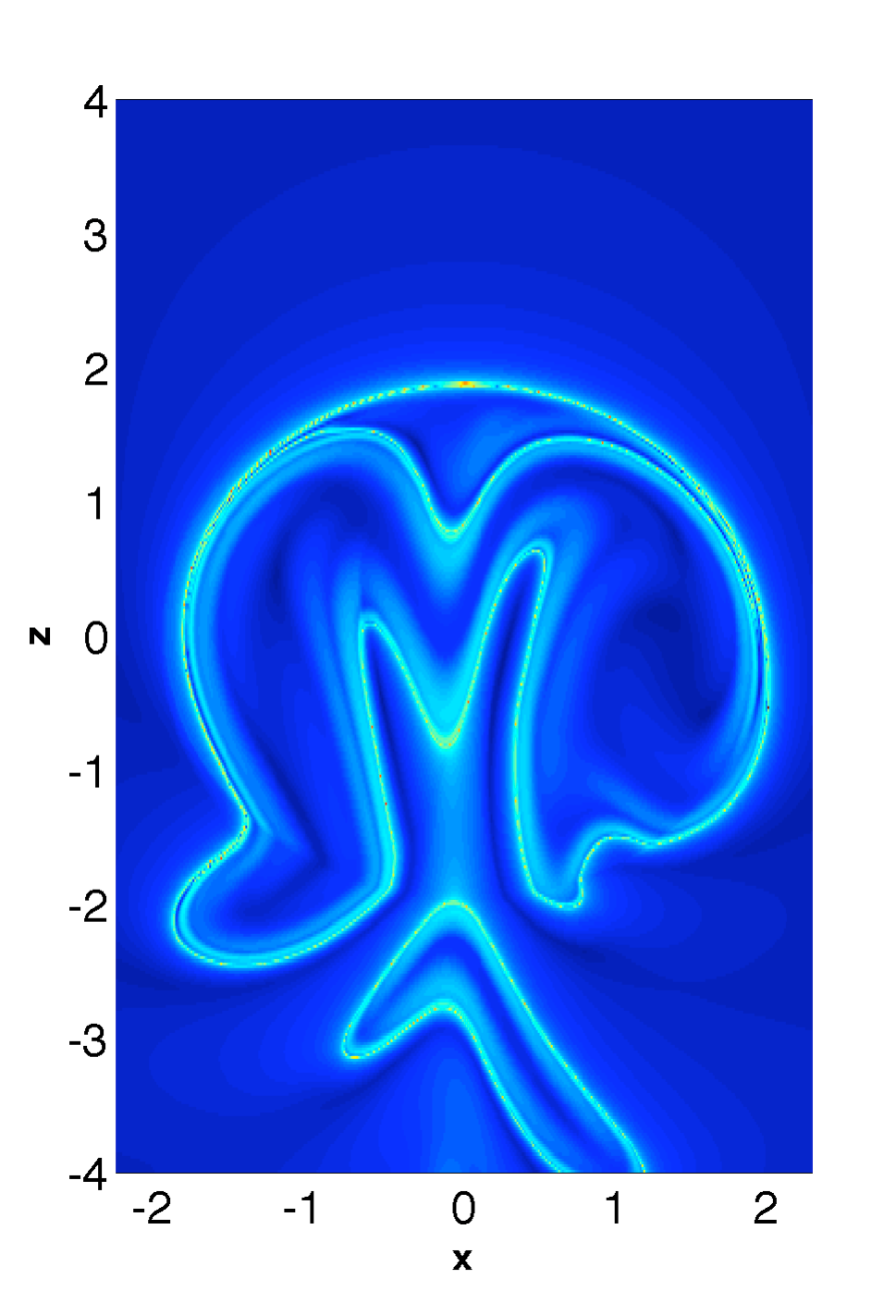} e)\includegraphics[width=4.5cm]{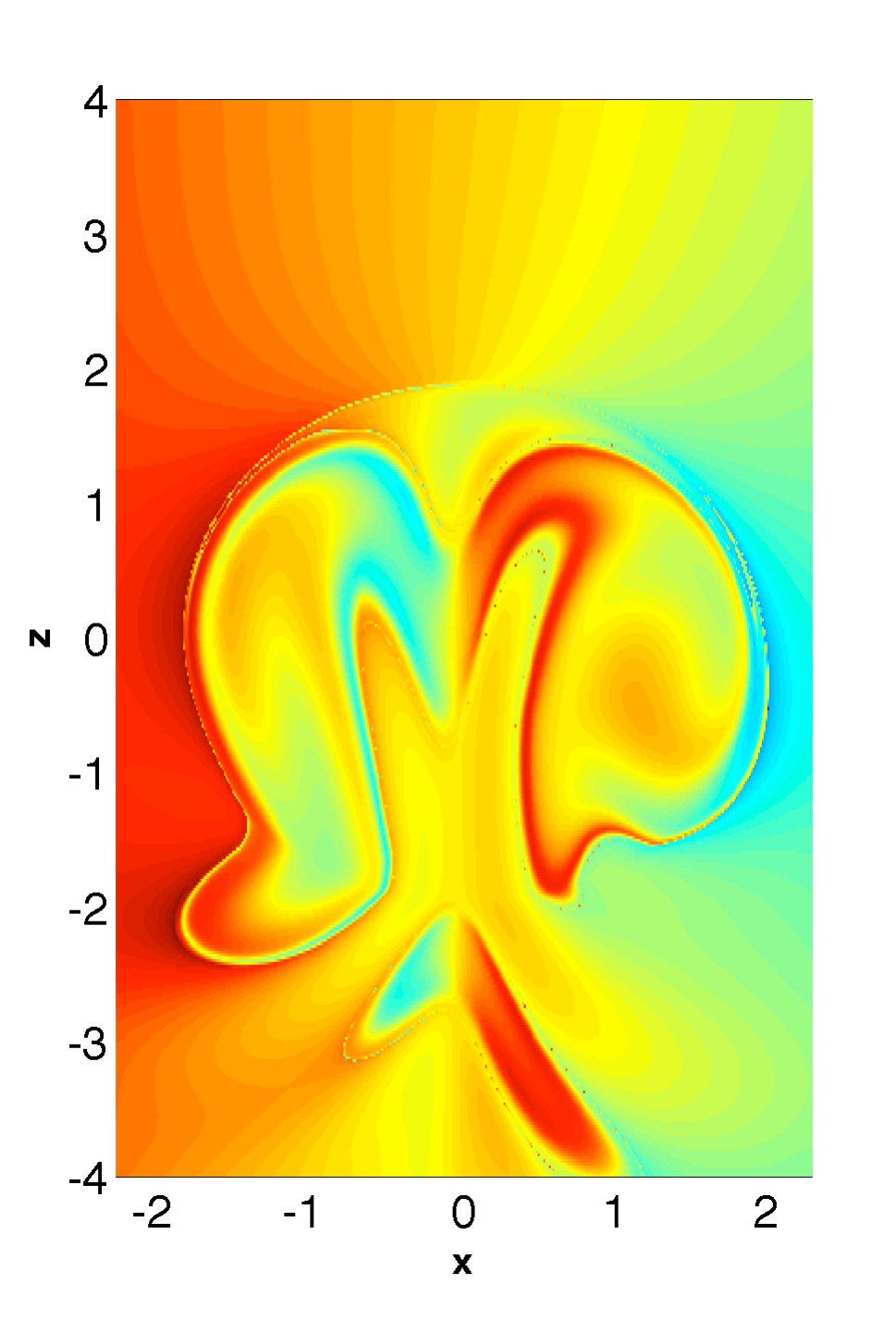} \\
b)\includegraphics[width=4.5cm]{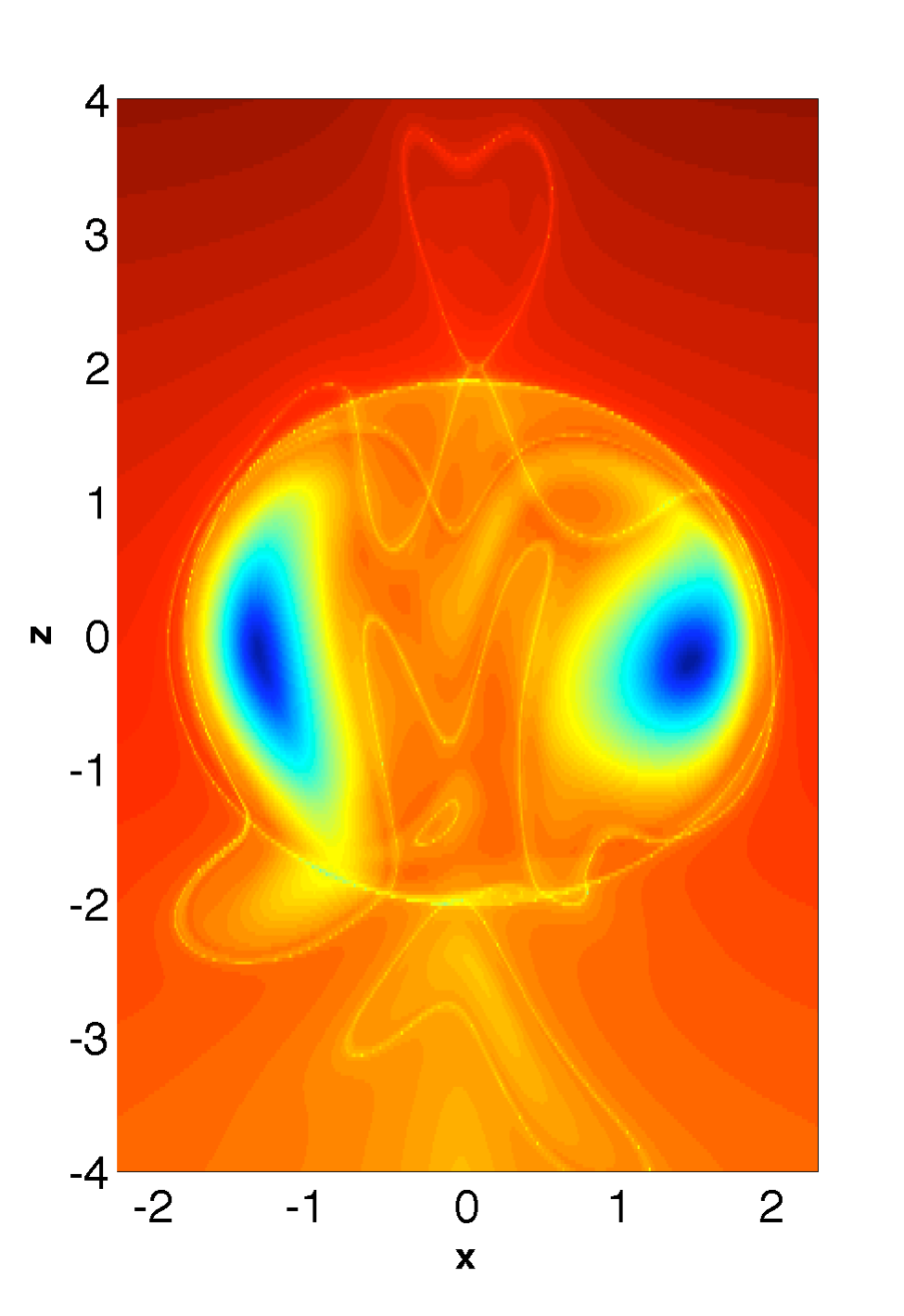}d)\includegraphics[width=4.5cm]{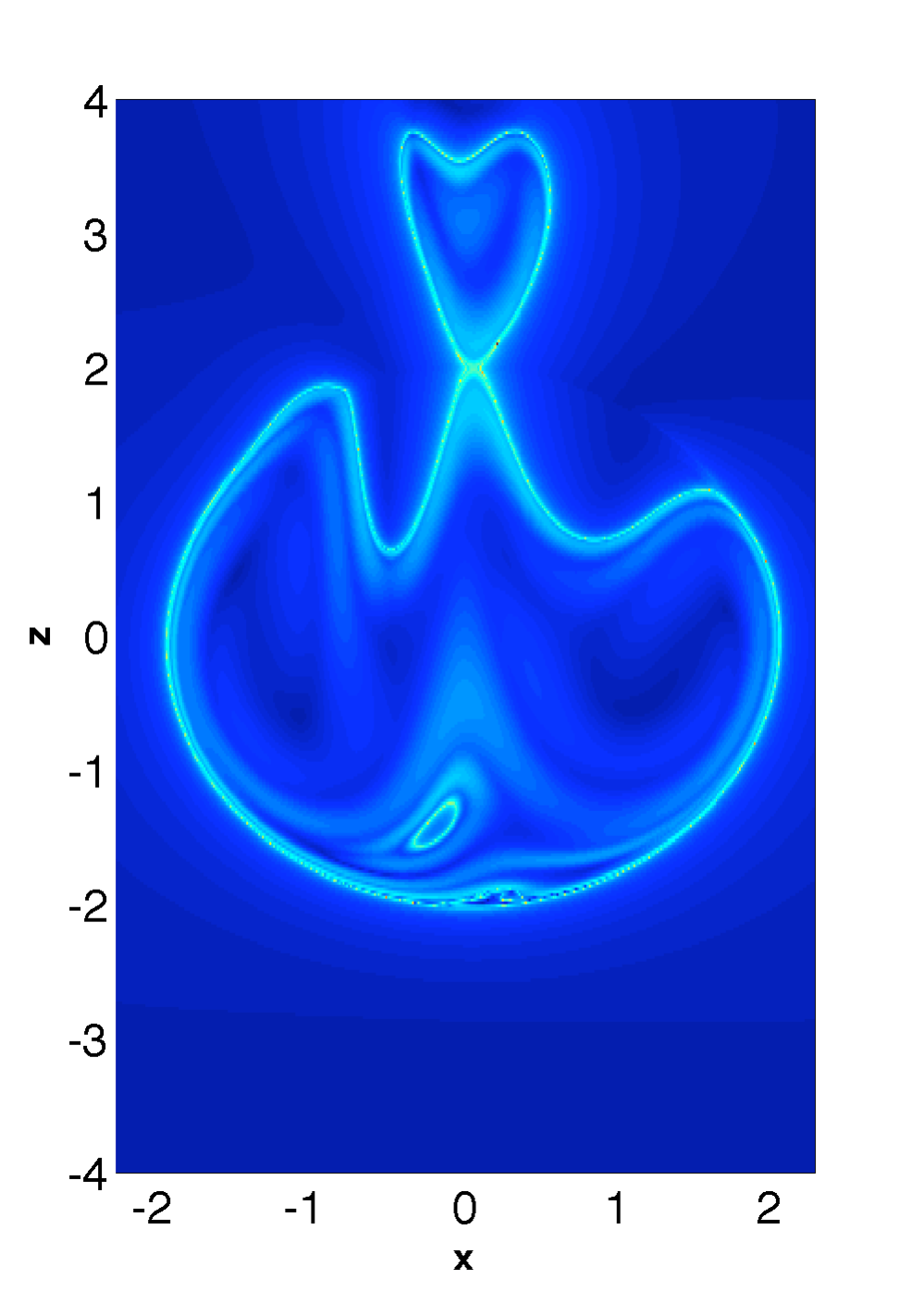}f)\includegraphics[width=4.5cm]{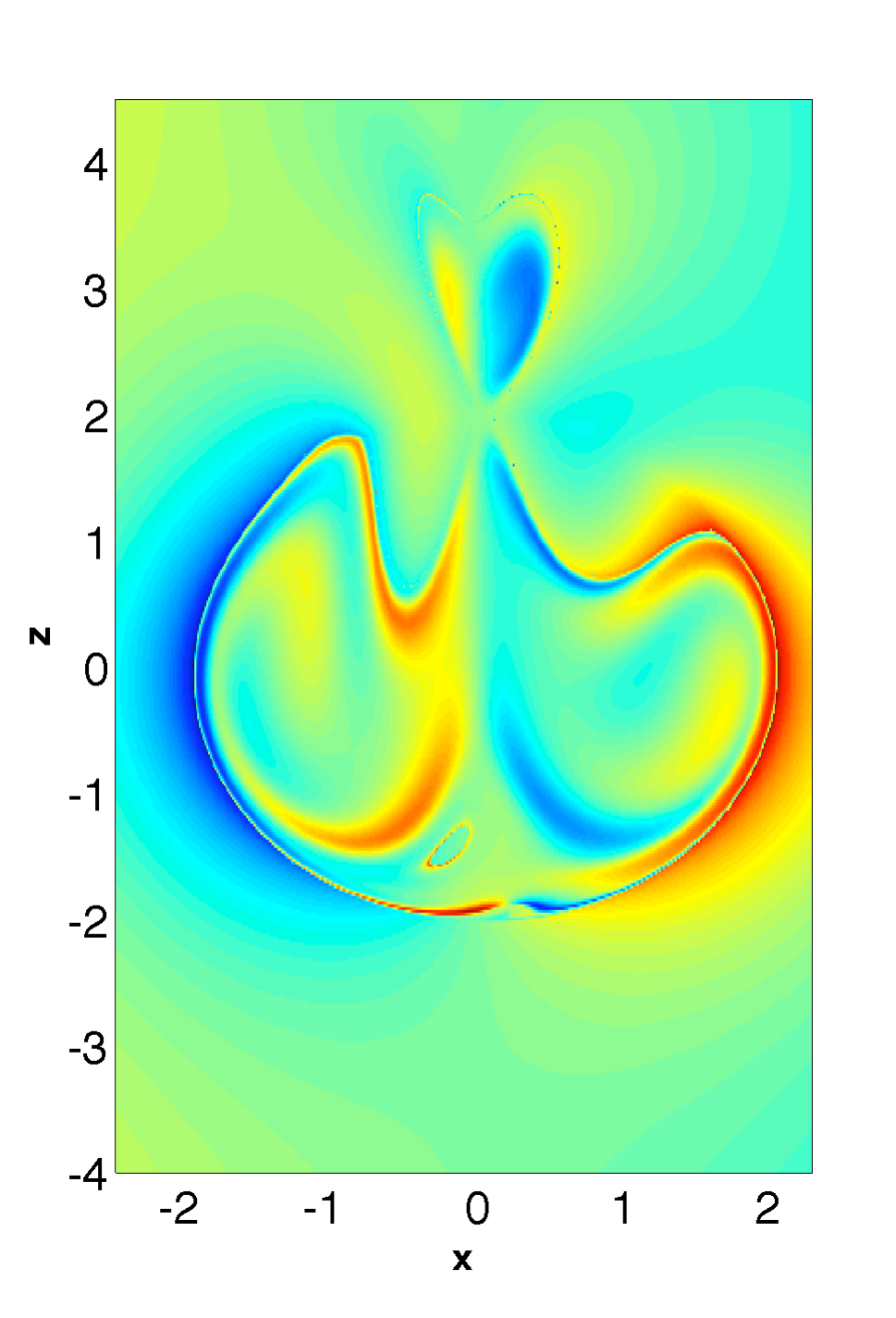}
 \caption{\label{M23dz} a) Evaluation of $M_2$ on the vertical plane  $(x,y=-0.3,z)$ at $t=5.3$ for $\tau=10$; b) the structure provided by $M_1$ also on the vertical plane  $(x,y=-0.3,z)$ at $t=5.3$ for $\tau=10$; c) the forward FTLE field on the vertical plane  $(x,y=-0.3,z)$ at $t=5.3$ for $\tau=10$; d) the backward FTLE on the vertical plane  $(x,y=-0.3,z)$ at $t=5.3$ for $\tau=10$; e)  the forward  average of the horizontal component of the velocity;  f) the backward  average of the horizontal component of the velocity.}
\end{figure}

 The Lagrangian descriptor $M_2$ provides many details on the Lagrangian structure of the flow as Figure \ref{M23dz}a) confirms. It shows the results on the vertical section displayed in
 Figure \ref{BW_manifolds}b) obtained with $\tau=10$. Similar slices are displayed in Figure \ref{M23dz}b) for the descriptor $M_1$. The output is similar to $M_2$, but with a lower contrast  which make more difficult the visualization.
 The structure is clean and sharp in contrast to that provided by  methods discussed next.
FTLE computations, similarly to what happened in the 2D case, distorts the image by introducing numerous structures that do not have Lagrangian interpretation. The output of forwards and backwards FTLE is displayed in figure \ref{M23dz}c) and d) confirms this extreme. Alternative averages as those proposed in section 2.3 provide the output shown    in figure \ref{M23dz}e) and f). These averages also present  spurious structures that distort the output and blur the  true Lagrangian information.

\section{Application  of Lagrangian Descriptors to Velocity Fields Defined as Data Sets}
\label{sec:veldata}

In this section we discuss  the performance and application  of  Lagrangian descriptors  to velocity fields defined as data sets. In particular, we will  consider the finite time data set obtained  from the numerical simulation of a wind driven,
quasi-geostrophic (QG) 3-layer model  in a rectangular basin geometry on a $\beta$-plane (\cite{r}).
The basic circulation pattern in the upper layer is an unsteady  double-gyre structure, \emph{i.e.} a counterclockwise gyre in the north and a clockwise gyre in the south separated by a jet. The velocity data set is obtained on a  1000 km $\times$ 2000 km rectangular domain with spatial and temporal resolutions of 12.5 km $\times$ 12.5 km and 2 hours (although the data is only saved every 24 hours), respectively, and spans over 4000 days. The 4000 day interval of the data set is considered after the fluid is started from rest and allowed to spin up for 25000 days. More details of the specific parameters used and the numerical method used for solving the QG equations can be found in \cite{r} or \cite{coulliette}.

\begin{figure}
a)\includegraphics[width=7cm]{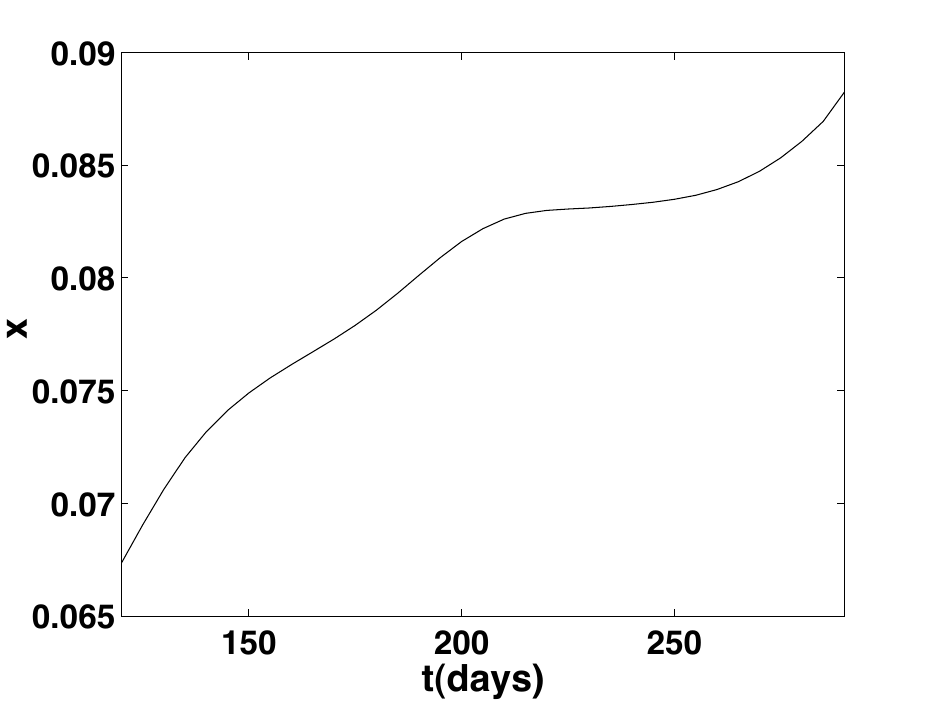}
b)\includegraphics[width=7cm]{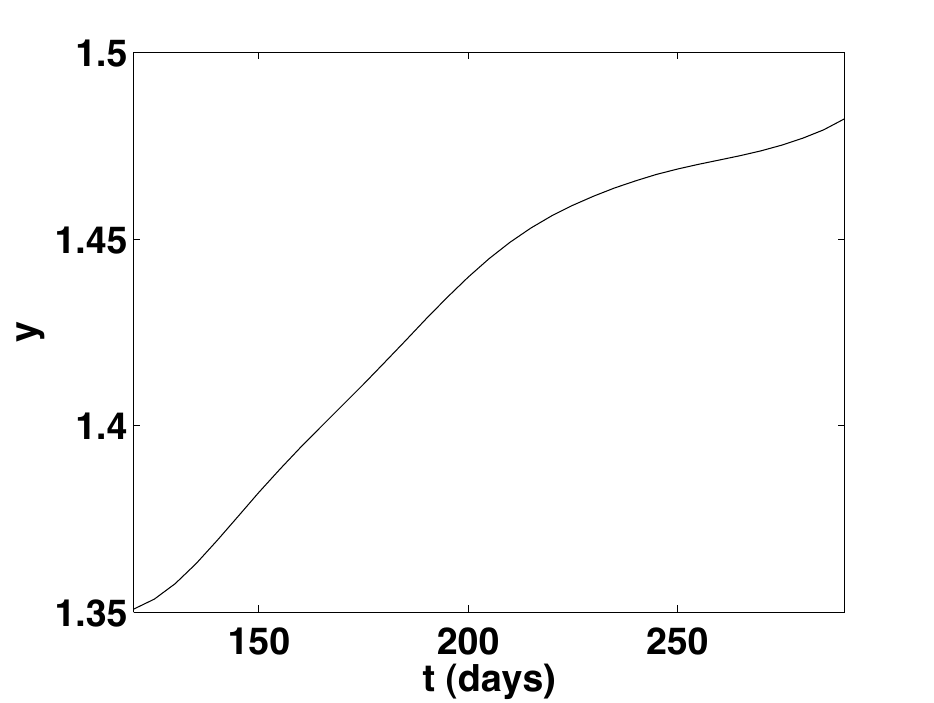}
\caption{\label{dhtqgn} The time evolution of a DHT; a) the $x$-coordinate;  b) the $y$-coordinate. Units are in thousands of kilometers.}
\end{figure}

We have chosen this data set because it is  aperiodic and results on the computation of distinguished hyperbolic trajectories (DHTs) and their stable and unstable manifolds have previously been reported in  \cite{msw04,physrep, chaos}. In particular, we choose three hyperbolic trajectories that have previously been identified in a range of days between 0 and  900 in \cite{msw04, physrep}.
Figure \ref{dhtqgn} shows one of these DHTs  located in the northern gyre, which has been  shown to possess the ``distinguished property'' in the  interval  between 120 days  and  300 days  (more details on this can be found in \cite{chaos}).   Beyond this time interval the trajectory still exists,  but it no longer  has the  ``distinguished'' property. The loss of the distinguished property has been linked in \cite{chaos} with the loss of hyperbolicity. We remark that in complex time varying  geophysical flows  one frequently observes trajectories whose finite time stability characteristics change from hyperbolic to elliptic, and vice versa. This is a particularly challenging phenomenon for the dynamical systems approach and has been discussed in \cite{brawigg,jpo,chaos, br}.

\begin{figure}
a)\includegraphics[width=7cm]{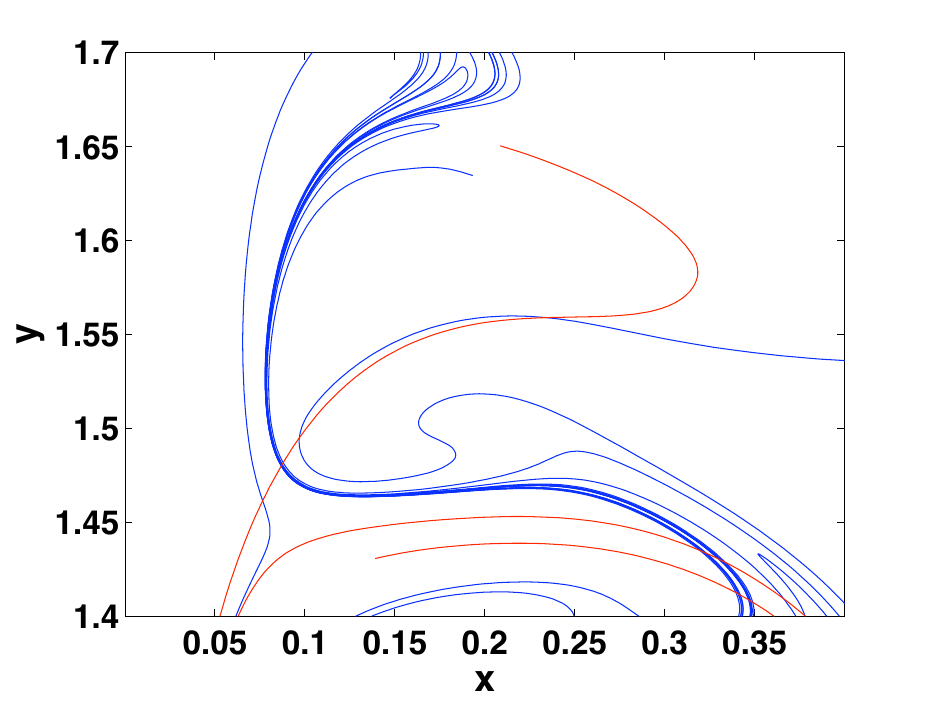}
b)\includegraphics[width=7cm]{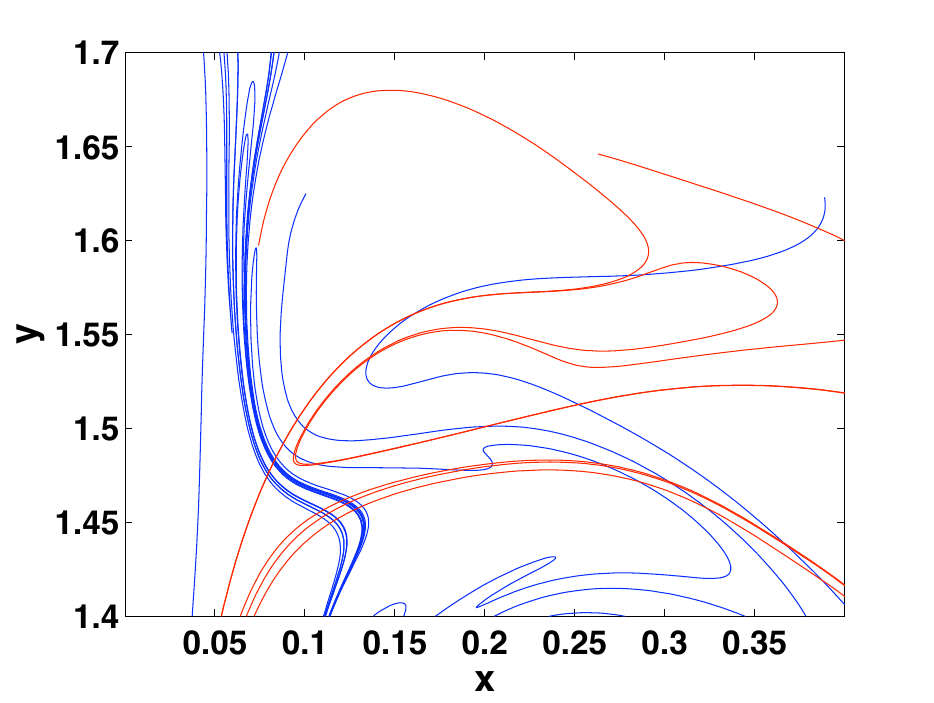}\\
c)\includegraphics[width=7cm]{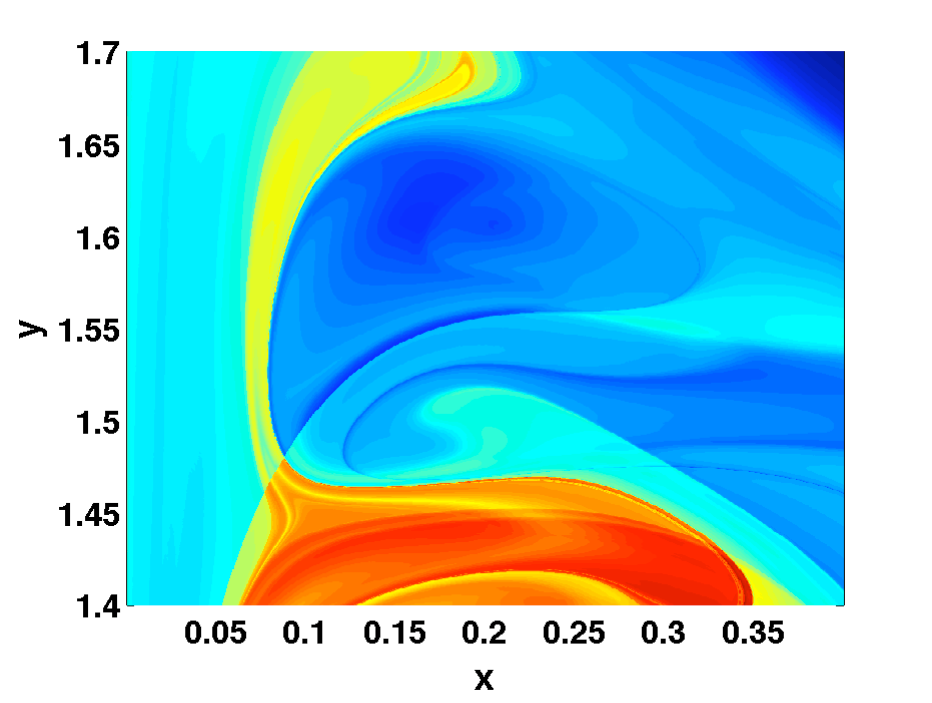}
d)\includegraphics[width=7cm]{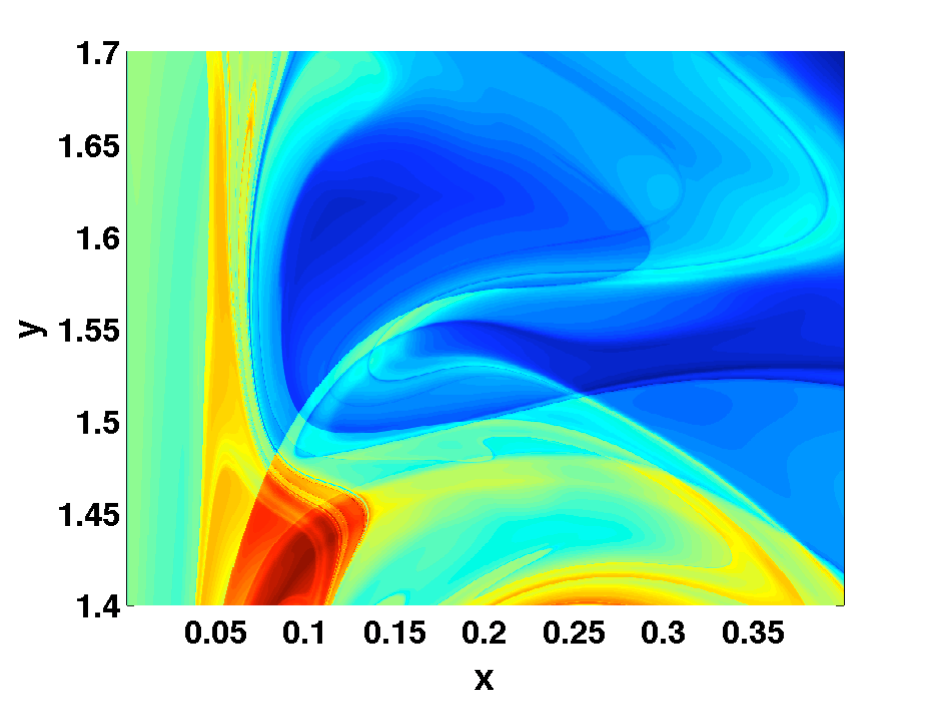}\\
e)\includegraphics[width=7cm]{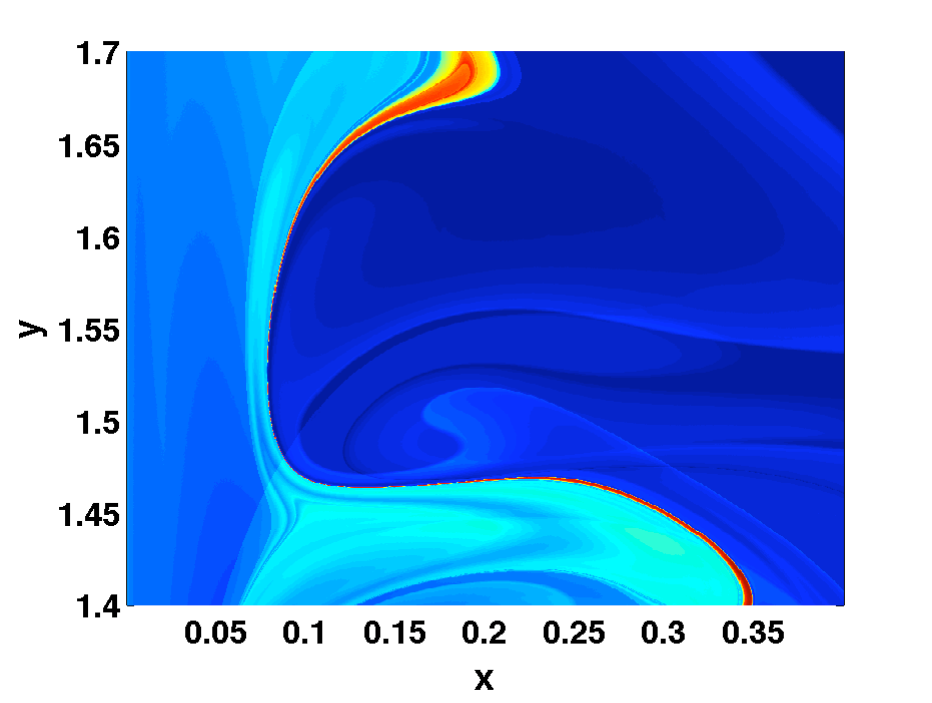}
f)\includegraphics[width=7cm]{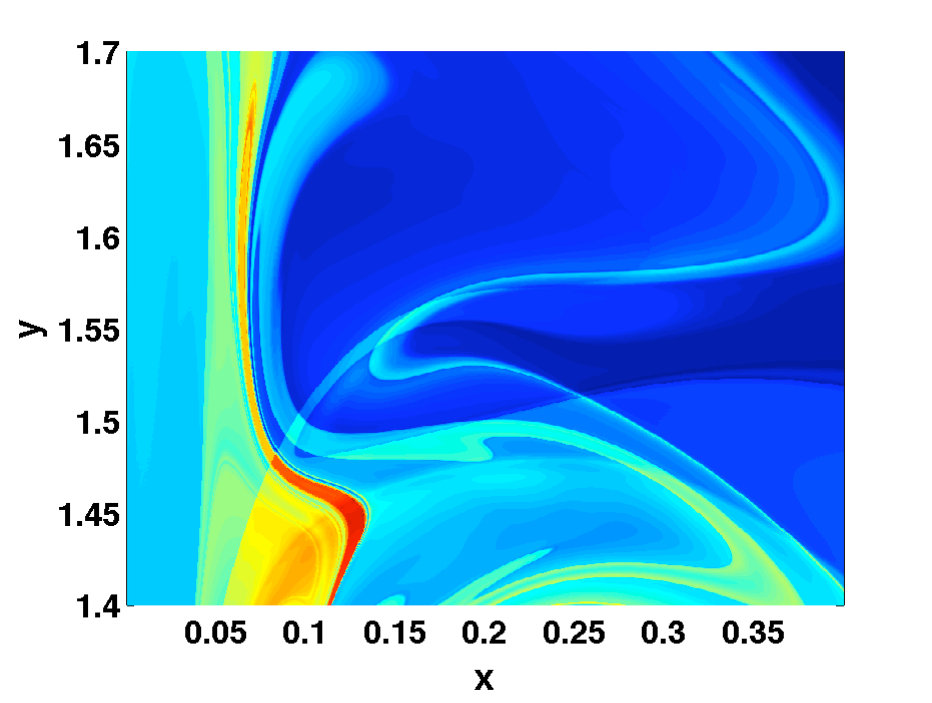}
\caption{\label{qgntaulong} Pieces of stable manifold (blue) for two DHTs and  and unstable manifolds (red) of the remaining DHT shown on  a) day 290; b) day 320. Contour plots for a selection of Lagrangian descriptors evaluated over the quasigeostrophic flow  with $\tau=150$.  c) $M_3$ at day 290; d) $M_1$ at day 320; e) $M_5$ at day 290; f) $M_2$ at day 320. The units on the axis are in thousands of kilometers.}
\end{figure}

Figures \ref{qgntaulong}a) and b) show the results of the direct computation of  the stable manifold for two of the DHTs  and the unstable manifold for the remaining DHT for days 290 and 320 using the algorithms described in \cite{msw04}.  This figure also shows contour plots for a selection of Lagrangian descriptors computed for
$\tau=150$.

We note that there is an essential difference between the figures showing the direct computation of DHTs and their stable and unstable manifolds and the contours of Lagrangian descriptors. In order to compute manifolds of DHTs, we must first identify and compute the DHTs.  Lagrangian descriptors, on the other hand, compute the stable and unstable manifolds for all DHTs present during a chosen time interval, \emph{i.e.}, there is no need to identify the DHTs a priori.

Typically the Lagrangian structure revealed by different Lagrangian tools  contains more details for longer integration times $\tau$. However, it is also of much interest  in applications to obtain accurate Lagrangian information for small $\tau$. We will address how Lagrangian descriptors perform in this context, as well as make comparisons with the performance of FTLEs, by studying the
convergence rate of different Lagrangian descriptors with $\tau$ towards the singular lines that reveal the Lagrangian structure of the flow.
Our attention is not on the manifold structure for a large spatial region, but on the stable and unstable subspaces tangent to the stable and unstable manifolds of a particular DHT. We remark that these subspaces are related to the also called {\it Lyapunov vectors} \cite{legraslv} that indicated dispersive directions  in the fluid flow, backwards and forwards in time.

\begin{figure}
\centering
\includegraphics[width=7cm]{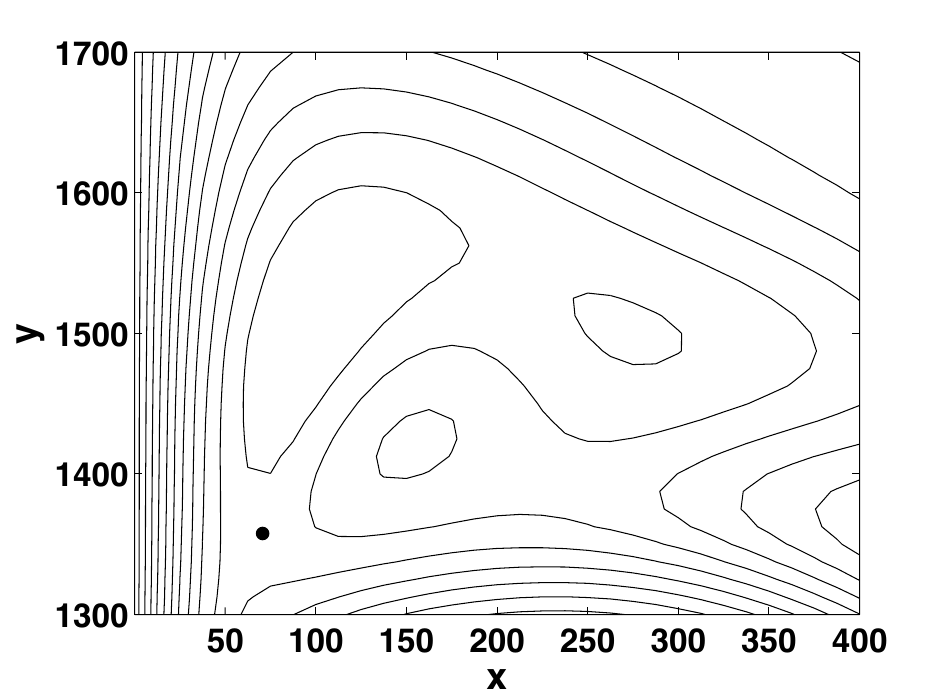}
\caption{\label{streamfun} Stream function of the quasigeostrophic model at day 130. The position of a DHT at this day is marked with a circle.  The units on the axis are in kilometers.}
\end{figure}

 We place our study at day 130,  far from the transition of the DHT. Figure \ref{streamfun} shows the stream function on this date with
 the position of the DHT indicated at (70.631,   1357.660) km. We use Lagrangian descriptors to approximate the stable and unstable subspaces of the DHT as follows. We  consider a ``small'' circle centred at the DHT. In particular, we consider the  circle centred at the DHT parametrized as $(r\cos(\alpha), r\sin(\alpha))$ and of radius $r=1.5$ km. We then compute the Lagrangian descriptor for trajectories starting in this circle for different $\tau$, and consider the singular contours of the Lagrangian descriptor near the DHT. These are approximations to the stable and unstable subspaces of the DHT.

\begin{figure}
a)\includegraphics[width=4.6cm]{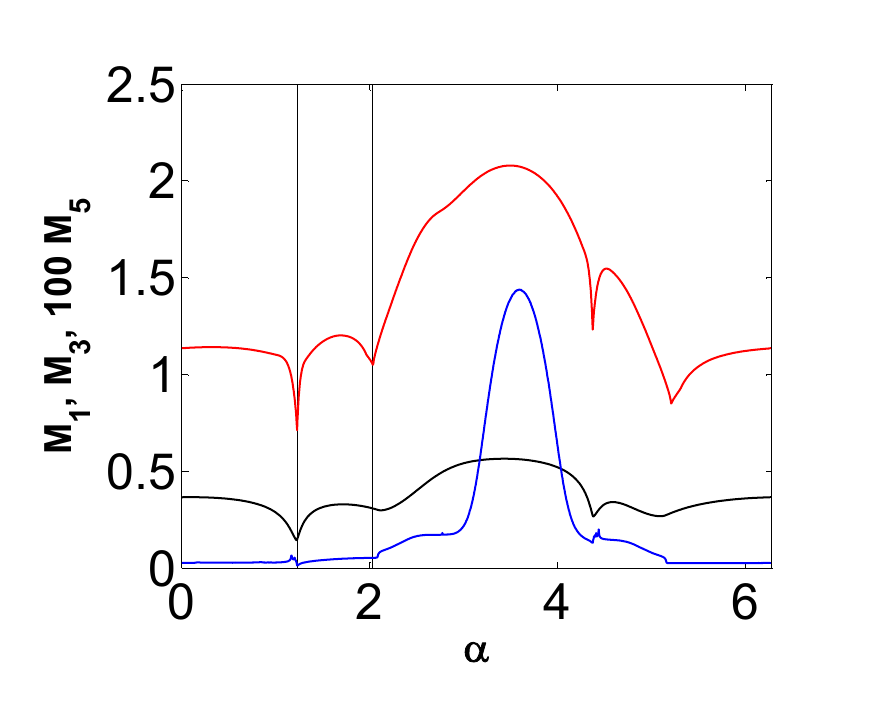}
b)\includegraphics[width=4.6cm]{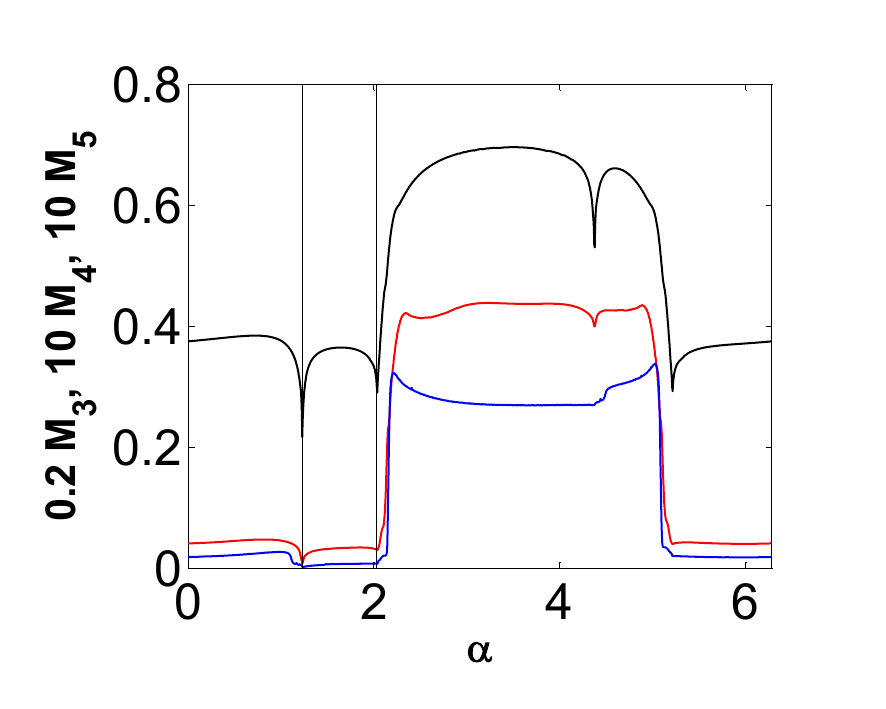}
c)\includegraphics[width=4.6cm]{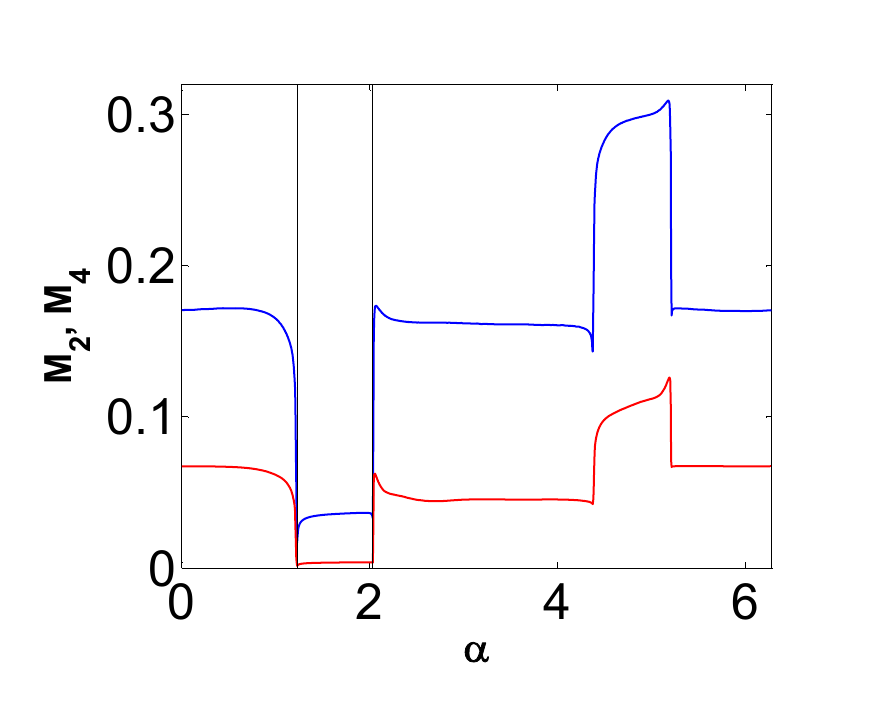}
\caption{\label{qgncir} a) Lagrangian descriptors $M_1$ (black), $M_3$ (red) and $100 \times M_5$ (blue) versus $\alpha$ for $\tau=45$; b) descriptors $0.2 \times M_3$ (black), $10 \times M_4$ (red) and $10 \times M_5$ (blue) versus $\alpha$ for $\tau=65$; c)  $M_2$ (blue) and $M_4$ (red) versus $\alpha$ for $\tau=95$}
\end{figure}

Figure \ref{qgncir} summarizes the convergence of different Lagrangian descriptors towards the stable and unstable subspaces of the  DHT at day 130. The angle  positions of the stable and unstable subspaces are marked with vertical lines.
Panel \ref{qgncir}a) shows $M_1$, $M_3$ and $M_5$   versus $\alpha$ for $\tau=45$, which for this specific vector field is
not a large value. At this stage singular features are
only developed by $M_3$, which has been obtained based on the magnitude of  acceleration and using $\gamma=1/2$ in the exponent of its definition. The Lagrangian descriptor $M_1$ shows a slower convergence  towards the subspace aligned along $\alpha=2.032$ and  $\alpha=2.032+\pi/2$.
The features displayed by $M_5$ are not yet sharp enough at this stage.
Panel \ref{qgncir}b) confirms a finer structure for $\tau=65$ for the Lagrangian descriptor $M_3$.  The Lagrangian descriptors $M_4$ and $M_5$ perform worse at this stage, while  the output of $M_1$ and $M_2$ is similar to that of $M_3$. Panel \ref{qgncir}c) shows $M_2$ and $M_4$ at  $\tau=95$, when convergence has been reached  for all of the Lagrangian descriptors considered.

These results indicate that the Lagrangian descriptor $M_3$ for this example performs the best in the sense that it indicates  the subspaces  for smaller  $\tau$.  It is followed by  $M_1$, $M_2$,  and $M_5$, respectively.  The Lagrangian descriptor $M_4$, also develop sharp features along the same  structures, but it requires longer time intervals of integration for the situations considered here.

\begin{figure}
a)\includegraphics[width=4.6cm]{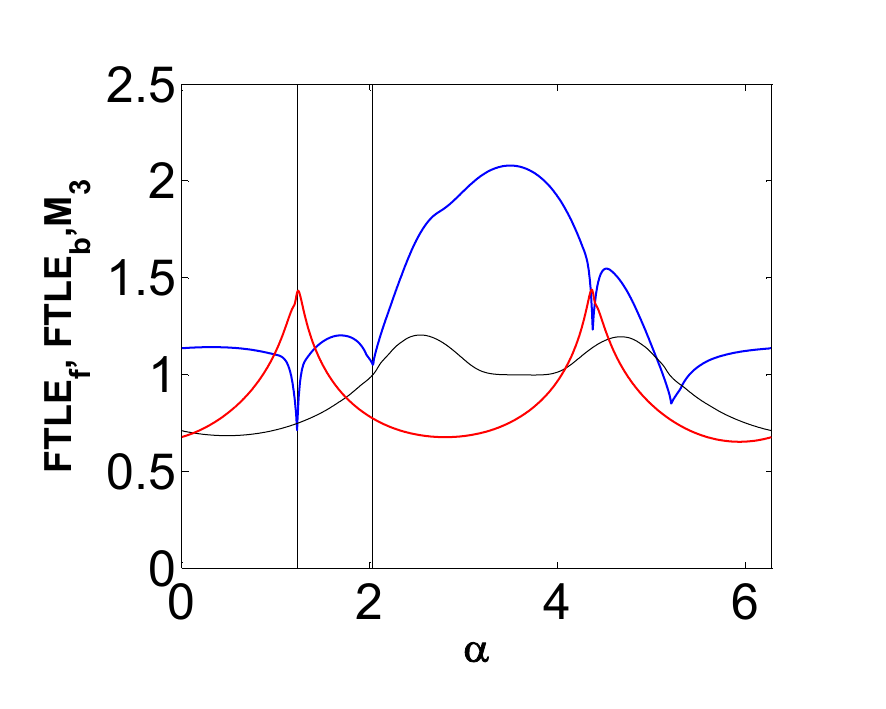}
b)\includegraphics[width=4.6cm]{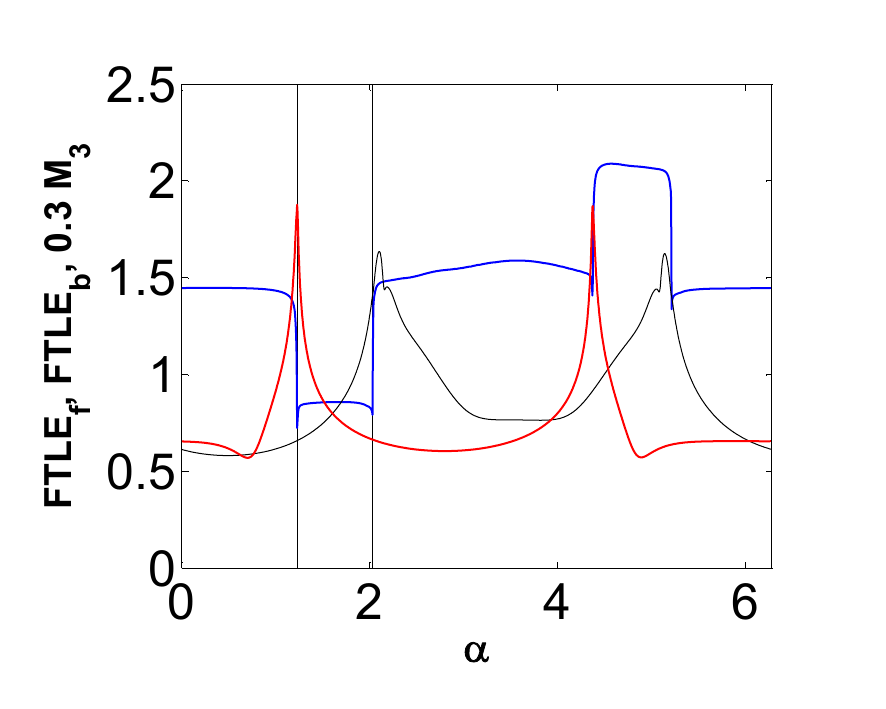}
c)\includegraphics[width=4.6cm]{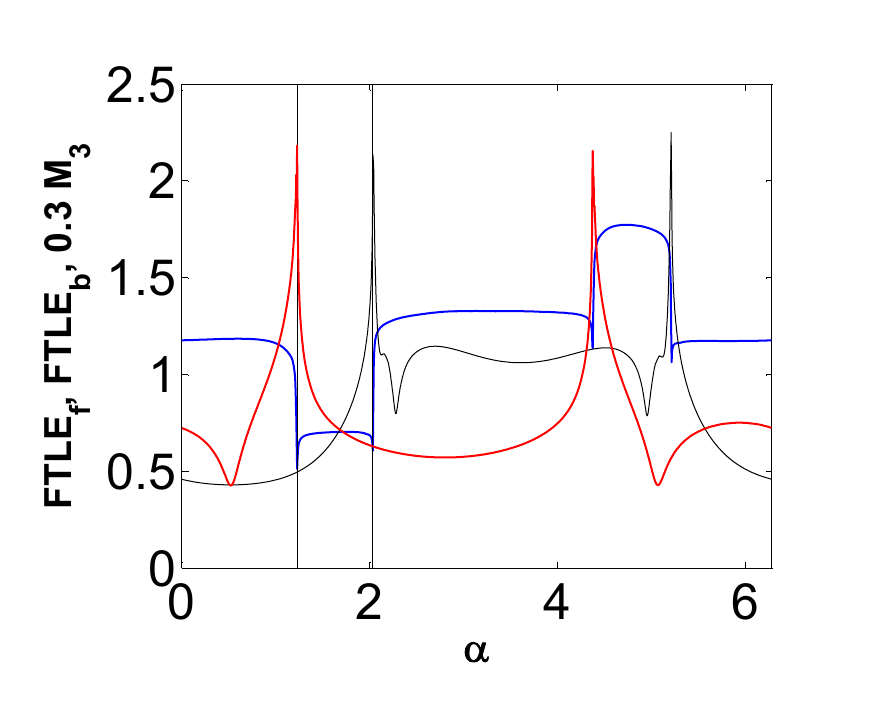}
\caption{\label{qgncirL} a) Forwards FTLE (black), backwards FTLE  (red) and $M_3$ (blue) versus $\alpha$ for $\tau=45$; b) forwards FTLE (black), backwards FTLE  (red) and $M_3$ (blue) versus $\alpha$ for $\tau=65$; c)  forwards FTLE (black), backwards FTLE  (red) and $M_3$ (blue)
 versus $\alpha$ for $\tau=95$}
\end{figure}

 Figure    \ref{qgncirL} contrasts the ability of FTLE to locate the stable and unstable subspaces of the DHT with that of the Lagrangian descriptors.   For comparison purposes, we choose $M_3$.
For reference, the angle  positions of the stable and unstable subspaces are marked with vertical lines.
Panel \ref{qgncirL}a) shows the forward and backward FTLE and $M_3$   versus $\alpha$ for $\tau=45$. The Lagrangian descriptor $M_3$ is represented by a blue line.
Its sharp features  mark  accurately the position of the stable and unstable subspaces.
The forward FTLE, represented by the thin black line, at this stage presents pronounced features, although not yet sharp.
 Deviations from the correct position of the stable subspace are significant. and the performance of forward  FTLE  is clearly lower than that of $M_3$.
The backward FTLE, represented by the red line,  accurately detects the unstable subspace with a similar performance to that of $M_3$. Panel \ref{qgncirL}b) presents results for $\tau=65$. However, deviations of the forward FTLE field (black line)
with respect to the stable subspace are still noticeable in this  panel.   Panel \ref{qgncirL}c) confirms that at $\tau=95$  both FTLE and Lagrangian descriptors perform well at locating the angles of the stable and unstable subspaces.

 \section{Computational performance}
 \label{sec:comp_perf}

Computationally the evaluation of Lagrangian descriptors  offer  numerous advantages in comparison with the computation of FTLE, as we discuss next. We explain these advantages in 2D flows, but the argued aspects  are enhanced in 3D flows.
 Calculation of FTLE requires the evaluation of the gradient of the flow map, which is numerically realized by expression
(\ref{flowgran}) in the Appendix.  As indicated there for obtaining the  $\sigma$ field at a grid point  $ {\bf x}^*_{i,j}$ in a 2D flow
it is requested to integrate the dynamics  of four neighbour points $ {\bf x}^*_{i-1,j}, {\bf x}^*_{i+1,j}, {\bf x}^*_{i,j-1},{\bf x}^*_{i,j+1} $ which must be a  small distance
$\epsilon$  from $ {\bf x}^*_{i,j}$ in order to keep valid expression (\ref{flowgran})
in the range of integration $(t^*, t^*+\tau)$.
Thus in order to be accurate, the numerical approach (\ref{flowgran}) must keep a balance between the distance of
grid points $\epsilon$ and the integration time $\tau$. These parameters must be appropriately tuned.
This tuning is not necessary in the computation of the Lagrangian descriptors, thus making their computation more direct.
On the other hand if the grid size  in which the $\sigma$ field is evaluated is bigger than   $\epsilon$, then for obtaining
its value in a point $ {\bf x}^*_{i,j}$ expression (\ref{flowgran}) indicates that four integrations are necessary, while Lagrangian descriptors require only one. This makes Lagrangian descriptors computationally much more efficient, specially  in geophysical
flows given as  finite time datasets, because in them integrations   require interpolations which are computationally
expensive. Some approaches compute  FTLE taking $\epsilon$ as the grid size in which the  $\sigma$ field
is evaluated. In this case integrations of neighbours are stored to prevent multiple performaces of the same trajectory, and
then only one integration per point $ {\bf x}^*_{i,j}$ is requested.
However this approach has also several computational  disadvantages with respect to the use of Lagrangian descriptors.
One is that it is harder to program because the evaluation of the FTLE at each point requires access to stored information, and
the second is that the grid size is   doomed to be small enough to keep  expression (\ref{flowgran}) accurate in the range $(t^*, t^*+\tau)$. This size can be much smaller that the grid size required to have a smooth representation for the Lagrangian descriptor, and in this case the computation of FTLE is again more expensive.

Another difference between  Lagrangian descriptors and FTLE is that the former provide the stable and unstable manifolds simultaneously in the same output, while the latter require post-processing of the results  into one picture. In the post-processing of the output  typically the results are 'filtered' by using a threshold in the field values (see \cite{brawigg})  and in this processes spurious structures may be removed. In the case of  Lagrangian descriptors an accurate image of the manifold features is obtained
without the need of any 'filtering' process. \cite{amism11} have demonstrated that in applications having both the stable and unstable manifolds simultaneously in one picture is an interesting feature that is essential for the successful  analysis of transport events.

\section{Conclusions and Outlook}
\label{sec:concl}

In this paper we have proposed a general method for revealing geometrical structures in phase space that is valid for aperiodically time-dependent dynamical systems.  The central quantities in our method are referred to as Lagrangian descriptors and are based on finite time integrals of some positive, bounded intrinsic geometrical or physical properties of the dynamical system along trajectories. In particular, these properties are integrated forward and backward in time  on some  interval $(t-\tau, t+\tau)$.  In this way, the Lagrangian descriptors have the capability of revealing both the stable and unstable manifolds in the same calculation. We have shown that the  convergence to geometrical structures of the dynamical system requires the use of a long enough $\tau$. For $\tau$ values below a  threshold, which  depends on each particular dynamical system,
the pattern provided by  the  Lagrangian descriptors is far from the geometrical  structures. However beyond that threshold, for increasing $\tau$, the output of the Lagrangian descriptors becomes more and more complex, revealing more details
of the geometrical structures of the dynamical system. We have given an analytical  justification for these observations for the benchmark case of a linear vector field having a hyperbolic saddle point.

In all examples that we have considered we have carried out simultaneous comparisons of FTLEs and  finite time averages of components of the vector field along trajectories with Lagrangian descriptors. We have shown that Lagrangian descriptors provide superior performance in all cases, in the sense that
they more accurately reveal  the geometrical structures and they require a shorter time to converge to the geometrical structures. For example, we have shown that FTLE give spurious structures for the integrable Duffing equation and fail completely to provide any structure for the linear saddle. Finite time averages of the horizontal component of the vector field exhibit the same deficiencies as FTLEs for these particular two examples. We have also shown that Lagrangian descriptors accurately provide the geometrical structures for a three dimensional flow. We have compared these results to results obtained from FTLE  calculations and finite time averages of a component of the vector field along trajectories and we have seen that Lagrangian descriptors provide a more accurate representation of the geometrical features of the flow than these techniques provide.
We have examined the issue of the convergence time to the stable and unstable subspaces of a DHT for a velocity field given as a data set and have shown that Lagrangian descriptors converge more rapidly to these subspaces than FTLE. This is promising from the applications point of view, in which  reductions for the forecast time of the velocity field is desirable. Lagrangian descriptors can give more accurate predictions with less information in time.
Finally we have shown that computationally Lagrangian descriptors provide several advantages over FTLE. In general, their use leads to a reduction in CPU requirements by a factor of 4, their implementation in code is simpler, and they require no ''parameter tuning''. Moreover, Lagrangian descriptors provide both the stable and unstable manifolds in one output. In order to achieve this with FTLE post-processing of the stable and unstable is required in order to combine the two outputs into one picture at the same time.

%Finally, we note that we have recently learned of work in the computer graphics community that bears some similarity to some of the ideas described in our work; see \cite{acar12, shi08, cabral93}. It is particularly interesting that this work uses the technique of ''domain filling'', which was used in the atmospheric science community in the early 90s, see  \cite{oneill94, sutton94}.

\section*{Acknowledgements}
Authors acknowledge  CESGA for  support with the SC  FINIS TERRAE and to Centro de Computacion Cientifica of UAM for computing support.
This research was supported by the Spanish Ministry of Science under grants MICINN-MTM2008-03754, MTM2011-26696, I-Math C3-0104 and ICMAT Severo Ochoa project SEV-2011-0087. Authors acknowledge support from CSIC under grants ILINK-0145 and OCEANTECH.
SW  acknowledges the support of the  Office of Naval Research (Grant No.~N00014-01-1-0769).

\bibliography{LD}

\appendix

\section{Numerical computation of Lagrangian descriptors}
\label{appA}

In this appendix we describe the numerical computation  of general     Lagrangian  descriptors:

\begin{equation}
M({\bf x^*},t^*)_{ {\bf v},\tau}=  \int^{t^*+\tau}_{t^*-\tau} \mathcal{F}({\bf x}(t)) \, dt. \label{ld2}
\end{equation}

\noindent
where  $\mathcal{F}({\bf x}(t))$ is a positive bounded scalar representing a  geometrical or physical property that is evaluated along a trajectory. This scalar  properties introduced in this paper depend on
vector quantities such as velocity, acceleration, the time derivative of acceleration or  combinations of these quantities (e.g. as in the example using curvature).  For example, for the Lagrangian  descriptor $M_1$,  Eq. \eqref{def:Mgen}, is evaluated by integrating the function  $\|{\bf v}({\bf x}(t),t)\|$ along a trajectory, $\bf{x} (t), \, {\bf x}(t^*) = {\bf x}^*$ from  from $t^*-\tau$ to $t^*+\tau$.

Geometrically, the computation of $M_1$  is equivalent to evaluating the area $A$ below the graph $\|{\bf v}({\bf x}(t),t)\|$ in the specified time interval. The area $A$ is obtained from the following one dimensional dynamical system:

\begin{equation}
\frac{d Y}{dt}=\|{\bf v}({\bf x}(t),t)\|. \label{1dds}
\end{equation}

\noindent
For the initial condition $Y(t^*)=0$, the  area $A$ is given by the value of $Y$ at $t^*+\tau$ minus  the value of $Y$ at $t^*-\tau$ ,  \emph{i.e.},   $Y(t^*+\tau)-Y(t^*-\tau)=A$.

The integration of the system (\ref{1dds}) is performed with a 5th order variable time step Runge-Kutta method, in particular with the subroutine {\tt rkqs} described in \cite{nr}. A peculiarity of this differential equation  is that it depends on $t$ both explicitly and implicitly through the trajectory of \eqref{gds}. A Runge-Kutta step from $t_0$ to $t_1$ applied to Eq. (\ref{1dds}) requires evaluating $\|{\bf v}\|$ along trajectories of \eqref{gds}  at intermediate steps $t_0+\Delta t$. To this end
the argument ${\bf x} (t)$ that must be passed to  $\|{\bf v}\|$ at time $t_0+\Delta t$
must be obtained by evolving the trajectory of \eqref{gds}  from ($t_0, {\bf x}(t_0)$) to ($t_0+\Delta t, {\bf x}(t_0+\Delta t)$).
We remark that this adaptive   method can be faster than the method proposed in  \cite{chaos} which is based on fixed time step integrations. Moreover, this method is quite versatile, since  from one descriptor to another it is only the right hand side in Eq. (\ref{1dds}) which needs to be modified.

In order to evaluate  the entire family of descriptors presented in this paper three different types  of vectors must be obtained:  velocity, acceleration and the time derivative of acceleration.
 Once these are obtained individually, their  combinations relevant to the particular type of descriptor of interest  are straightforwardly computed. Velocity is a vector which is directly provided if  Eqs. (\ref{gds}) are given; the others are obtained as we explain next.
 The   general expression for ${\bf a}({\bf x},t )$  is as follows:

   \begin{equation}
{\bf a}({\bf x},t)=\frac{d {\bf v}}{dt}=\frac{\partial {\bf v}}{\partial t}+\frac{\partial {\bf v}}{\partial x}v_x+\frac{\partial {\bf v}}{\partial y}v_y\label{eq1}
\end{equation}

\noindent
which is easily computed if there is an exact expression for the velocity field   {\bf v}. If the velocity field is given
 as a finite time data set, Eq. (\ref{eq1}) is evaluated  by providing a discrete  version of the equation:

 \begin{equation}
{\bf a}=\frac{d {\bf v}}{dt}=\frac{d^2 {\bf x}}{dt^2} \label{acc}
\end{equation}

\noindent
as follows:

 \begin{equation}
{\bf a}({\bf x}(t),t)=\frac{{\bf v}({\bf x}(t+h),t+h)-{\bf v}({\bf x}(t-h),t-h)}{2h} + \mathcal{O}(h^2).\label{accn}
\end{equation}

\noindent
The time derivative of the acceleration is given by:

\begin{eqnarray}
\frac{d{\bf a}({\bf x},t)}{dt}&=&\frac{d}{dt}\frac{d {\bf v}}{dt}=\frac{\partial^2{\bf v}}{\partial t^2}+\frac{\partial^2 {\bf v}}{\partial t \partial x}v_x+\frac{\partial {\bf v}}{ \partial x}\frac{\partial v_x}{\partial t} +\frac{\partial^2 {\bf v}}{\partial t\partial y}v_y+\frac{\partial {\bf v}}{\partial y}\frac{\partial v_y}{\partial t}+  \nonumber \\ &&
\frac{\partial^2{\bf v}}{\partial x \partial t}+\frac{\partial^2 {\bf v}}{\partial x^2}v_x+\frac{\partial {\bf v}}{ \partial x}\frac{\partial v_x}{\partial x} +\frac{\partial^2 {\bf v}}{\partial x\partial y}v_y+\frac{\partial {\bf v}}{\partial y}\frac{\partial v_y}{\partial x}+   \nonumber  \\
&&\frac{\partial^2{\bf v}}{\partial y \partial t}+\frac{\partial^2 {\bf v}}{\partial y \partial x}v_x+\frac{\partial {\bf v}}{ \partial x}\frac{\partial v_x}{\partial y} +\frac{\partial^2 {\bf v}}{\partial y^2}v_y+\frac{\partial {\bf v}}{\partial y}\frac{\partial v_y}{\partial y}.\label{eq2}
\end{eqnarray}

\noindent
If the velocity is supplied  as a finite time data set, Eq. (\ref{eq2}) is evaluated  by providing a discrete  version to the equation:

\begin{equation}
\frac{d{\bf a}}{dt}=\frac{d^2 {\bf v}}{dt^2}.
\label{dtacc}
\end{equation}

\noindent
which reads as follows:

\begin{equation}
\frac{d{\bf a}({\bf x}(t),t)}{dt}=\frac{{\bf v}({\bf x}(t+h),t+h)-2{\bf v}({\bf x}(t),t)  +{\bf v}({\bf x}(t-h),t-h)}{2h} + \mathcal{O}(h^2).\label{dtaccn}
\end{equation}

\noindent
For the results reported in this article  Eqs. (\ref{accn}) and (\ref{dtaccn}) are implemented  where needed using $h=10^{-4}$.

\section{Computation of FTLE's}
\label{appB}

In this appendix we give a brief discussion of the computation of FTLEs, following \cite{lsm07}. The FTLE field is denoted by $\sigma_{t^*}^{\tau}({\bf x}^*)$, and is computed from the maximum eigenvalue $\lambda _{\rm   max}$  of the matrix $\Delta$:

\begin{equation}
\sigma_{t^*}^{\tau}({\bf x}^*)=\frac{1}{\tau}\ln \sqrt{  \lambda _{\rm   max} (\Delta) },
\end{equation}

\noindent
where the matrix $\Delta$ is defined from  the flow map, as we now describe. Consider a point ${\bf x}^*$ at time $t^*$, and let it be advected by the flow map, $\phi$,   from $t^*$ to $t^*+\tau$:

\begin{equation}
{\bf x} ^*\to \phi^{t^*+\tau}_{t^*}({\bf x}^*).
\end{equation}

\noindent
The gradient of the flow map defines the matrix $N$:

\begin{equation}
N=\frac{d \phi^{t^*+\tau}_{t^*}({\bf x}^*)}{d {\bf x}^*}
\end{equation}

\noindent
Using the convention that $N^T$ denotes the transpose of $N$, $\Delta$ is the symmetric matrix:

\begin{equation}
\Delta=N^T N
\end{equation}

 Note that in two dimensions  the  gradient of the flow map   at a grid point  $ {\bf x}^*_{i,j}=(x_{i,j},y_{i,j})$
can be computed central differences as follows (see \cite{shaddenweb}):

\begin{equation}\label{flowgran}
\frac{d \phi^{t^*+\tau}_{t^*}({\bf x}^*)}{d {\bf x}^*}\Bigg \vert_{{\bf x}^*_{i,j}}=\left( \begin{array}{cc}  \frac{x_{i+1,j}(t^*+\tau)-x_{i-1,j}(t^*+\tau)}{x_{i+1,j}(t^*)-x_{i-1,j}(t^*)} &  \frac{x_{i,j+1}(t^*+\tau)-x_{i,j-1}(t^*+\tau)}{y_{i,j+1}(t^*)-y_{i,j-1}(t^*)} \\
 \frac{y_{i+1,j}(t^*+\tau)-y_{i-1,j}(t^*+\tau)}{x_{i+1,j}(t^*)-x_{i-1,j}(t^*)}&\frac{y_{i,j+1}(t^*+\tau)-y_{i,j-1}(t^*+\tau)}{y_{i+1,j}(t^*)-y_{i-1,j}(t^*)} \end{array} \right)
\end{equation}

\noindent
where $ (x_{ij}(t^*+\tau),y_{ij}(t^*+\tau))=\phi^{t^*+\tau}_{t^*}( {\bf x}^*_{i,j}(t^*))$.

\section{Regularity of Lagrangian descriptors}
\label{appC}

In this appendix we discuss some issues  associated with the regularity of Lagrangian descriptors, especially when dealing with velocity fields defined as data sets, as discussed in Section \ref{sec:veldata}.

In this case, since  the data is discrete in space and time,
in order to have a continuous representation for the dynamical system,

\begin{eqnarray}
\frac{d{\bf x}}{dt} &=& {\bf v}({\bf x},t),  \, \,{\bf x} \in \, \mathbb{R}^2, \, t \in \, \mathbb{R},
\label{advf}
\end{eqnarray}

\noindent
the velocity field ${\bf v}$  must be interpolated in space and time in order to compute  particle trajectories. In \cite{cf} it was shown that
 bicubic interpolation in space and 3rd order Lagrange polynomials in
time is sufficiently accurate for our purposes.  This interpolation is $C^1$ continuous in space and $C^0$ continuous in time.

\begin{figure}
a)\includegraphics[width=4.6cm]{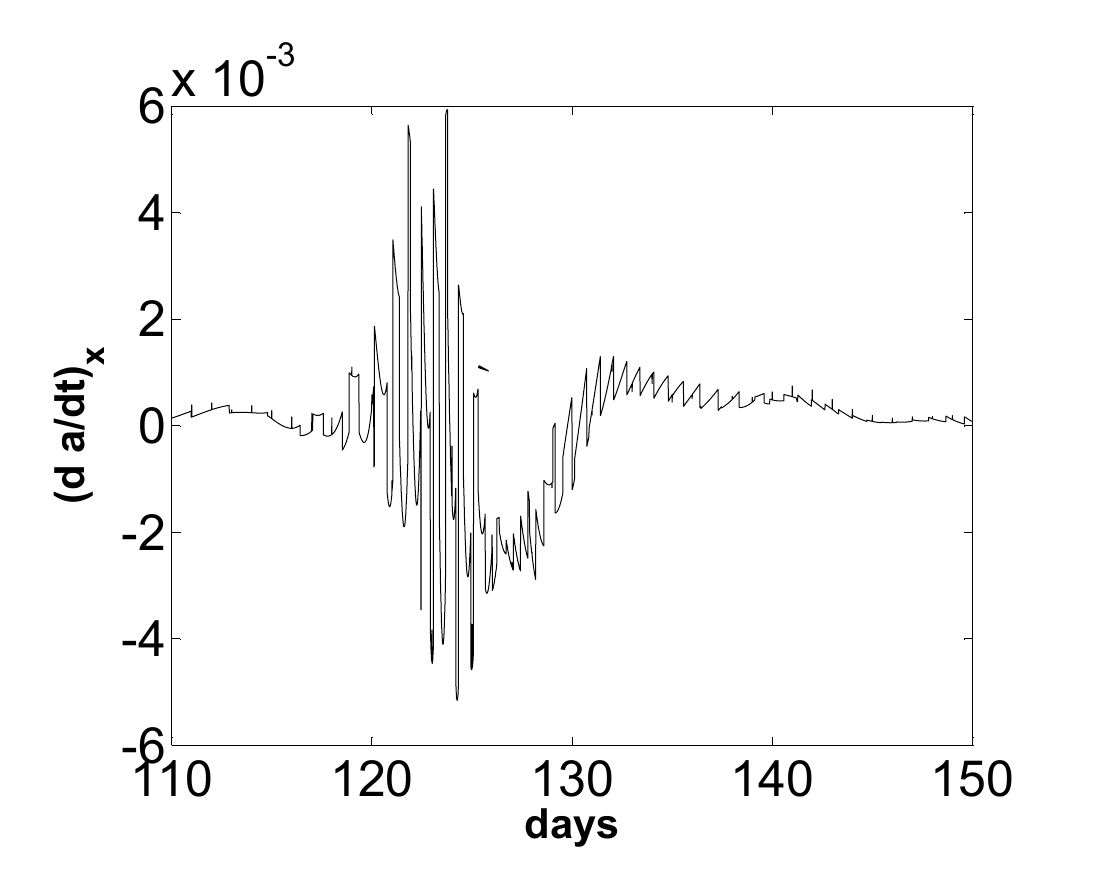}
b)\includegraphics[width=4.6cm]{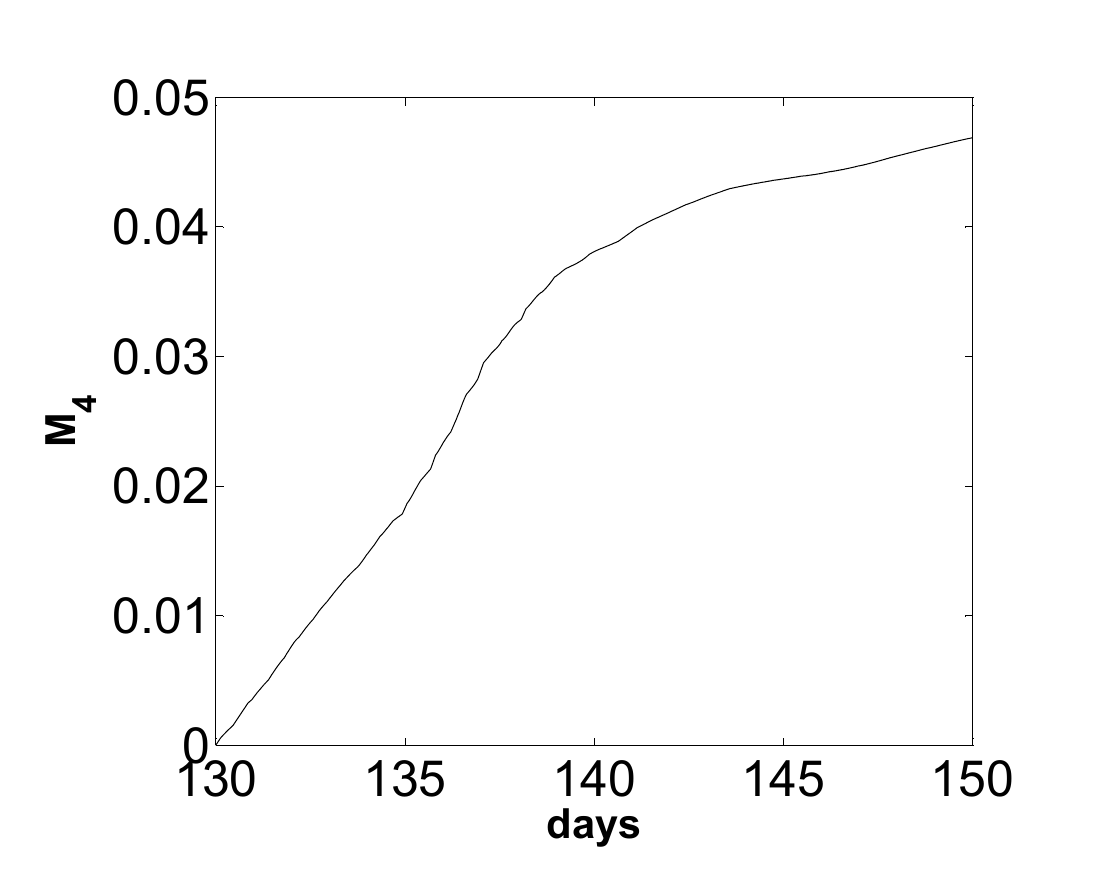}
c)\includegraphics[width=4.6cm]{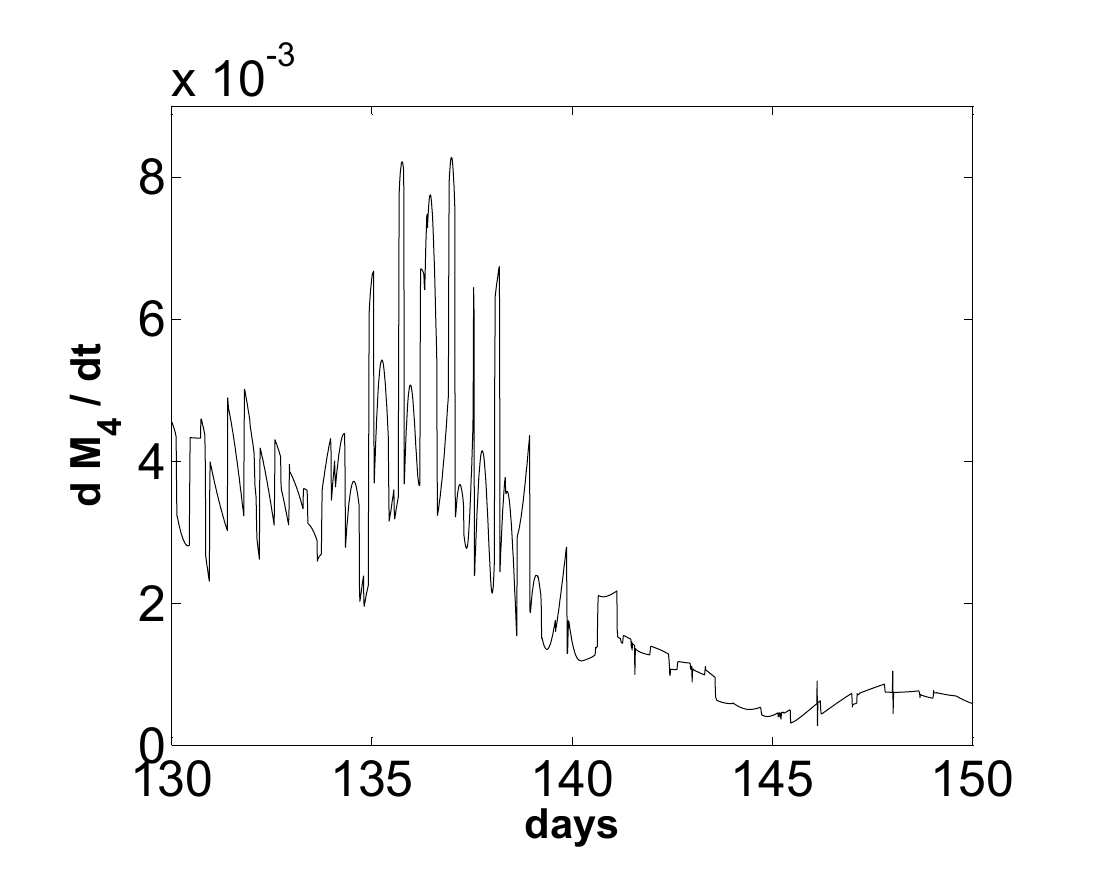}
\caption{\label{dadttry} a) Evaluation of the $x$ component of the  vector   $d{\bf a}/dt$  along a trajectory from   $t-\tau$ to $t+\tau$;  b) evolution of  $M_4$ in the same time interval; c) evolution of  the time derivative of $M_4$.}
\end{figure}

Lagrangian descriptors are sensitive to the quality of the interpolation of the data field, as are FTLE.
Regarding Lagrangian descriptors, we focus first on those involving the time derivative of the acceleration.
Following Eq. (\ref{eq2}), the time derivative of acceleration along trajectories involves second order spatial derivatives of a velocity field   (\ref{advf}) which are not so regular.
 Figure  \ref{dadttry}a) illustrates the evaluation of the time derivative of acceleration along a trajectory
 for the system   (\ref{advf}) with initial condition $t=130\,\, {\rm days}, x=410\,\, {\rm km}, y=1100 \,\,{\rm km}$. The integration is performed  backwards
from $t$ to  $t-\tau$ and forwards from $t$ to $t+\tau$, with $\tau=20$. In particular, this figure shows the evolution of the horizontal component of this vector.

\begin{figure}
a)\includegraphics[width=4.6cm]{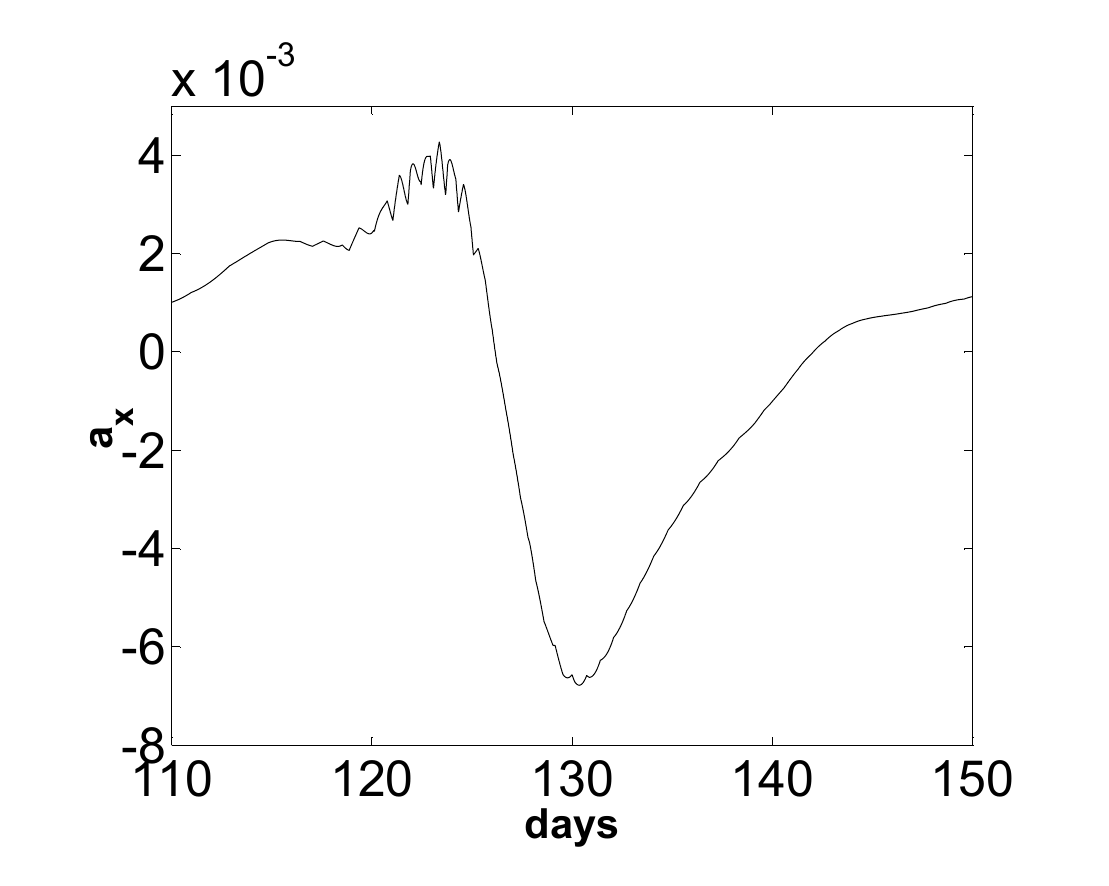}%acx.fig
b)\includegraphics[width=4.6cm]{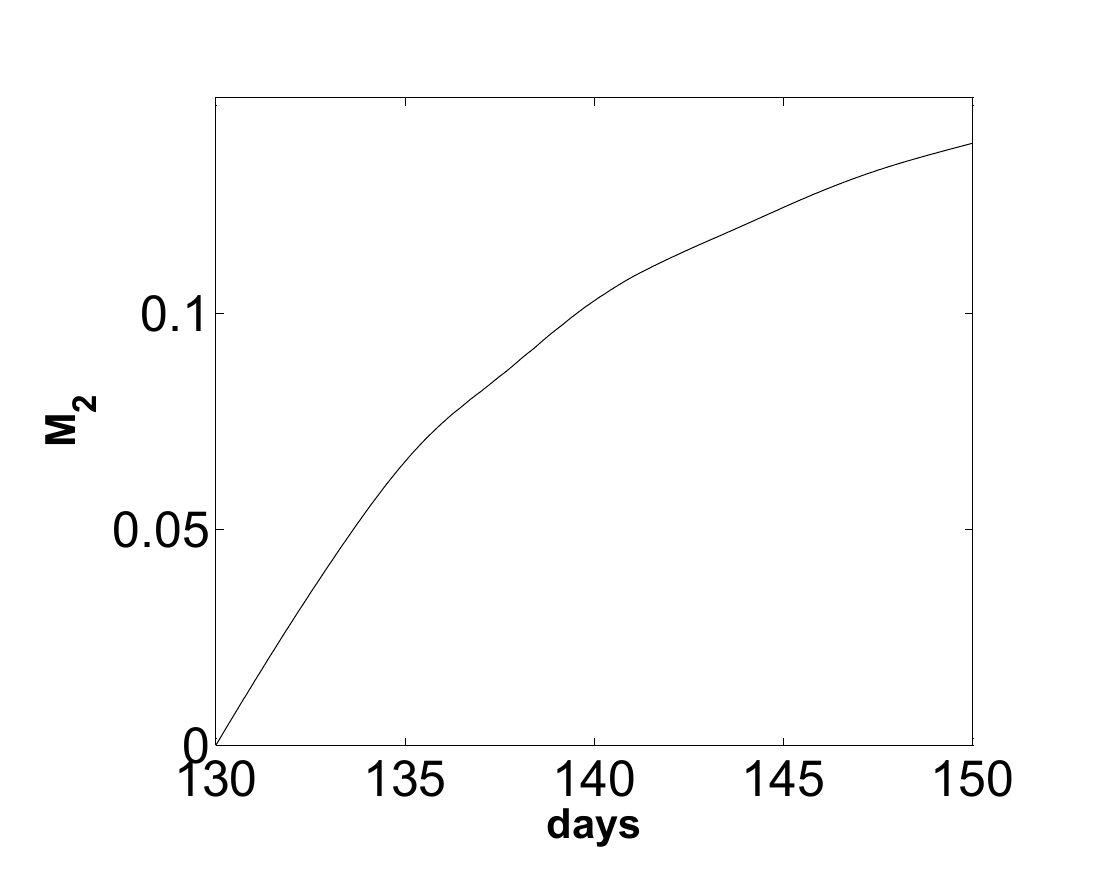}
c)\includegraphics[width=4.6cm]{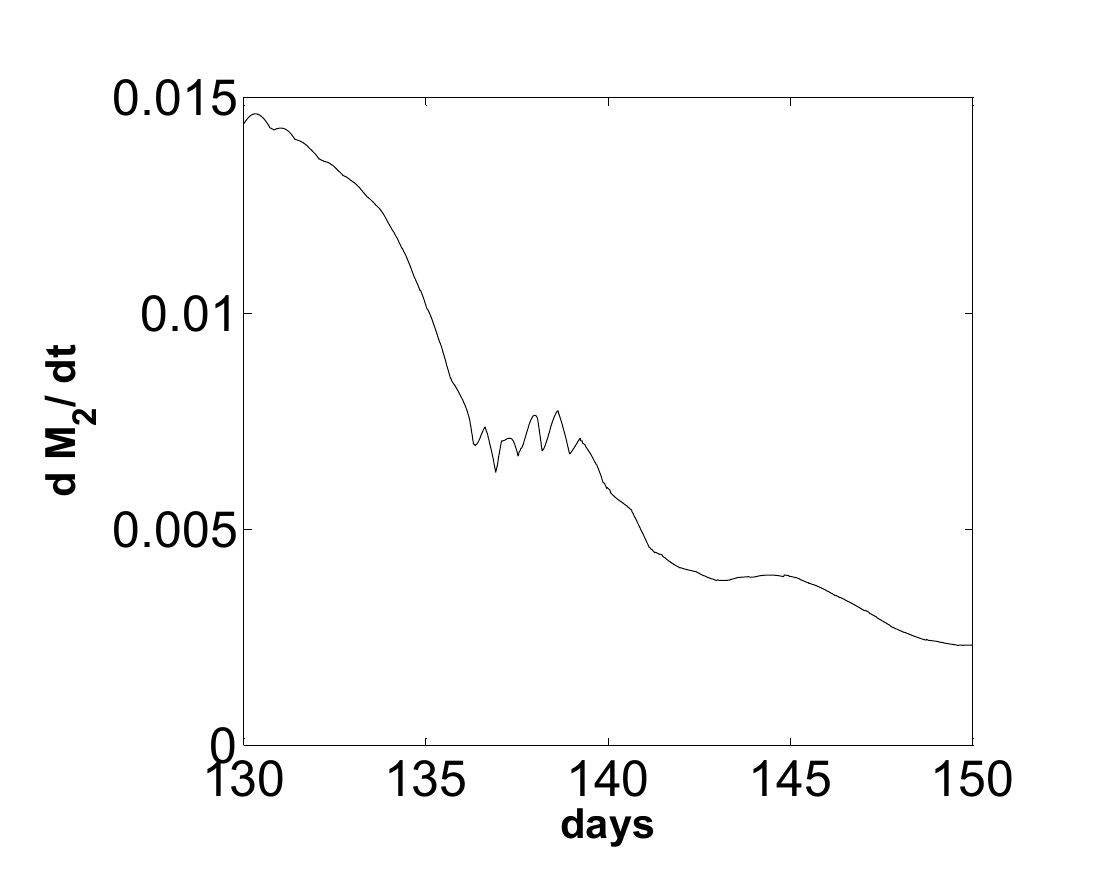}\\
\caption{\label{adttry}  Evaluation of the $x$ component of the  vector   ${\bf a}$  along a trajectory from   $t-\tau$ to $t+\tau$;  b) evolution of  $M_2$ in the same time interval; c) evolution of  the time derivative of $M_2$.}
\end{figure}

\noindent
Despite the lack of regularity displayed in Figure \ref{dadttry}a), the actual descriptor is more regular since it is based on integrals over these functions. Figure \ref{dadttry}b) shows the evolution versus $\tau$ of $M_4$. Figure \ref{dadttry}c)
displays the time derivative of $M_4$, confirming that its regularity is comes from that of  $d{\bf a}/dt$.
Expression (\ref{eq1}) shows that the evaluation of acceleration along trajectories involves first order spatial derivatives of the velocity field   (\ref{advf}). Figure \ref{adttry}a)
shows the evolution of the horizontal component of acceleration, which  confirms that the regularity of the interpolator of  the underlying velocity field is observable.  The actual regularity of $M_2$
is larger than that of acceleration as it is based on integrals over these quantities. Figure \ref{adttry}b)  shows  the evolution versus $\tau$ of $M_2$ as a more regular curve, and in c) the time derivative of $M_2$, confirms
that its regularity  comes from that of  ${\bf a}$.

\begin{figure}
a)\includegraphics[width=4.6cm]{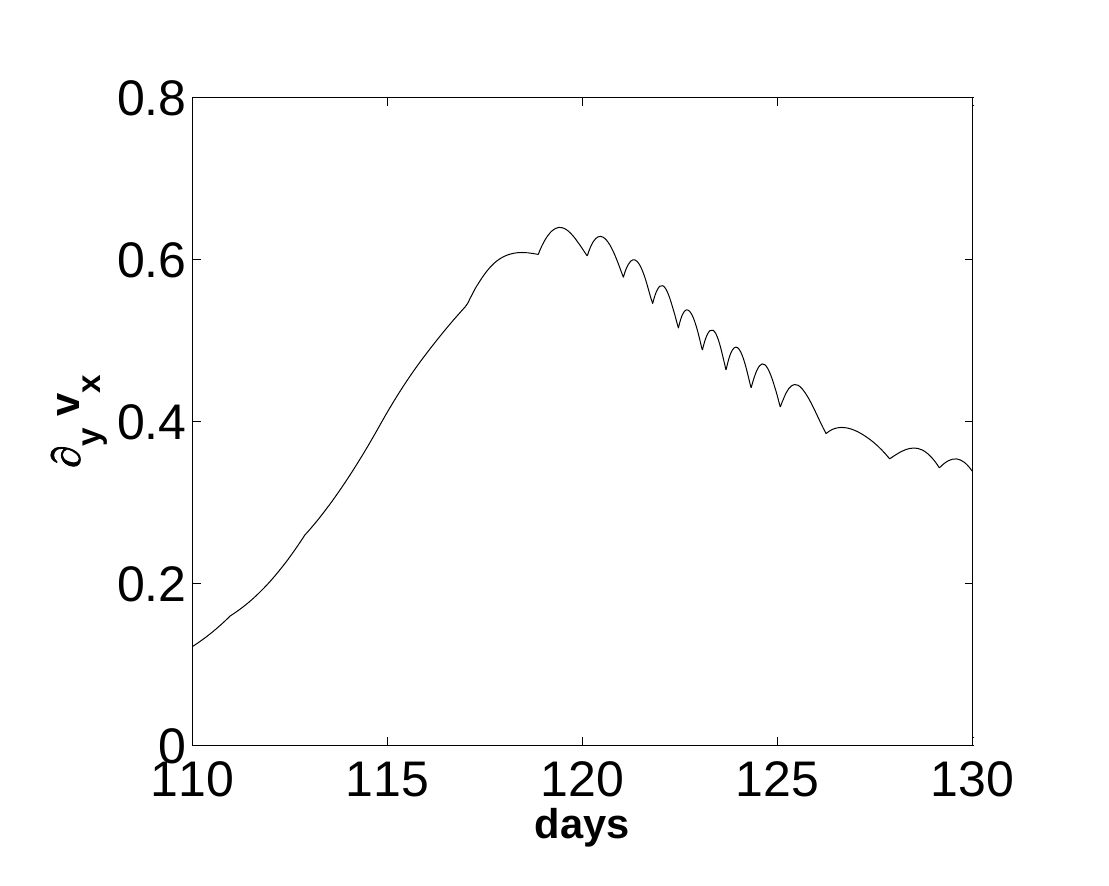}
b)\includegraphics[width=4.6cm]{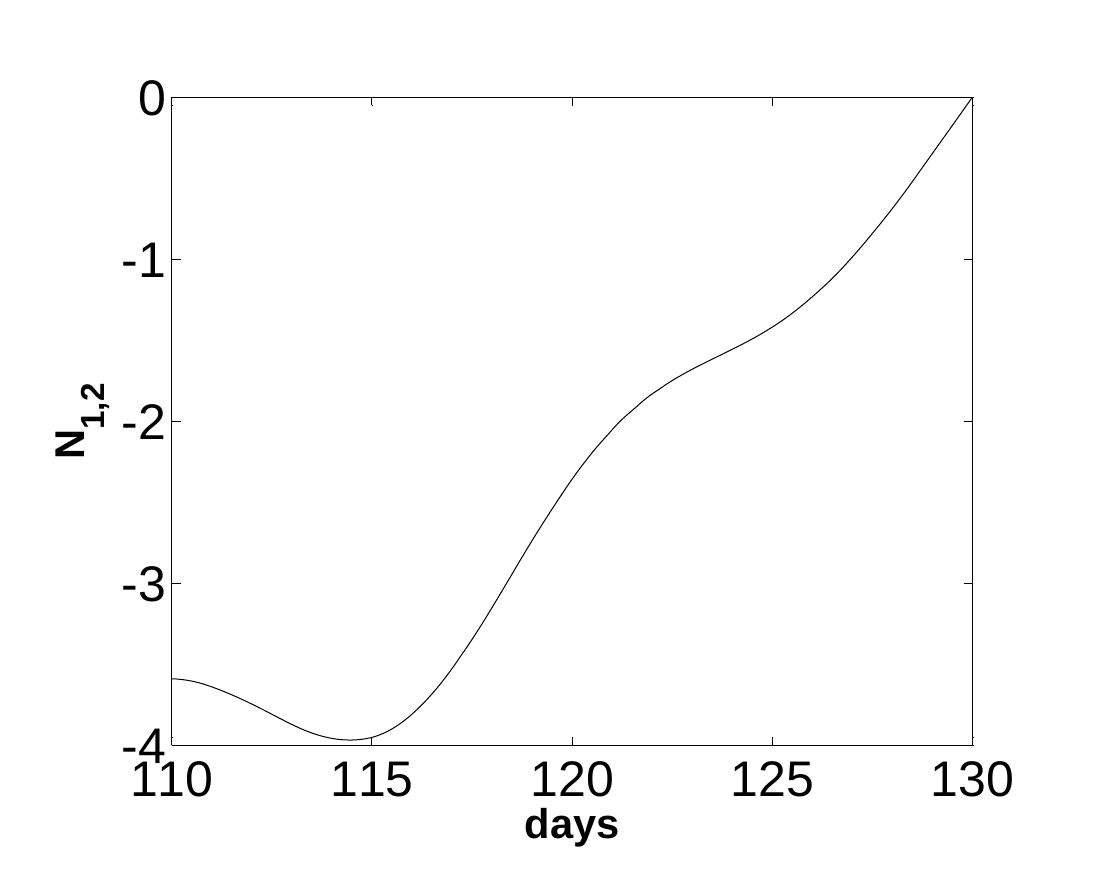}
c)\includegraphics[width=4.6cm]{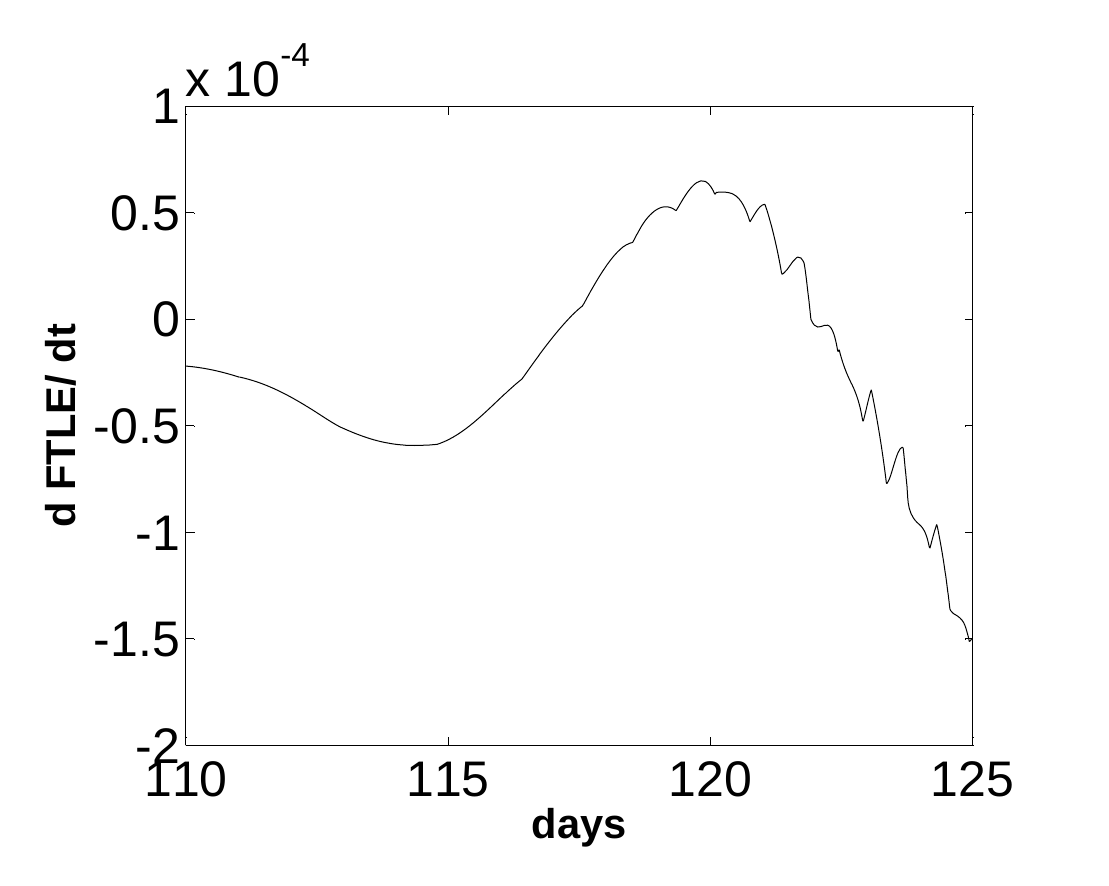} \\
\caption{\label{Ly_tr} a) Evolution of $\partial_y v_x ({\bf x}(t))$ versus $t$ for backwards intergation;  b) evolution of  the  $N_{1,2}$ component versus $t$ for backwards integration; c) evolution of the time derivative of the backwards FTLE.}
\end{figure}

The performance of FTLE is similar to the performance of Lagrangian descriptors involving the evaluation of acceleration, as they require the evaluation of the gradient of the flow map. The gradient of the
flow map is based on integrations on linear equations which are obtained by linearizing the velocity field    (\ref{advf})  along a trajectory. This involves evaluating first order spatial derivatives, as in the case of  acceleration.
Figure \ref{Ly_tr}a) shows the partial derivative of  $v_x$ with respect to $y$  along the same  trajectory as above, for backwards integration. The lack of regularity is noticed, however as the  FTLE computation is based on integrations along this linearized field, it gains regularity and for instance in Figure  \ref{Ly_tr}b) we show the evolution of one of the components of the matrix $N$ used to compute the FTLE. Again in Figure  \ref{Ly_tr}c) illustrates how the regularity of the backwards FTLE is governed  by the regularity of linearized velocity fields.
Finally we note that descriptors based on the integration of the velocity are the most regular. They are advantageous in this regard not only to those LD involving acceleration or its time derivative, but also to FTLE.

The discussed  reasons suggest that for analyzing velocities fields  given as  data sets, the choice of Lagrangian descriptors  involving $d{\bf a}/dt$ may be less appropriate  than those involving ${\bf v}$ or ${\bf a}$ since they require interpolators with a higher order of regularity than is required by  the latter quantities. On the other hand, for data sets interpolated with not too regular interpolators, Lagrangian descriptors based on ${\bf v}$ are the better choice, even with respect to  FTLE.

\end{document}